\documentclass[apj]{emulateapj}
\usepackage{apjfonts}

\newcommand{\bvi}{\mbox{$BV\!I$}}		% BVI
\newcommand{\bvik}{\mbox{$BV\!IK$}}		% BVIK
\newcommand{\ubvri}{\mbox{$U\!BV\!RI$}}		% UBVRI
\newcommand{\beff}{\mbox{$B_{\rm eff}$}}	% B_eff
\newcommand{\veff}{\mbox{$V_{\rm eff}$}}	% V_eff
\newcommand{\btot}{\mbox{$B_{\rm T}$}}		% B_T
\newcommand{\vtot}{\mbox{$V_{\rm T}$}}		% V_T
\newcommand{\itot}{\mbox{$I_{\rm T}$}}		% I_T
\newcommand{\ktot}{\mbox{$K_{\rm T}$}}		% K_T
\newcommand{\bvzero}{\mbox{$(B\!-\!V)_0$}}	% (B-V)_0
\newcommand{\bvtot}{\mbox{$(B\!-\!V)_{\rm T}$}}	% (B-V)_T
\newcommand{\bvtotzero}{\mbox{$(B\!-\!V)_{\rm T,0}$}}
\newcommand{\bvedzero}{\mbox{$(B\!-\!V)_{\rm eD,0}$}}
\newcommand{\vi}{\mbox{$V\!-\!I$}}		% V-I
\newcommand{\vizero}{\mbox{$(V\!-\!I)_0$}}	% (V-I)_0
\newcommand{\vitot}{\mbox{$(V\!-\!I)_{\rm T}$}}	% (V-I)_T
\newcommand{\vitotzero}{\mbox{$(V\!-\!I)_{\rm T,0}$}}
\newcommand{\viedzero}{\mbox{$(V\!-\!I)_{\rm eD,0}$}}
\newcommand{\vk}{\mbox{$V\!-\!K$}}		% V-K
\newcommand{\vktotzero}{\mbox{$(V\!-\!K)_{\rm T,0}$}}
\newcommand{\dvindex}{\mbox{$\Delta V_{0.5-3}$}}
\newcommand{\czhel}{\mbox{$cz_{\rm hel}$}}	% Heliocentric radial velocity
\newcommand{\czlg}{\mbox{$cz_{_{\rm LG}}$}}	% Velocity rel. to Local Group
\newcommand{\dvlos}{\mbox{$\Delta v$}}		% Relative LOS velocity diff.
\newcommand{\etal}{et al.}			% Present AJ style
\newcommand{\hbet}{\mbox{${\rm H}\beta$}}	% H-beta
\newcommand{\hgama}{\mbox{${\rm H}\gamma_A$}}	% H-gamma_A
\newcommand{\hdela}{\mbox{${\rm H}\delta_A$}}	% H-delta_A
\newcommand{\hi}{\ion{H}{1}}			% Symbol for neutral hydrogen
\newcommand{\hii}{\ion{H}{2}}			% Symbol for H~II region
\newcommand{\hst}{{\em HST}}			% Abbrev. for Hubble Space Tel.
\newcommand{\kms}{km~s$^{-1}$}			% kilometers per second
\newcommand{\lir}{\mbox{$L_{\rm IR}$}}		% Infrared luminosity
\newcommand{\lsun}{\mbox{$L_{\odot}$}}		% Solar luminosity
\newcommand{\magarcs}{mag arcsec$^{-2}$}	% magnitudes/arcsec^2
\newcommand{\magarcstab}{mag/$\square$\arcsec}  % magnitudes/arcsec^2 f/Tables
\newcommand{\mhi}{\mbox{$M_{\rm H\,I}$}}	% Neutral-hydrogen-gas mass
\newcommand{\mhtwo}{\mbox{$M_{\rm H_2}$}}	% Molecular-gas mass
\newcommand{\msun}{\mbox{$M_{\odot}$}}		% Solar mass
\newcommand{\msunyr}{\mbox{$M_{\odot}$ yr$^{-1}$}}
\newcommand{\mgb}{Mg~$b$}			% Mg b index
\newcommand{\mgfe}{\mbox{${\rm [MgFe]}$}}	% Index [MgFe]
\newcommand{\mub}{\mbox{$\mu_B$}}		% B surface brightness
\newcommand{\muv}{\mbox{$\mu_V$}}		% V surface brightness
\newcommand{\muvhst}{\mbox{$\mu_{V, HST}$}}	% V_HST surface brightness
\newcommand{\muvlco}{\mbox{$\mu_{V, {\rm LCO}}$}}
\newcommand{\mueff}{\mbox{$\mu_{\rm eff}$}}	% Effective surface brightness
\newcommand{\muzero}{\mbox{$\mu_0$}}		% Central surface brightness
\newcommand{\n}{NGC~}				% NGC = New General Catalogue
\newcommand{\nclust}{\mbox{$N_{\rm cl}$}}	% No. of clusters per bin
\newcommand{\nuclust}{\mbox{$\nu_{\rm cl}$}}	% No. of clusters per magnitude
\newcommand{\rdV}{\mbox{$r^{1/4}$}}		% de Vaucouleur's r^1/4
\newcommand{\reff}{\mbox{$r_{\rm eff}$}}	% Effective radius
\newcommand{\reffcl}{\mbox{$r_{\rm eff,cl}$}}	% Eff. radius of cluster system
\newcommand{\rhole}{\mbox{$r_{\rm h}$}}		% Hole radius
\newcommand{\rmax}{\mbox{$r_{\rm max}$}}	% Maximum radius
\newcommand{\sphed}{Sph\,+\,eD}			% Sph+eD (model)
\newcommand{\signcl}{\mbox{$\Sigma_{\rm cl}$}}	% Surface number density of cl
\newcommand{\sigvel}{\mbox{$\sigma_v$}}         % Velocity dispersion
\newcommand{\tnm}[1]{\tablenotemark{#1}}	% Abbreviation used in Table 8
\newcommand{\tnt}[1]{\tablenotetext{#1}}	% Abbreviation used in Table 8
\newcommand{\vnad}{\mbox{$v_{\rm Na~I}$}}	% Velocity from Na D lines
\newcommand{\zsun}{\mbox{$Z_{\odot}$}}		% Solar metallicity
\slugcomment{To appear in {\em The Astronomical Journal}, Vol.\ 133 }

\shorttitle{Remnant of a ``Wet'' Merger: NGC~34}
\shortauthors{Schweizer \& Seitzer}

\begin{document}

\title{REMNANT OF A ``WET'' MERGER:\ \ NGC 34 AND ITS YOUNG MASSIVE
                                                           CLUSTERS, \\
       YOUNG STELLAR DISK, AND STRONG GASEOUS OUTFLOW\altaffilmark{1}}
\altaffiltext{1}{
Based in part on observations with the 6.5 m Magellan Telescopes located
at Las Campanas Observatory, Chile.
}

\author{Fran\c cois Schweizer}
\affil{Carnegie Observatories, 813 Santa Barbara Street, Pasadena, CA
91101; schweizer@ociw.edu}

\and

\author{Patrick Seitzer}
\affil{Department of Astronomy, University of Michigan, 818 Dennison Building,\\
Ann Arbor, MI 48109; pseitzer@umich.edu}

% Notice that each of these authors has alternate affiliations, which
% are identified by the \altaffilmark after each name.  The actual alternate
% affiliation information is typeset in footnotes at the bottom of the
% first page, and the text itself is specified in \altaffiltext commands.
% There is a separate \altaffiltext for each alternate affiliation
% indicated above.
%

% The abstract environment prints out the receipt and acceptance dates
% if they are relevant for the journal style.  For the aasms style, they
% will print out as horizontal rules for the editorial staff to type
% on, so long as the author does not include \received and \accepted
% commands.  This should not be done, since \received and \accepted dates
% are not known to the author.

\begin{abstract}
This paper presents new images and spectroscopic observations of \n34
(Mrk 938) obtained with the du Pont 2.5-m and Baade 6.5-m telescopes at
Las Campanas, plus photometry of an archival $V$ image obtained with
{\em Hubble Space Telescope}.
This $M_V = -21.6$ galaxy has often been classified as a Seyfert 2, yet
recently published infrared spectra suggest a dominant central starburst.
We find that the galaxy features a single nucleus, a main spheroid
containing a blue central disk and much outer fine structure, and tidal
tails indicative of two former disk galaxies.
At present these galaxies appear to have completed merging.
The remnant shows three clear optical signs that the merger was gas-rich
(``wet'') and accompanied by a starburst:
(1) It sports a rich system of young star clusters, of which 87 have
absolute magnitudes $-10.0 \geq M_V \geq -15.4$.
Five clusters with available spectra have ages in the range 0.1--1.0 Gyr
and photometric masses of $2\times 10^6 \la M \la 2\times 10^7\,\msun$;
they are gravitationally bound young globulars.
(2) The blue central disk appears to be young.  It is exponential, can be
traced to $\ga$10 kpc radius, and has a smooth structure and colors
suggesting that its optical light is dominated by a $\sim$400~Myr old
poststarburst population.
And (3), the center of \n34 drives a strong outflow of cool, neutral gas,
as revealed by broad blueshifted \ion{Na}{1} D-lines.
The center-of-line velocity of this gas is $-$620 \kms, while the maximum
detected outflow velocity reaches $-$1050 \kms.
Assessing all available evidence, we suggest that \n34 stems from two
recently merged gas-rich disk galaxies with an estimated mass
ratio of $1/3 \la m/M \la 2/3$.
The remnant seems to have first experienced a galaxy-wide starburst that
then shrank to its current central and obscured state.
The strong gaseous outflow came last.
\end{abstract}

% The different journals have different requirements for keywords.  The
% keywords.apj file, found on aas.org in the pubs/aastex-misc directory, 
% contains a list of keywords used with the ApJ and Letters.  These are 
% usually assigned by the editor, but authors may include them in their 
% manuscripts if they wish. 


\section{INTRODUCTION}
\label{sec1}

Over the past decade it has become widely accepted that hierarchical
mergers play an important role in galaxy evolution \citep{white78}.
Most such mergers involve gas and the formation of stars since, even
at the present epoch, the great majority of galaxies contain cold gas
\citep{tt72,lt78,schw83}.
Thus, gas-rich, ``wet'' mergers are an integral part of galaxy formation
and growth \citep{bh92,bh96,mh96}.
Many interesting processes occur during these mergers, leading to
phenomena such as starbursts and active galactic nuclei (AGN).

Whereas star and star-cluster formation during merger-induced starbursts
are beginning to be better understood, the processes associated with the
formation and growth of central black holes, and with their feedback on the
surrounding gas, remain mysterious \citep[e.g.,][]{spri05}.
Yet, the existence of a strong correlation between black-hole mass and
spheroid mass \citep{ferr00,gebh00,trem02} compels us to believe that, no
matter what the details of gas-rich mergers and accretions, the physical
processes dictating black-hole growth may be surprisingly uniform and well
defined \citep{hopk06}.

Some questions of interest are:  What is the sequence of events leading up
to the formation of a central starburst and AGN?
When does gaseous outflow begin and when does it peak?
And what are the conditions under which quenching of further star formation
may occur?
Observational answers to these questions can be sought through statistical
studies of large samples of relatively distant objects or through more
detailed studies of individual nearby mergers and merger remnants, where the
relevant processes can be observed in detail.

The present paper takes the latter approach and addresses mainly the first
two of the above questions.
The galaxy \n34 was first pointed out to us as a likely merger remnant
by Dr.\ Christopher Mihos.
It is a relatively nearby representative of the class of Luminous
Infrared Galaxies (LIRGs)---defined as having infrared luminosities \lir\
(from $\sim$8--1300 \micron) in the range
$11.0\leq \log(\lir/\lsun) < 12.0$ \citep{soif87}---and has, therefore,
been studied extensively in the infrared.
Its logarithmic infrared luminosity of $\log(\lir/\lsun) = 11.54$
\citep{chin92} places it in the mid-range of this class.\footnote{
Smaller values of $\log(\lir/\lsun)\approx 11.3$--11.4 are often
quoted for \n34, but are usually based only on the 60 \micron\ and
100 \micron\ fluxes observed by {\em IRAS}.}
The galaxy has been noted for its exceptionally high central luminosity
density in the $H$ passband \citep{vdma00,miho01} and has recently
attracted attention for its feature-rich IR spectrum.
In the near IR (0.8--2.4 \micron), relatively weak forbidden emission
lines from [\ion{C}{1}], [\ion{S}{3}], and [\ion{Fe}{2}] are present in
addition to the usual permitted lines of \ion{H}{1}, H$_2$, and \ion{He}{1}
\citep{riff06}, while in the mid-IR (5--20 \micron) the spectrum is
dominated by strong emission lines from polycyclic aromatic hydrocarbons
(PAHs) and a broad, deep Silicate absorption trough centered at
$\sim$10 \micron\ \citep{buch06}.
In fact, based on the latter features Buchanan et al.\ declare \n34
to be the archetype of a large group of (mostly Seyfert) galaxies with
very red nuclear continua suggestive of cool dust and strong PAH emission
lines.

The exact nature and origin of \n34's {\em optical} nuclear spectrum have
long been controversial.
The present tendency is to place \n34 in a transition category of objects
with nuclear spectra between starburst-dominated and of type Seyfert 2
(= Sey 2; e.g., \citealt{gonc99}).  This transition category is
occasionally called AMB for ``ambiguous'' \citep[e.g.,][]{corb03}.

In the past, many observers have classified \n34 as a Sey 2
\citep[e.g.,][]{afan80,daha85,vero86,gold97a}.
Yet, others have emphasized the apparent weakness of the
[O III]\,$\lambda$5007 emission line relative to either H$\beta$ or H$\alpha$
and have classified \n34 as a narrow-emission-line galaxy
\citep{oste83,veil87,gold97b}
or starburst galaxy
(e.g., \citealt{mazz91}; \citealt{mazz93}; \citealt{mulc96}; near IR:
\citealt{riff06}).
In one illustrative example of the ambivalence of spectral classifiers,
\citet{vero86} describe \n34 as ``indeed a Seyfert 2 galaxy''
based on their ESO 3.6-m telescope CCD spectrum, yet display this spectrum
under the non-Seyfert category ``Miscellaneous'' and label it with ``\hii''
(for starburst).

Modern studies of the possible compositeness of \n34's nuclear spectrum
have led to estimates for the relative contributions to the bolometric flux
of the starburst and AGN ranging between about 75\%/25\% \citep{iman04}
and 90\%/10\% \citep{gonc99}.
What all modern studies involving infrared flux measurements agree on
is that the central star formation rate (SFR) in \n34 is high, with
estimates ranging from about 50 \msunyr\ \citep{vald05} to
80--90 \msunyr\ \citep{prou04}.

There is independent evidence for the likely presence of an AGN in
\n34 from the measured X-ray luminosity which, corrected for strong and
uncertain absorption, is
$L_{\rm X,\,2-10 keV}\approx 2.2_{-0.9}^{+2.8}\times 10^{42}$ erg s$^{-1}$
(\citealt{guai05}; 1-$\sigma$ errors computed by Dr.\ Jane Rigby, private
commun.).
This luminosity, measured from {\em XMM-Newton} observations, is comparable
to that of the classical Seyfert 1.5 galaxy \n4151 and is too high to
be explained in terms of a pure starburst and its associated X-ray binaries.
Thus, it seems to support the presence of a weak AGN, even though there
is no high-resolution observation by {\em Chandra} to directly confirm the
presence of a dominant central X-ray point source.

\n34 is definitely gas rich and can, therefore, sustain its strong central
starburst and present mild AGN activity for some time to come.
Compilations of available measurements yield a neutral-hydrogen gas mass of
$\mhi\approx 5.3\times 10^9 \msun$ \citep{kand03} and a molecular-gas
mass (from CO observations) of
$\mhtwo= (7\pm 3)\times 10^9 \msun$ \citep{kand03,krug90,chin92}.
Both mass values have been adjusted to the distance scale adopted here
($H_0 = 70$, see below).

The galaxy \n34 is also known as \n17, Mrk 938, VV 850, MCG$-$02-01-032,
IRAS F0085$-$1223, and 2MASX J00110661$-$1206283.
It is located at $\alpha_{\rm J2000}=00^{\rm h}11^{\rm m}06\fs54$,
$\delta_{{\rm J}2000}=-12\degr06\arcmin27\farcs4$ (see \S~\ref{sec31})
and has a recession velocity relative to the
Local Group of $\czlg = +5961\pm 15$ \kms\ (\S~\ref{sec33}),
which places it at a distance of 85.2 Mpc for $H_0 = 70$ \kms\ Mpc$^{-1}$.
At that distance, adopted throughout the present paper, $1\arcsec = 413$ pc.
The corresponding true distance modulus is $(m-M)_0 = 34.65$. The Milky Way
foreground extinction is small, with values in the literature ranging
between $A_V=0.053$ (\citealt{rc3}) and 0.089 \citep{schl98}.  We adopt
the latter value, with which the absolute visual magnitude of \n34 becomes
$M_V = -21.57$ (\S~\ref{sec33}).

In the following, \S~\ref{sec2} describes the observations and reductions,
including imaging, aperture and surface photometry, and spectroscopy.
Section~\ref{sec3} presents results concerning the morphology and photometric
structure of \n34, its nuclear spectrum, and its system of young massive
clusters.
Section~\ref{sec4} then discusses the structure and nature of this galaxy
as a likely remnant of a ``wet'' unequal-mass merger, the nature of its
clusters, and the properties of its gaseous outflow.
Finally, \S~\ref{sec5} summarizes our main conclusions.

\section{OBSERVATIONS AND REDUCTIONS}
\label{sec2}

The observations of \n34 described in the present paper include a series
of \bvi\ exposures obtained with the direct CCD camera of the Ir\'en\'ee
du Pont 2.5-m telescope at Las Campanas Observatory, an archival $V$ image
taken with the Wide Field and Planetary Camera 2 (WFPC2) of the
{\em Hubble Space Telescope} (\hst\,), 
and spectra of the nucleus and of five bright star clusters taken with
the Low-Dispersion Survey Spectrograph (LDSS-2) of the Baade 6.5-m telescope
at Las Campanas.  Table~\ref{tab01} presents a log of these observations.

\subsection{Imaging}
\label{sec21}

Direct images of \n34 in \bvi\ were obtained with the CCD camera of the
du Pont 2.5-m telescope on 2000 Sep 30 (see Table~\ref{tab01}).
The camera was equipped with the chip Tek\#5 (format 2048$\,\times\,$2048),
which yielded a field of view of $8\farcm9\times 8\farcm9$ and a scale of
$0\farcs2607$/pixel.  A standard \ubvri$_{\rm KC}$ filter set was used, with
the $I_{\rm KC}$ filter designed for the Kron-Cousins system and the
corresponding passband hereafter called $I$ for short.  Conditions were
photometric, and the seeing was in the range $0\farcs7$--$0\farcs9$ (FWHM).

\begin{figure*}
  \centering
  \includegraphics[scale=0.962]{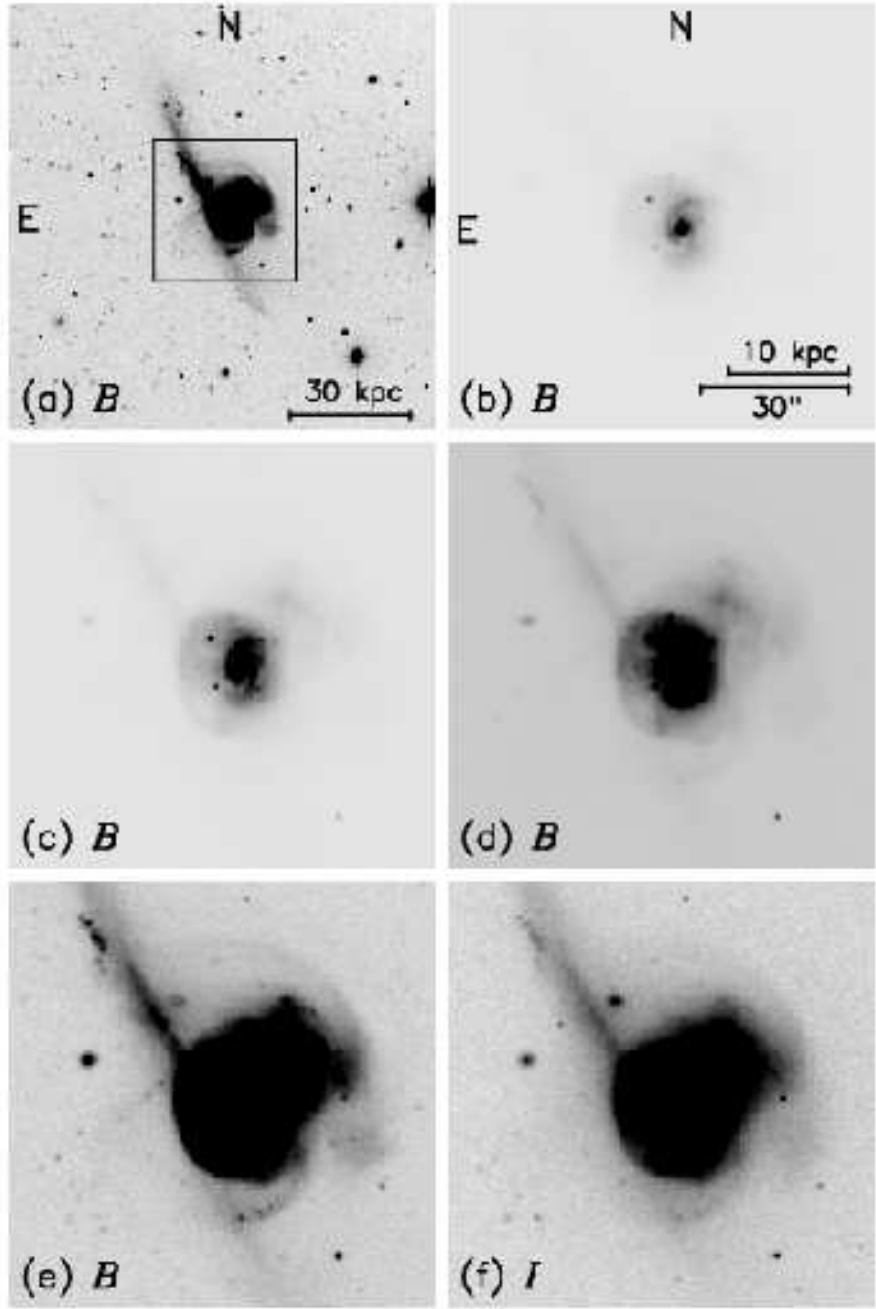}
  \caption{
Groundbased $B$ and $I$ images of \n34 obtained with the du Pont 2.5-m
telescope.
({\em a}) High-contrast display of $B$ image, showing $4\farcm3\times
4\farcm3$ field of view; the box marks the $86\arcsec\times 86\arcsec$
field of view of the other five panels.  Notice the two tidal tails emerging
from the system in opposite directions.
({\em b}) to ({\em e}) Displays of $B$ image at four different contrasts,
enlarged 3$\times$ relative to Panel~{\em a}.  Notice bright nucleus, central
disk with spiral structure and dust lanes, two bright star clusters E of
nucleus, and faint disk-like material NW of main body.
({\em f}) High-contrast display of $I$ image, for comparison with similar
display of $B$ image in Panel~{\em e}.
  \label{fig01}}
\end{figure*}

Figure~\ref{fig01} shows portions of the $B$ image (Panels a--e) and
$I$ image (Panel f) at various contrasts, with scale bars indicating the
angular and projected linear scales.
The box in Fig.~\ref{fig01}a identifies the area shown 3$\times$ enlarged
in Figs.~\ref{fig01}b--\ref{fig01}f. 
Notice the two near-linear tidal tails to the NE and SW of the system and
the various sharp-edged streamers, all typical of interacting and merging
{\em disk} galaxies.
A comparison of Figs.~\ref{fig01}e and \ref{fig01}f shows that the
general appearance of \n34 changes relatively little from $B$ to $I$, but
that the tidal features appear significantly brighter and more knotty in
blue light ($B$).

\begin{figure*}
  \centering
  \includegraphics[angle=-90,scale=0.863]{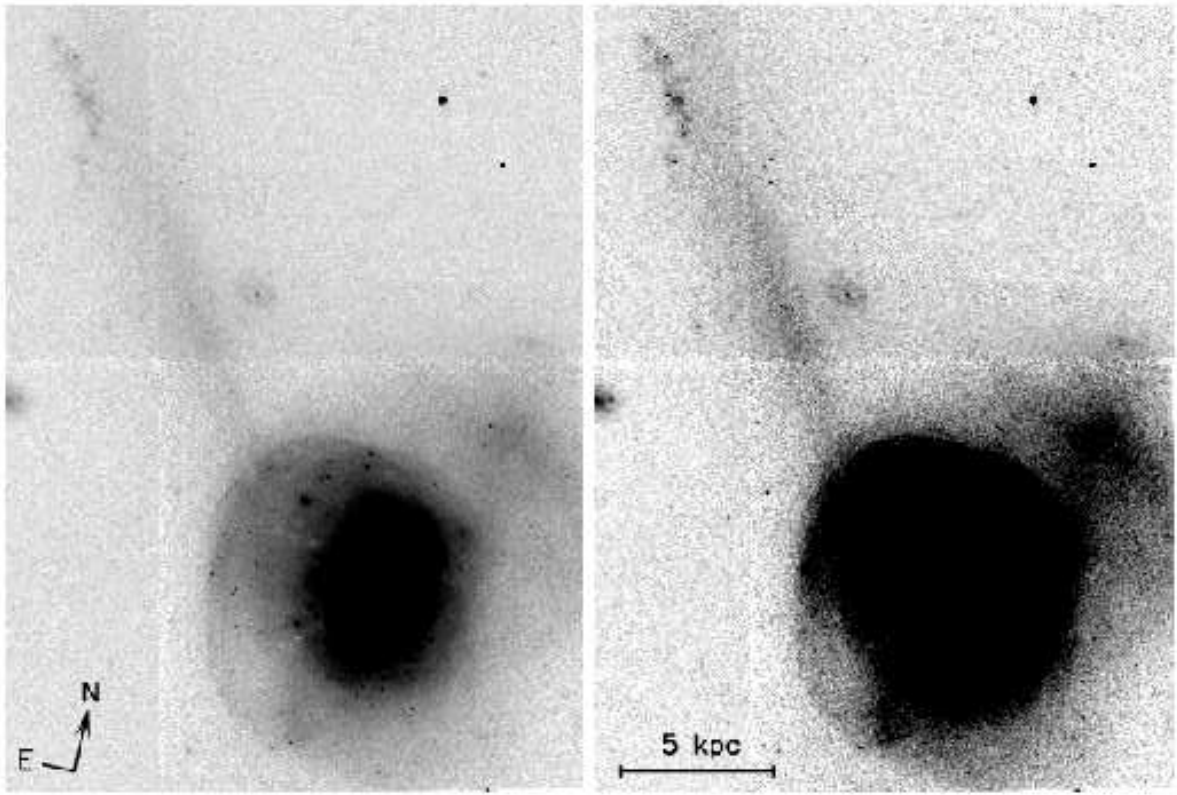}
  \caption{
Main body and inner part of N tail of \n34 as imaged with WFPC2 in $V$
(filter F606W, exposure of 500 s) and displayed at moderate and high
contrasts.  The part of the mosaicked $V$ image shown measures
$45\farcs8\times 62\farcs8$ ($\approx$ 19$\,\times\,$26~kpc), and the North
arrow is 5$\arcsec$ long. Notice the sharp ripples to the N, NE, and E of
the main body, and the knotty group of young star clusters in the N tail at
$\sim$19~kpc projected distance from the center.
  \label{fig02}}
\end{figure*}

\begin{figure*}
  \centering
  \includegraphics[angle=-90,scale=0.900]{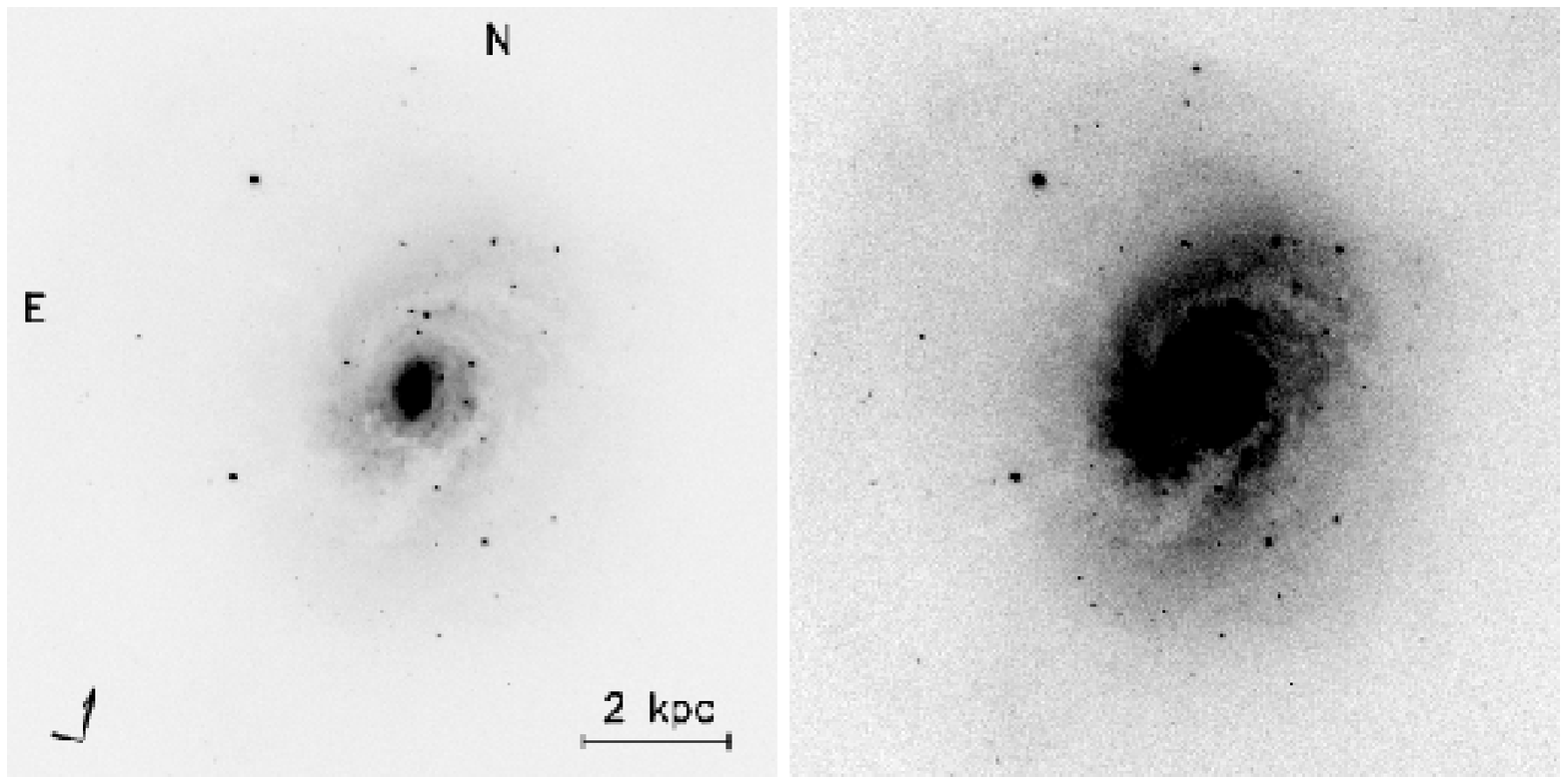}
  \caption{
Central region of \n34 as imaged with the Planetary Camera of WFPC2 in $V$
(filter F606W, exposure of 500 s), displayed at two different contrasts.
The field shown measures $25\farcs5\times 25\farcs5$ (= $10.5\times 10.5$ kpc)
and is centered slightly ESE of the nucleus.  The arrow to the lower left
measures 2\arcsec\ and points North, while the 1\arcsec\ leg points East.
Notice the many point-like sources, which are candidate star clusters,
and the intricate, spirally system of central dust lanes.
  \label{fig03}}
\end{figure*}

A search of the \hst\ Archive showed that, besides many NICMOS observations,
a single optical broad-band image of \n34 existed (as of 2006 July 1).
This image, obtained during \hst\ Cycle~5 with the WFPC2 camera and
single-star guidance (GO-5479, PI: M.\ Malkan), is a 500 s exposure through
the F606W filter;  it was taken with the nucleus centered on the PC chip
\citep{malk98}.  The image being a single exposure with many cosmic-ray
events, we cleaned it using the IRAF script {\em glacos.cl\,} kindly provided
by Dr.\ Pieter van Dokkum.  This script is similar to the script
{\em lacos\_im.cl\,} described in \citet{vdok01}, but handles the four frames
of a \hst/WFPC2 image (GEIS-format) in a single operation.

Figure~\ref{fig02} shows (at two different contrasts) a portion of the
WFPC2 mosaic image covering the main body of \n34 and the bright inner part
of the N tail, while Fig.~\ref{fig03} shows (also at two different
contrasts) an enlargement of the central regions as imaged on the PC.
Note the spiral-shaped dust lanes and the many bright point-like sources
that are candidate young star clusters.
Photometry and spectroscopy of these candidate star clusters is presented
in \S\S~\ref{sec23} and \ref{sec24} below.

\subsection{Aperture and Surface Photometry of \n34}
\label{sec22}

\begin{figure}
  \epsscale{0.85}
  \plotone{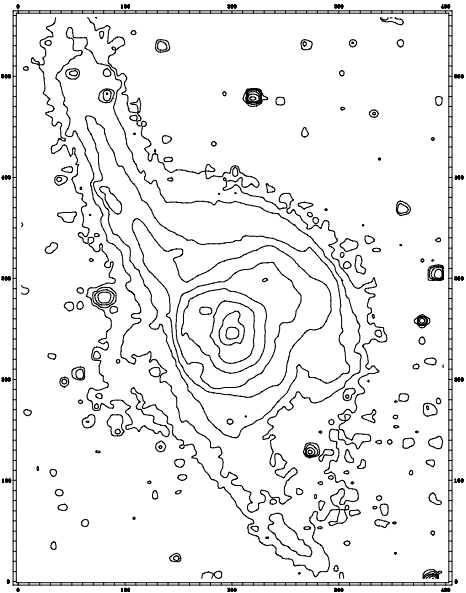}
  \caption{
Isophotal map of \n34, produced from $B$-image obtained with du Pont CCD
camera.  The faintest isophote corresponds to $\mub = 26.5$ \magarcs, and
subsequent isophotes increase in surface brightness by 1 mag.  To diminish
noise, the image has been smoothed with a 5-pix ($1\farcs3$) square boxcar.
North is up, and East to the left.  The field of view measures
$104\farcs3\times 146\farcs0$ ($\approx 43\times 60$ kpc).
  \label{fig04}}
\end{figure}

Figure~\ref{fig04} displays a calibrated blue isophotal map of \n34.
As this and more detailed maps suggest, the isophotes are roughly concentric
around the nucleus out to a surface brightness of
$\mub\approx 21.5$ \magarcs, but then bulge toward the NW and increasingly
trace tidal debris and tails.
The irregularity of the isophotes, especially in the blue and visual,
precludes any detailed analysis of the surface-brightness distribution in
terms of elliptical isophotes.
Hence, we perform photometry only in concentric apertures and then derive
{\em mean} surface-brightness and color-index profiles from it.

Table~\ref{tab02} presents $V$ magnitudes and \bv, \vi\ color indices
measured in 10 apertures of 5\arcsec\ to 150\arcsec\ diameter centered on
the nucleus.
These magnitudes and color indices were extracted from the CCD images
of \n34 obtained with the du Pont telescope and were calibrated
via measurements of 327 \citet{stet00} standard stars in the fields L98
and L110 \citep{land73,land92} observed during the same night.
The quoted 1$\,\sigma$ errors reflect uncertainties in the mean sky
levels, which amount to $\pm$0.14\% in $B$, $\pm$0.18\% in $V$, and
$\pm$0.08\% in $I$.
The measured magnitudes and colors are used in \S~\ref{sec33} to derive
global photometric parameters for \n34. 

To derive the mean surface brightness in $V$ and the mean color indices
as functions of radius we performed photometry in many concentric apertures
(41 apertures for the groundbased \bvi\ images and 48 apertures for the
\hst\,\ $V$ image), using the differences in flux measured in successive
apertures.  This simple method gives a rough preview of the mean light
distribution that the merger remnant is likely to develop as perturbations
caused by the merger average out in phase.

Table~\ref{tab03} presents a partial, but representative listing of our
measurements at selected radii.
Successive columns give the median-area radius $r$ of
the selected ring zone, the fourth root \rdV\ and logarithm of this radius,
the mean surface brightness \muv\ measured in $V$ from the \hst\ Planetary
Camera (PC, part of WFPC2) image
(\muvhst) and from the du Pont image (\muvlco), and the mean color indices
\bv\ and \vi. The \muvhst\ surface brightnesses are based on the calibrations
and color transformations given by \citet{voit98} and \citet{holt95},
with a mean value of $\vi = 1.10$ from the Las Campanas photometry adopted
for the color terms.
The \muvlco\ surface brightnesses and color indices are calibrated in the
same manner as described above for Table~\ref{tab02}. 
When one excludes the central two
groundbased measurements at $r < 0\farcs5$, which clearly suffer from
seeing effects, the agreement between \muvhst\ and \muvlco\ is reasonably
good, with a mean systematic difference of
$\langle\muvhst - \muvlco\rangle = +0.05$ mag and rms scatter of $\pm$0.04 mag
over the radius range $0\farcs7$--$6\farcs5$.

The combined surface-brightness and color-index profiles are shown in
Fig.~\ref{fig06} and discussed in \S~\ref{sec32}.

\subsection{Cluster Photometry}
\label{sec23}

\n34 hosts a relatively rich system of point-like sources that
are candidate young star clusters (Fig.~\ref{fig03}, also Figs.~\ref{fig01}
and \ref{fig02}).
To explore the properties of these clusters, we searched the one $V$ image
currently available from the \hst\ archive for point sources and performed
aperture photometry on them.
For the brightest two candidate clusters, we also measured color indices
from the \bvi\ images obtained at Las Campanas (Fig.~\ref{fig01}).
The archival \hst/WFPC2 image being a single, undithered 500 s exposure
in $V$, both our search for clusters and the subsequent photometry had to
be relatively simple.
We ignored the few fuzzy associations and candidate clusters in the N tail,
imaged on Chip WF3 of WFPC2 (Fig.~\ref{fig02}), and restricted our search
to the PC image, which appears to contain all other point sources likely
to be associated with \n34 (Fig.~\ref{fig03}).

To select a complete sample of candidate clusters on the PC image, we used
a simplified version of the search algorithm described by \citet{whit02}.
The simplified algorithm first invokes the task {\em daofind} of the
IRAF/DAOPHOT package \citep{stet87} to identify all potential sources
$>\,$3$\sigma$ above background and then filters the resulting source list
via various object-shape and photometric parameters.
In a final step, the algorithm performs photometry in apertures of 0.5 and
3 pixel radius and selects candidate clusters by requiring a
$\geq\,$5$\sigma$ detection in the larger of the two apertures and a
concentration index, defined as the difference between the magnitudes
measured in the two apertures, of $\dvindex > 1.5$ mag.
This choice of \dvindex\ filters out hot pixels and noise spikes (see
\citealt{mill97} for an example and more details).
We checked by visual inspection that the algorithm found nearly all
objects that we would have classified as likely cluster candidates, but
rejected hot pixels as well as parts of cosmic-ray events that survived
the original cosmic-ray cleaning of the image (\S~\ref{sec21}).
The adopted search algorithm yielded a total of 117 candidate clusters on
the PC image, of which 98 have apparent magnitudes $19.4 \la V < 25.0$
and the faintest has $V = 25.42$.

\begin{figure*}
  \centering
  \includegraphics[scale=0.863]{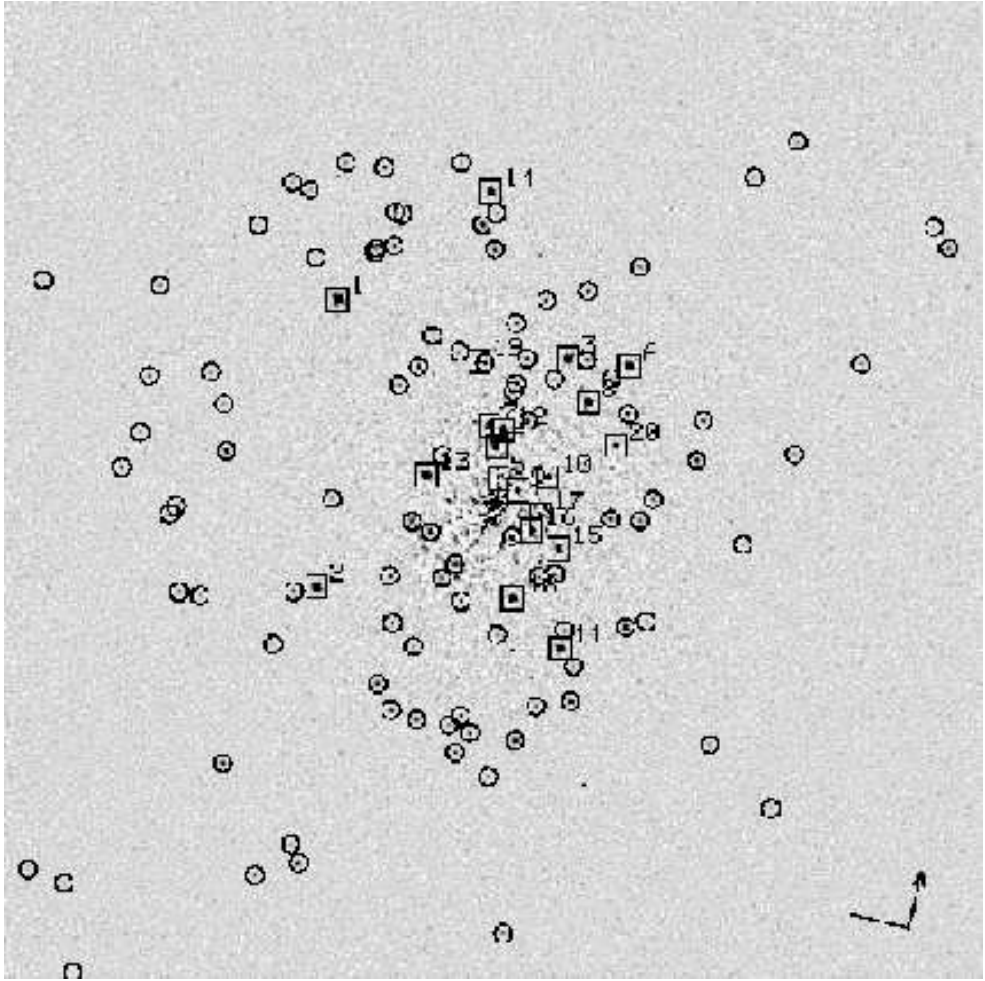}
  \caption{
Median-masked \hst/WFPC2 image of \n34 in $V$ showing 117 candidate
clusters, marked
mostly by circles.  The 20 brightest clusters are marked by squares and
ID numbers, with positions and magnitudes given in Table~\ref{tab04}.
The field of view is that of the Planetary Camera ($34\arcsec\times
34\arcsec$), and the arrow pointing north and bar pointing east each
measure $2\farcs0$.
  \label{fig05}}
\end{figure*}

Figure~\ref{fig05} shows a median-masked\footnote{
See \citet[Appendix A]{migh95} for a description of the masking
algorithm, which uses a ``low-pass difference filter.''}
version of the PC image with all 117 candidate clusters marked, mostly
by circles.  The 20 brightest
candidate clusters, with apparent visual magnitudes in the range
$19.4 \la V \la 22.3$, are marked by squares with identification numbers.
Table~\ref{tab04} gives their coordinates, projected distance from the
nucleus of \n34, approximate apparent magnitude $V$ on the Johnson passband
system, the corresponding absolute magnitude $M_V$ corrected for Milky-Way
foreground extinction, and the measured concentration index \dvindex.
For the two most luminous clusters, reddening-corrected color indices
\bvzero\ and \vizero\ measured from the Las Campanas \bvi\ images are given
in the footnotes.

Note that the apparent magnitudes $V$ on the Johnson system are only
approximate because we had to compute them from the magnitudes measured on
the instrumental $V_{606}$ system of WFPC2 without detailed knowledge of the
colors for most clusters.
We solved the task by assuming that most candidate clusters have the same
color as the weighted mean apparent \vi\ of the two brightest clusters
measured from the Las Campanas images, $\langle\vi\rangle_{1,2} = 0.67$,
and using the transformation equation between the F606W passband and the
Johnson $V$ band derived by \citet[eq. (9) and Table 10]{holt95}.
If some clusters were to have \vi\ colors deviating by as much as
$\pm$0.5 mag from the adopted value, their Johnson $V$ magnitudes would
differ by about $\pm$0.13 mag from the values given in Table~\ref{tab04}.
The photometric standard deviations $\sigma_V$ listed there do not reflect
this transformation uncertainty.

The luminosity function of the cluster system is illustrated and
discussed in \S~\ref{sec352}.

\subsection{Spectroscopy}
\label{sec24}

Spectra of the nucleus and of five candidate star clusters in \n34\ were
obtained with the Low-Dispersion Survey Spectrograph (LDSS-2;
\citealt{alli94}) on the Baade 6.5-m telescope during two
observing runs in late summer and fall of 2002 (see Table~\ref{tab01}).
All spectra were meant to be exploratory and have, therefore, relatively
low signal-to-noise ratios, except for the nuclear spectrum.

The five targeted clusters are objects 1, 2, 3, 7, and 13 of
Table~\ref{tab04}.
During the observations, the spectrograph slit of
$1\farcs03\times 330\arcsec$ was positioned across pairs of bright
clusters, yielding spectra of two additional faint clusters in one case
(Table~\ref{tab01}).
The seeing was in the range $0\farcs7$--$0\farcs9$ (FWHM) throughout the
observations.
The spectrograph was equipped with a 600 g mm$^{-1}$ grism blazed at
5000 \AA\ and a SITe 2K$\,\times\,$2K chip as detector.
This combination provided a reciprocal dispersion of 2.41 \AA\ pixel$^{-1}$
at $\lambda$\,5000 \AA, a spectral resolution of $\sim$5.3 \AA, and a
wavelength coverage of about 3700--6850 \AA.
The scale perpendicular to the dispersion was $0\farcs380$ pixel$^{-1}$.

All reductions were performed with IRAF\footnote{
The Image Reduction and Analysis Facility (IRAF) is distributed by the
National Optical Astronomy Observatories (NOAO), which are operated by
the Association of Universities for Research in Astronomy (AURA), Inc.,
under a cooperative agreement with the National Science Foundation.}
and its spectral-extraction tasks.  First the data frames were debiased,
flat-fielded, cleaned of cosmic-ray defects, and wavelength calibrated.
Then the spectra of individual clusters were traced on each frame and
extracted in $\sim${}$2\farcs1$ (= 5.5 pixel) wide bands with simultaneous
subtraction of sky and galaxy background.  For the nuclear spectrum,
the extraction band was $2\farcs00$ wide, and only sky was subtracted.
Next, each spectrum was corrected for atmospheric extinction and flux
calibrated via observed standard stars from the list by \citet{hamu92}.
Each spectrum was then corrected for Earth's orbital motion
to heliocentric, and all spectra for any given object were coadded into
one sum spectrum per object.

Figure~\ref{fig09} in \S~\ref{sec34} shows the resulting flux-calibrated
spectrum of the strongly reddened nucleus, while Fig.~\ref{fig13} in
\S~\ref{sec354} shows the spectra of the two brightest clusters, all
plotted versus rest wavelength.\notetoeditor{
Please do not insert figs.\ 9 and 13 here; the references to these figures
here are just a preview.  Full discussion is later, in referenced
subsections.}
In the spectra of the two clusters, note the strong Balmer absorption lines
indicative of A-type main-sequence stars; on expanded plots, these lines
are visible up to H12. 
The spectra of the remaining three clusters are of lower quality and
were used only to determine radial velocities (\S~\ref{sec353}).

\section{RESULTS}
\label{sec3}

\subsection{Morphology of \n34}
\label{sec31}

The morphology of \n34 is complex (Figs.~\ref{fig01} and \ref{fig02}),
especially at levels of surface brightness fainter than
$\muv\approx 21$ \magarcs\ ($\mub\approx 21.5$), where tidal debris
increasingly dominate.
Above this limit, the central part of the galaxy appears dominated by a
bluish disk with some spiral structure, and a very bright nucleus.
The following morphological description proceeds from this nucleus outwards.

In the $V$-image obtained with \hst/WFPC2, the nucleus appears centered at
$\alpha_{\rm J2000}=$\break
$00^{\rm h}11^{\rm m}06\fs536$,
$\delta_{{\rm J}2000}=-12\degr06\arcmin27\farcs42$, which is the position of
reference we adopt.  It agrees to within $1\farcs1$ with the astrometric
position measured by \citet{argy90} from blue photographic
plates obtained at Herstmonceux. The difference is well within the combined
errors of this earlier astrometry ($\sim${}$0\farcs3$) and of \hst/WFPC2
positions ($\la${}$1\farcs5$).  Given the extinction caused by the strong
central dust lanes and the deleterious effects of seeing, it is not clear that
the groundbased position in the blue is superior to the approximate space-based
position in the visual, which is why we adopt the latter.

In the visual, the bright nucleus shows structure on all scales down to the
resolution limit of the \hst/PC.  A dust lane at $r\approx 0\farcs10$ (41 pc)
to the NW of the semi-stellar nucleus seems to either flank this nucleus or
perhaps interrupt a small nuclear bulge of semi-major axis
$a\approx 0\farcs15$ (62 pc) slightly elongated in the same direction
(P.A.\ $\approx 318\degr$). 
Surrounding this nucleus or nuclear bulge is a small high-surface-brightness
disk of semi-major axis $a\approx 0\farcs80$ (330 pc) oriented nearly exactly
north--south (P.A. $= 4\degr$).  This central disk can be seen as the
black-saturated area in the left panel of Fig.~\ref{fig03}. 

The main disk with spiral structure extends out to $a\approx 8\arcsec$
(3.3 kpc), has an apparent axial ratio of $b/a\approx 0.72$ (corresponding to
an inclination of $i\approx 44\degr$ for an infinitesimally thin disk), and
has its apparent major axis oriented at P.A.\ $\approx 351\degr$.  Note that
the first increasing and then decreasing position angles of the various nested
structures (nuclear bulge, small central disk, and main disk) may reflect
the presence of a warp.  Although there appear to be two relatively smooth
main spiral arms, the spiral structure of the disk seems defined mostly
by an extensive system of dust lanes rather than by excess luminous matter.
The two strongest dust lanes define the inner edges of the two spiral arms,
with the southern lane especially broad and prominent.

The entire main disk appears peppered with bright point-like sources,
most of which are likely star clusters.
Visual inspection offers few clues as to whether these candidate clusters
belong to a disk population or a bulge/halo population.
A few of them seem to be associated with dust lanes or patches, while the
two most luminous clusters (for which we have spectra, \S~\ref{sec354})
appear to lie outside the main disk.

Beyond this main disk there is a bewildering variety of structures, including
an envelope that contains many sharp-edged ripples and dust lanes, a cloud
of luminous debris to the NW, fans and jets of luminous matter, and the two
main tidal tails.

Among the envelope's various ripples are four major ones that we shall
designate R$_1$ to R$_4$:
R$_1$ represents the sharp drop-off in surface brightness marking the
NNW end of the main disk at $r\approx 7\farcs7$ (3.2 kpc);
R$_2$ the next-out, fainter ripple at $r\approx 10\farcs9$ (4.5 kpc) NNE
(with candidate Cluster~14 superposed on it to the north and Cluster~1 just
inside it);
R$_3$ the giant arc-like ripple curving from north through
$r\approx 13\farcs5$ (5.6 kpc) NE all the way to the south, which forms
the envelope's eastern boundary;
and R$_4$ the faint, more fuzzy and knotty ripple at $r\approx 23\arcsec$
(9.5 kpc) SSW (Figs.~\ref{fig01}e and \ref{fig01}f). 
Note that long sections of R$_2$  and R$_3$ appear surprisingly sharp
($\la 0\farcs2$) even on the \hst/PC $V$-image (Fig.~\ref{fig02}).

The cloud of luminous debris to the NW, which we shall call the ``NW Cloud''
for short, reaches its diffuse brightness peak at a projected
$r\approx 15\farcs8$ (6.5 kpc) NNW of the nucleus (Fig.~\ref{fig01}d).
This cloud itself features much interesting structure
(Figs.~\ref{fig01}d--\ref{fig01}f) and may consist of remnant material
from the smaller of the two recently merged galaxies (\S~\ref{sec41}).
Besides several inner looping structures of $\sim$\,4\arcsec--\,6\arcsec\
diameter seen especially well on the du Pont $B$-image (but not visible in
Fig.~\ref{fig01}), the NW Cloud features an anvil-shaped extension about
$13\farcs8$ (5.7 kpc) to the WSW, from which emerges a $\ga${}70\arcsec\
(29 kpc) long, curved structure (tidal tail?) to the north and then bends
eastward.
This anvil-shaped extension shows best in Fig.~\ref{fig01}e.
Sitting about 15\arcsec\ (6 kpc) SSW of it is an extended cloud of
faint debris ($\mub\approx 24$--25.5 \magarcs) that appears to connect
to both the anvil and the curved structure emerging from it. 

The southern part of the NW Cloud appears strongly extincted by a major
aggregate of semi-continuous dust patches
($\Delta\alpha\times \Delta\delta\approx 6\farcs5\times 16\arcsec \approx
2.7 \times 6.6$ kpc), which suggests that the NW Cloud itself may lie
{\em behind} a veil of dust and gas perhaps more associated with the main
body of \n34.

Besides the two main tidal tails and the long curved structure emerging
from the NW Cloud, there are at least two other interesting fans or
jets of luminous matter.  One is the relatively bright, roughly
triangular-shaped protrusion emerging from the region of the main disk
to the SSE and best seen in Fig.~\ref{fig01}d.
This protrusion resembles a bent ribbon of material whose western leg
{\em may} be connected with a weak dust lane seen in absorption against
the southern-most part of the main disk.
If so, the western leg might lie in front of the main body of \n34, and
the ribbon itself might loop behind this body on its eastern leg.

The other interesting fan or jet appears to protrude eastward from near the
base of the northern tail and extends $\ga$20\arcsec\ (8.4 kpc) ESE from the
limit of \n34's envelope defined by Ripple R$_3$ (Fig.~\ref{fig01}a).
In its brighter inner part, this feature seems to contain two slightly
fuzzy objects that could conceivably be background galaxies, although that
would seem rather fortuitous.
This luminous ``jet'' {\em may} be an extension of very faint material seen
north of Ripple R$_3$, but it is definitely not an extension of R$_3$ itself.
It can also be traced in Fig.~\ref{fig04} in the form of ``appendices'' to
the $\mub = 23.5$, 24.5, and 25.5 \magarcs\ isophotes.

Perhaps of greatest value for future model simulations of \n34
are the two main tidal tails, of which the N tail is clearly both the
brighter and the longer one.  It can be traced to a projected distance
of at least $r_{\rm max} = 92\arcsec$ (38 kpc) from the nucleus, while the
fainter S tail is optically traceable only out to
$r_{\rm max}\approx 63\arcsec$ (26 kpc).

\begin{figure}
  \epsscale{1.05}
  \plotone{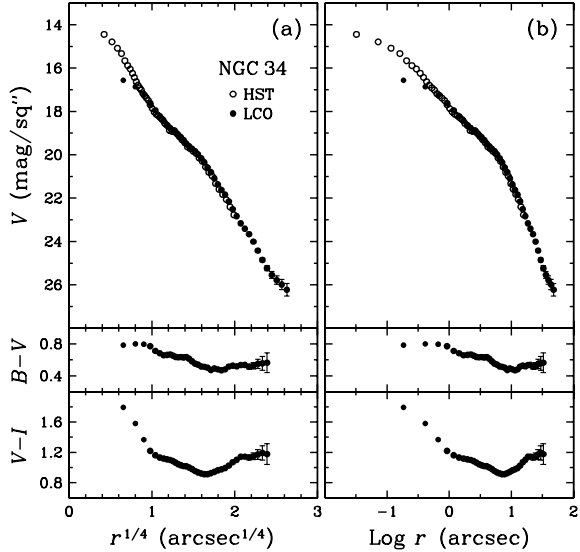}
  \caption{
Mean radial surface-brightness and color-index profiles of \n34 plotted
(a) vs fourth root of radius, \rdV, and (b) vs $\log r$.  Top panels show
mean surface-brightness in $V$ measured with \hst/PC and from Las Campanas
(``LCO''), while middle and bottom panels show color indices \bv\ and \vi,
respectively, measured from Las Campanas; points most affected by seeing
(LCO, $r<0\farcs8$) are marked by smaller dots.  Note the extended ``bump''
in the $V$ vs \rdV\ profile centered on $\rdV\approx 1.65$ ($r\approx
7\farcs4\approx 3.0$ kpc), representing extra light from the blue main disk.
The sharp ``knee'' in the $V$ vs $\log r$ profile at $\log r\approx 0.87$ is
caused by the same disk.  For details, see \S~\ref{sec32}.
  \label{fig06}}
\end{figure}

The N tail features a bright, remarkably straight inner part ending in
a group of 5--6 blue knots, the most distant of which lies at a
projected distance of 52\arcsec\ (21.5 kpc) from the nucleus.
On the \hst/WFPC2 $V$-mosaic image, these knots resolve into a number of
relatively point-like sources (likely star clusters) and more extended,
unresolved stellar associations (Fig.~\ref{fig02}).
The fact that the inner half of the tail appears straight and narrow
suggests that it consists of former disk material seen nearly edge-on.
Further out, the N tail becomes slightly curved and more fan-like,
suggesting that westwardly this former disk material may warp.

Following the brighter part of the N tail inward one can see a dust lane
in absorption against the luminous background of Ripple R$_3$, from where
it passes south of Cluster~1 and apparently ends in a major
dust patch at $r = 5\farcs8$ (2.4 kpc) from the nucleus (P.A. = 66\degr).
Thus, the N tail clearly crosses {\em in front} of the main body of \n34.
It would be of interest to know whether and how this tail connects further
inward or to the east.  This cannot be established unambiguously from the
present images.  {\em If\,} some roughly parallel dust lanes seen further
south of the above dust patch also belong to this tail, then the tail may
connect to the very prominent dust lane spiraling toward the nucleus on
the inside of the southern spiral arm (Fig.~\ref{fig03}).

The S tail appears much less well defined than the N tail, being both
fainter and more diffuse.  It is well visible beyond Ripple R$_4$, but
difficult to trace further in.  On its way in, its western edge seems to
either hug, or coincide with, the eastern leg of the triangular-shaped
luminous protrusion described above, and the whole tail ``fades'' into
a low-surface-brightness, SE area of the envelope which may, or may not,
be obscured by extinction.  If it were to be extinction and dust in the
S tail causes it, then the S tail would enter the main body from the
front, like the N tail.  This is an unexpected, and perhaps impossible,
geometry. Unfortunately, the signal-to-noise ratio of the \hst/PC
$V$-image in this region is too low to settle the issue of possible
dust extinction morphologically.  To do so, deeper high-resolution
images (e.g., with \hst/ACS) in two or three passbands will be needed.

\begin{figure}
  \epsscale{0.95}
  \plotone{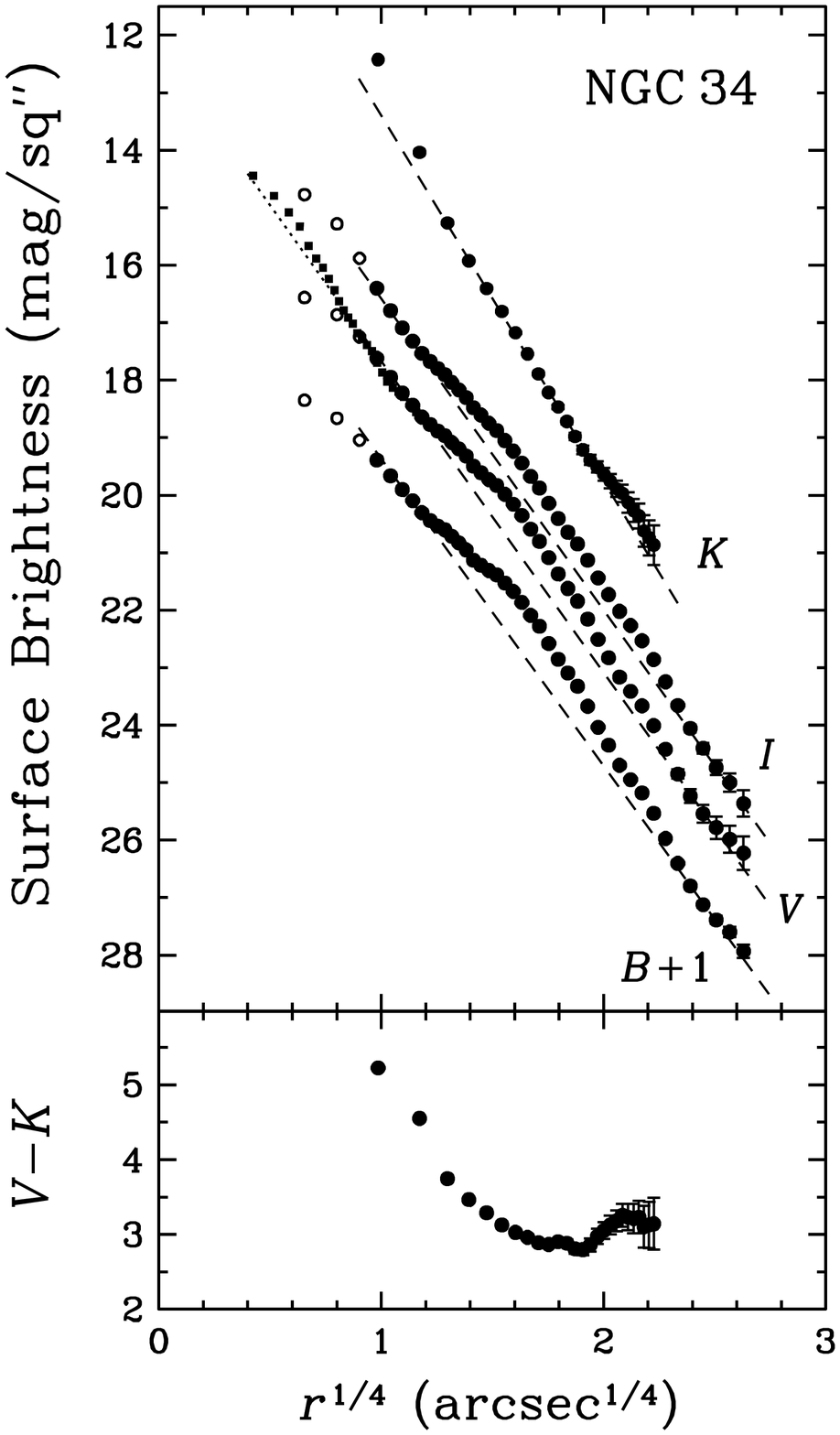}
  \caption{ 
Mean radial profiles of ({\em top}) surface brightness in \bvik\ and
({\em bottom}) color index \vk\ plotted vs fourth root of radius, \rdV.
The $K$ data are from \citet{rj04}, while the \bvi\ data are
from Las Campanas ({\em filled and open circles}) and the
\hst/PC ({\em small squares)}. The $B$ profile is shown shifted by +1 mag
for clarity.  The dashed lines represent least-squares fits (in $K$) or
lines drawn parallel to such fits (in \bvi); see text for details.
Note that the slope of the $K$-profile is significantly steeper than
the slopes of the $B$-, $V$-, and $I$-profiles, leading to the strong
inner color gradient seen in \vk.
  \label{fig07}}
\end{figure}

\begin{figure*}
  \includegraphics[angle=-90,scale=0.70]{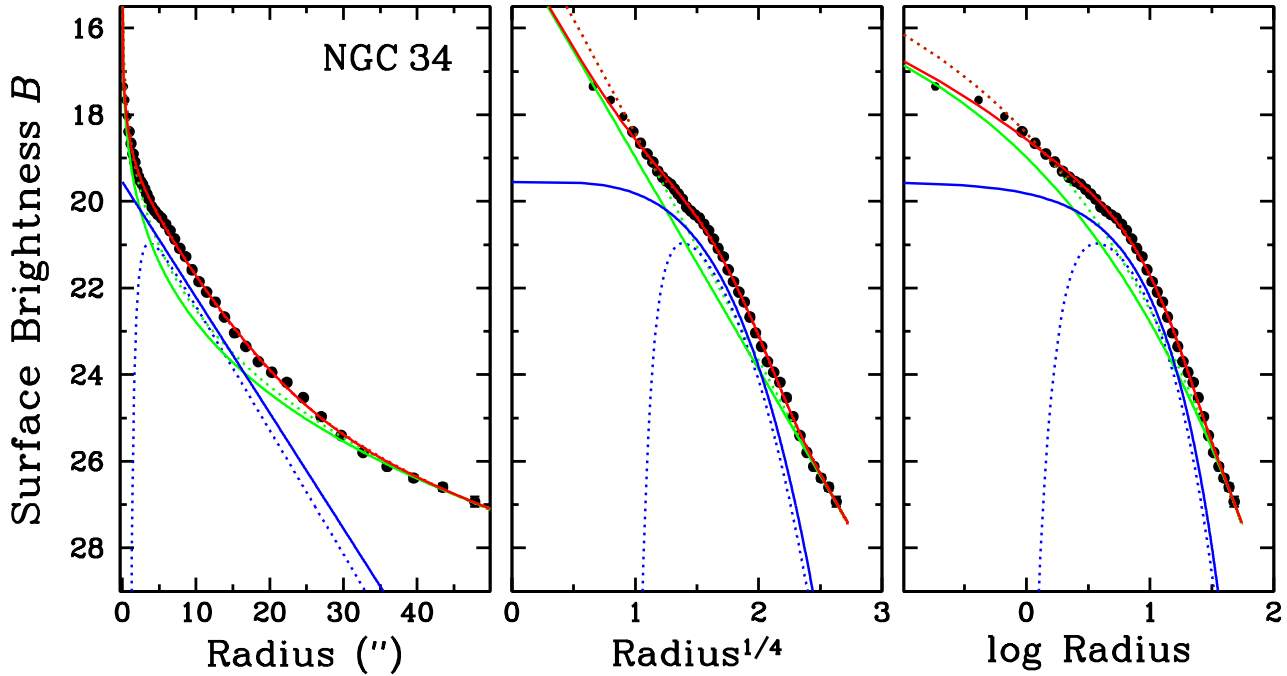}
  \caption{
\sphed\ model fits to $B$ profiles plotted vs ({\em left}) radius $r$,
({\em middle}) \rdV, and ({\em right}) $\log r$.  Data points represent
$B$-profile measured from Las Campanas, while the solid lines mark the
best \sphed\ fit ({\em red}) and its component spheroid ({\em green}) and
disk ({\em blue}) for the case of an exponential disk without central hole.
Dotted lines mark the corresponding best fits for an exponential disk with
a central hole.  Note how the exponential disk produces the ``bump'' in
the \rdV\ profile and the ``knee'' in the $\log r$ profile.
  \label{fig08}}
\end{figure*}

\subsection{Surface-Brightness Distribution}
\label{sec32}

Figure~\ref{fig06} shows the mean radial profiles of $V$ surface brightness
and \bv\ and \vi\ color (\S~\ref{sec22}) in \n34, plotted both vs the
fourth root of radius, \rdV, and vs $\log r$.
These various profiles show two related salient features.

First, although overall the $V$ profile appears to approximate an \rdV\ law
\citep{deva53} reasonably well, the deviations of the data points in
Fig.~\ref{fig06}a from a straight line are highly significant (most error
bars being smaller than the size of the data points). 
The main feature is an extended ``bump'' in the $V$ vs \rdV\ profile,
centered on $\rdV\approx 1.65$ ($r\approx 7\farcs4\approx 3.0$ kpc) and
showing up in the $V$ vs $\log r$ profile as a relatively sharp ``knee'' or
break in the slope at $\log r\approx 0.87$.
Note that the bump and knee are seen clearly in both the groundbased and
the \hst\,\ data (filled and open circles, respectively), whence they
represent a real feature.

And second, the color-index profiles display minima in both \bv\ and \vi\
near the radii of the bump and knee ($\sim$\,3--4 kpc), indicating that
the extra light causing these features in the $V$ profile must be quite
blue.

Comparisons with simple model light distributions that contain contributions
from both an \rdV-law spheroid and an exponential disk show that similar bumps
and knees trace relatively bright exponential disks.  Hence, the above
two features in \n34 suggest the presence of a significant, blue exponential
disk embedded in a redder spheroid.  This blue disk presumably corresponds
to the ``main disk with spiral structure'' noted morphologically as extending
out to $a\approx 3.3$ kpc (\S~\ref{sec31}).

As a first test of this hypothesis, Fig.~\ref{fig07} displays the individual
surface-brightness profiles in $B$, $V$, and $I$ plotted vs \rdV, and for
comparison also the mean surface-brightness profile in $K$ similarly
measured by \citet{rj04}.   To help assess
the amplitudes of the bump in \bvi, straight dashed lines with slopes equal
to least-squares fits to all the data not seriously affected by seeing
(filled circles) have been drawn with small vertical shifts so as to roughly
form lower boundaries to the observed profiles.  Note that the relative amount
of light in the bump increases significantly from $I$ through $V$ to $B$,
while no similar bump is observed in $K$ (where the fit is taken from
\citeauthor{rj04}).  Hence, this increasing amplitude with decreasing
wavelength indeed supports the view that the hypothetical disk is
significantly bluer that the host spheroid, suggesting a population of
young disk stars.

A second, more thorough test of the above hypothesis is to fit the observed
surface-brightness profiles with models consisting of an \rdV-law spheroid
plus an exponential disk \citep[e.g.,][]{korm77,bagg98} and to
see (1) how much such fits improve the mean residuals over fits with a pure
\rdV\ law and (2) how consistent the model parameters obtained independently
from $B$, $V$, and $I$ are.  Using the IRAF/STSDAS task {\em nfit1d}, we 
performed such ``\sphed'' (spheroid plus exponential disk) model fits
in both magnitude-vs-radius and linear-flux-vs-radius space, and with
exponential disks with and without central hole.  To obtain consistent
color information we ignored the \hst\ data in $V$ and fitted only the mean
\bvi\ surface-brightness profiles measured from Las Campanas data.

Figure~\ref{fig08} shows how these fits appear when the surface brightness
in $B$ is plotted vs linear radius $r$, vs \rdV, and vs $\log r$.
Superposed on
the observations (data points), the solid lines mark the best \sphed\ fit
(red) and its component spheroid (green) and disk (blue) for the case of an
exponential disk without central hole.  Correspondingly, dotted lines mark
the best fit for the case of an exponential disk with a central hole
\citep{korm77,bagg98}.  As a comparison of the solid and dotted red lines
shows, the \sphed\ model profiles without and with a central hole in the
exponential disk fit the data nearly identically well over the fitting range
(large data points from $r = 0\farcs92$ to $47\farcs8$), though---not
surprisingly---they diverge at $r \la 1\arcsec$.  The model profiles
represent the observed $B$ (and also $V\!I$) surface brightnesses remarkably
well, with rms residuals of only 0.07 mag in $B$, thus strongly supporting
the hypothesis that the bump in the \rdV-profile and the knee in the
$\log r$-profile are caused by the presence of an exponential disk.

Table~\ref{tab05} summarizes both the pure \rdV-law fits in \bvi\ mentioned
above and the new \sphed\ model fits.
As comparisons of the rms residuals in the table and of Figs.~\ref{fig07}
and \ref{fig08} suggest, allowing for the presence of an exponential disk
in addition to the spheroid brings a dramatic improvement in the fits,
reducing the rms residuals in the various passbands by factors of 2--4.
Also, the two scale lengths of the \sphed\ model, the effective radius
\reff\ of the spheroid and the scale length $\alpha$ of the exponential
disk, agree remarkably well among the three passbands, as does the
``hole radius'' \rhole\ in the case of the exponential disk with a central
hole.\footnote{
We note, however, that the model without a hole in its exponential
disk reproduces the total color index \vitot\ of \n34 slightly better
(to within 0.03 mag) than does the model with a central hole (to within
0.05 mag), while both models reproduce the measured \bvtot\ very well
(to $\la$0.01 mag).}
Thus, the model parameters obtained independently from $B$, $V$, and $I$ are
reasonably consistent, which again supports the hypothesis that an
approximately exponential disk is present in \n34.               

What are the relative luminosity and colors of this model disk?
If we pick the best-fit \sphed\ model without a hole in the disk, the
model disk contributes 51\%, 43\%, and 27\% of the total $B$, $V$, and $I$
light, respectively, within $r=15$ kpc (corresponding to about the $B=26.5$
isophote, see \S~\ref{sec33}).
Its absolute visual magnitude, corrected for Galactic foreground (but not
internal) extinction, is $M_{V,0} = -20.61$, and its colors are
$\bvedzero = 0.36\pm 0.02$ and $\viedzero = 0.52\pm 0.02$.
If, on the other hand, we pick the best-fit \sphed\ model {\em with} a hole
in its disk, then the model disk contributes about 34\%, 26\%, and 18\% of
the total $B$, $V$, and $I$ light, respectively, and has an absolute
luminosity $M_{V,0} = -20.11$ and color indices $\bvedzero = 0.28\pm 0.02$
and $\viedzero = 0.59\pm 0.02$.
Thus, in either case the exponential disk is very blue and contributes
significantly to the total optical luminosity of \n34.

An alternative way to state this is that the present-day bulge-to-disk
ratios of \n34 for the model without a central hole in its disk are
(B/D)$_B$ = 0.94, (B/D)$_V$ = 1.32, and (B/D)$_I$ = 2.7 (within $r=15$ kpc),
while those for the model with a central hole are (B/D)$_B$ = 1.9,
(B/D)$_V$ = 2.8, and (B/D)$_I$ = 4.6.
These bulge-to-disk ratios will, of course, change as the stellar
populations of the bulge and blue disk age.

\subsection{Systemic Velocity and Absolute Magnitude}
\label{sec33}

The heliocentric systemic velocity, measured from six absorption lines
in the spectrum of the nucleus (\S~\ref{sec34}), is
$cz_{\rm hel,sys} = 5870 \pm 15$ \kms.
This value agrees to within 1.1\,$\sigma$ (combined error) with the modern
optical measurement of $5821 \pm 44$ \kms\ by \citet{daco98} and
to within 0.7\,$\sigma$ with that of $5881\pm 2$ \kms\ (from the \ion{Ca}{2}
triplet) by \citet{rj06a}, but disagrees significantly with
many older optical measurements \citep[e.g.,][]{rc3}
and with the value of $5931 \pm 11$ \kms\ until recently adopted in the
NASA/IPAC Extragalactic Database (NED).

The likely reason for the disagreement with older optical measurements,
which tend to yield 100--150 \kms\ lower values, is that these
measurements often included the strong Na\,D lines, shown in \S~\ref{sec34}
to yield a velocity about 620 \kms\ lower than the above systemic velocity
due to the presence of a gaseous outflow.
The significantly higher systemic velocity until recently given in NED, on
the other hand, was based on the mean heliocentric velocity from compiled
integrated \hi\ observations \citep{bott90}.\footnote{
As of 2006 August, the new value adopted by NED for the heliocentric
velocity is that measured by \cite{rj06a}.}
In merging galaxies, the \hi\ distribution is often strongly asymmetric
\citep[e.g.,][]{hibb96}, and the mean \hi\ velocity can be a
poor measure of the galaxies' or remnant's systemic velocity.
Hence, and despite the high intrinsic accuracy of \hi\ velocities, the new
optical absorption-line velocity of the nucleus of \n34, supported by that
measured in the near-infrared from the \ion{Ca}{2} triplet by 
\citeauthor{rj06a}, is probably more accurate.

After correction to the barycenter of the Local Group via
$\Delta cz_{_{\rm LG\,-\,hel}} = 300 \sin l \cos b = +91$ \kms\ 
\citep{sand75}, the systemic velocity of \n34 is $\czlg = +5961$
\kms, leading to the adopted distance of 85.2 Mpc for $H_0 = 70$ \kms\
Mpc$^{-1}$ (\S~\ref{sec1}).

To calculate the absolute magnitudes of \n34 in different passbands, we
first need to measure total magnitudes, traditionally defined as integrated
magnitudes within the $\mub = 26.5$ \magarcs\ isophote.
From the isophotes shown in Fig.~\ref{fig04}, and ignoring their extensions
along the tidal tails, we determined that the area within the above isophote
corresponds to that of a circular aperture of radius
$r = 36\farcs5$ (15.1 kpc), and we approximated the total magnitudes by
interpolating the magnitudes measured in concentric apertures
(Table~\ref{tab02}) to this radius.
The resulting total magnitudes \btot, \vtot, and \itot\ are given in
Table~\ref{tab08} below.
For comparison, Table~\ref{tab08} also gives the total $K$ magnitude
measured by \citet{rj04} and here corrected to the same aperture (i.e.,
$\sim$ \mub~= 26.5 isophote) as the total optical magnitudes.\notetoeditor{
Please do not place Table 8 here, despite these two early refs.\ to it.
The table belongs in summarizing \S~\ref{sec36}.}

Based on these total apparent magnitudes and the adopted distance and
foreground extinction (\S~\ref{sec1}), the absolute visual magnitude and
foreground-reddening-corrected total color indices of \n34 are $M_V= -21.57$,
$\bvtotzero = 0.55 \pm 0.02$, $\vitotzero = 1.05 \pm 0.01$, and
$\vktotzero = 3.15 \pm 0.03$, respectively.

From the multi-aperture photometry given in Table~\ref{tab02} and the
isophotes, we estimate that the extra light {\em beyond} the above
$r = 36\farcs5$ aperture used for the total magnitudes is about
7\% $\pm$ 3\% in $V$ and originates mostly in the two main tidal tails.

\begin{figure*}
  \centering
  \includegraphics[scale=0.75]{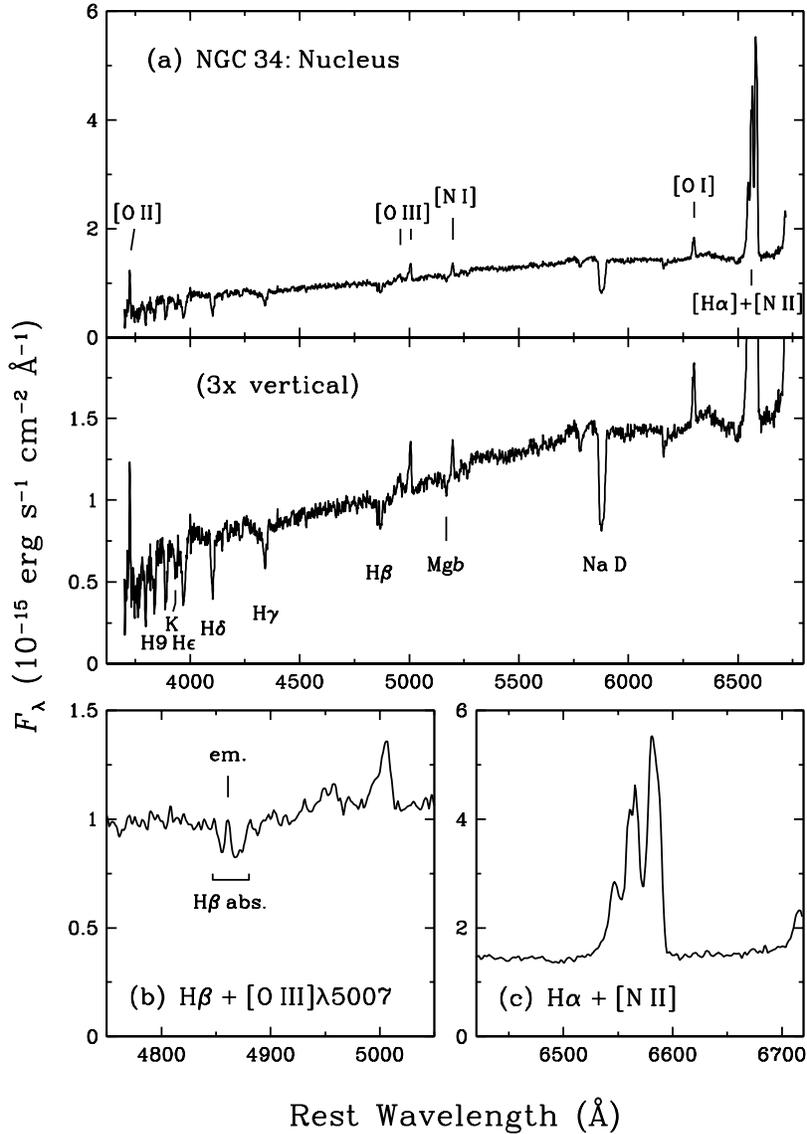}
  \caption{
({\em a}) Ultraviolet-to-red spectrum of the \n34 nucleus, obtained with
the Baade 6.5-m telescope and LDSS-2 spectrograph through a
$1\farcs03\times 2\farcs0$
slit aperture oriented at P.A.~= $144\fdg5$. The spectrum is flux calibrated
and plotted vs.\ rest wavelength.  Emission lines are identified in the upper
panel, absorption lines in the lower panel with 3$\times$ expanded flux scale.
Panels ({\em b}) and ({\em c}) show spectral details around H$\beta$ and
H$\alpha$, respectively, on a scale 5$\times$ enlarged in wavelength. 
Note the poststarburst nature of the spectrum in the UV--blue region, the
small blue-shift of the H$\beta$ emission relative to the absorption, the
somewhat broadened emission lines, and the very strong Na\,D absorption,
whose measured blueshift indicates an outflow with a mean velocity of about
$-$620 \kms.
   \label{fig09}}
\end{figure*}

\subsection{The Nuclear Spectrum: Strong Outflow}
\label{sec34}

As described in \S~\ref{sec1}, the appearance of the nuclear spectrum
of \n34 has variously been attributed to the presence of a Seyfert~2
nucleus, a nuclear starburst, or a combination of both.

The new nuclear spectrum shown in Fig.~\ref{fig09} does not contribute
toward settling this controversy, but does reveal that there are systematic
velocity differences between the ionized gas and the stars, and that some
of the cool gas---as traced by \ion{Na}{1} D absorption---is experiencing
a strong outflow of the kind often observed in IR-luminous galaxies
with star formation rates $\ga$10 \msunyr\ \citep{rupk05,veil05}.

After we discovered the strong blueshift of the Na D doublet, we
determined the systemic {\em stellar} velocity in two ways: by
measuring individually five Balmer lines between H$\gamma$ and H10
plus the \ion{Ca}{2} K line and computing the mean velocity, and
by cross-correlating the blue part of the nuclear spectrum
($\lambda\lambda$3820--5050), clipped of emission lines, with
the template spectrum of the massive young cluster \n7252:\,W3
described in \S~\ref{sec353} below.
The weighted average of the two methods yielded the adopted heliocentric
systemic velocity of $5870 \pm 15$ \kms\ (\S~\ref{sec33}).

Relative to this systemic stellar velocity, the emission lines from
ionized gas show small blueshifts with considerable scatter,\footnote{
The blueshift of the H$\beta$ emission line relative to the stellar
H$\beta$ absorption line is directly visible in the lower left panel of
Fig.~\ref{fig09}.}
corresponding to a mean velocity of about $-75 \pm 23$ \kms.  Although
this velocity likely indicates an outflow of ionized gas from the nucleus,
the evidence is not unambiguous since the mean position of the emitting gas
along the line of sight is unknown.

\begin{figure}
  \centering
  \includegraphics[scale=0.52]{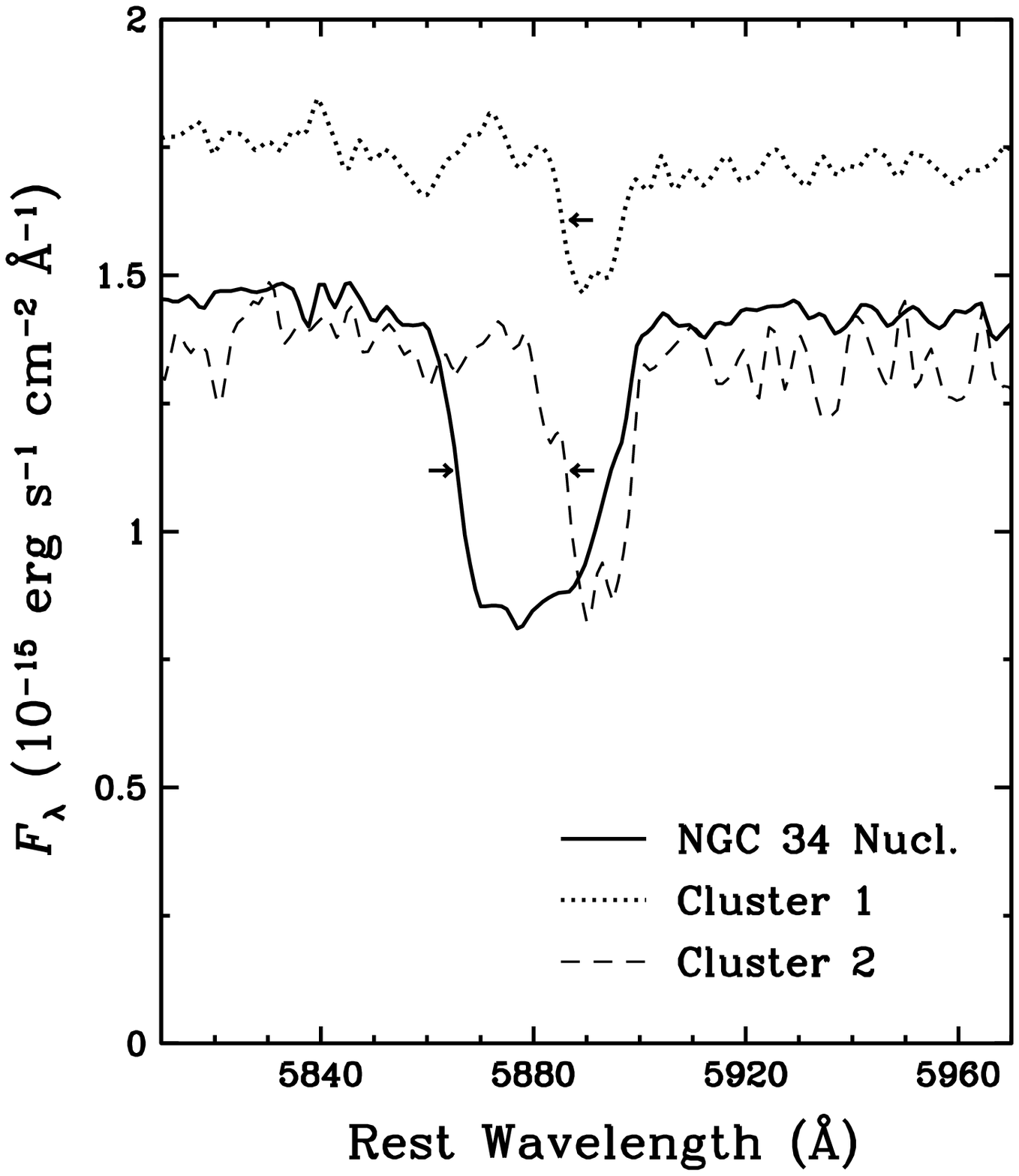}
  \caption{
Enlarged portions of the nuclear spectrum and of two cluster spectra around
the \ion{Na}{1} D doublet, plotted vs rest wavelength.  The flux scale
applies to the nucleus, while each cluster spectrum has been scaled
arbitrarily (but with zero level fixed at the bottom).  Note the larger
width and blueshift of the nuclear doublet,
indicating a gaseous outflow with a mean velocity of about $-$620 \kms\
and a maximum velocity of about $-$1050 \kms.  This maximum velocity has been
derived from the shifts between the blue edges of the nuclear gas outflow
and of the (mostly) stellar cluster doublets ({\em arrows}).
  \label{fig10}}
\end{figure}

Tracing the line-of-sight velocities of cool, neutral gas between us and
the nucleus of \n34, the \ion{Na}{1} D doublet appears both broad and
blueshifted relative to the stars.
This can be seen directly in Fig.~\ref{fig10}, which compares the line
profile of this doublet for the nucleus (solid line) with the profiles
for the two luminous clusters 1 and 2, all plotted vs {\em rest} wavelength.
The nuclear line profile is $\sim$2.3$\times$ as wide (FWHM) as the cluster
line profiles, and the measured mean position indicates a mean outflow of
$-620\pm 60$ \kms\ from the nucleus.
The maximum outflow velocity of the cool gas, estimated from the shift
between the blue edges of the nuclear and cluster line profiles at half
minimum (arrows in Fig.~\ref{fig10}) is about $-1050\pm 30$ \kms.
These mean and maximum outflow velocities indicate an outflow that is
strong, even for a galaxy with a star formation rate of
$70 \pm 20$ \msunyr\ like \n34 (\S~\ref{sec43}). 

The spatial extent of this outflow needs to be established before the
mass involved in the outflow can be calculated.  In principle, spectra of
the many young star clusters can be used to map the outflow.
However, the spectra presently available for five clusters
(\S\S~\ref{sec24} and \ref{sec353}) have too low signal-to-noise ratios
to lead to conclusive results.
Whereas the Na D doublets in clusters 1 and 2 appear to be mostly of
stellar origin, with perhaps a mildly blue-shifted extra component in
Cluster~2 (Fig.~\ref{fig10}), the doublets of clusters 3 and 7 do seem to
indicate major blue shifts relative to both the clusters and the galaxy,
even though the spectra are very noisy.
Relative to the systemic velocity of \n34, the blueshift of the Na D doublet
in cluster 3 ($\sim$2.3 kpc NNW of the nucleus) appears to be about
$-390\pm 100$ \kms, while for cluster 7 (1.1 kpc NNE of nucleus) it appears
to be about $-460\pm 200$ \kms.
Clearly, new  spectra for more clusters are needed to properly map the
full outflow.

Note that the shapes of the [\ion{O}{3}]\,$\lambda$5007 and
H$\alpha$\,+\,[\ion{N}{2}]\,$\lambda\lambda$6548,\,6583 emission lines
seem to indicate velocity broadening as well, even though the latter three
lines blend in part because of the limited, $\sim$5.3 \AA\ resolution of
our observations.  Yet, much higher-resolution observations made by
\citet{busk90} with a Coud\'e spectrograph confirm that the
H$\alpha$\,+\,[\ion{N}{2}] lines are significantly broadened
(FWHM = $264\pm 7$ \kms\ for H$\alpha$ and $384\pm 12$ \kms\ for
[\ion{N}{2}]\,$\lambda$6583) and {\em may} have weak underlying broader
components. 
Clearly, the apertures used to extract the nuclear spectrum by both
\citeauthor{busk90} ($1\farcs2\times 5\farcs0$) and ourselves ($1\farcs03
\times 2\farcs0$) cover nuclear areas large enough to comprise varied
ionized-gas motions, and new observations with higher {\em spatial}
resolution are needed to better isolate individual narrow- and broad-line
components.

Finally, it seems worth pointing out that the optical nuclear spectrum
appears not only strongly reddened, but also highly composite
(Fig.~\ref{fig09}).
Blueward of $\lambda\approx 4500$ \AA\ the spectrum is dominated by strong
Balmer absorption lines typical of a poststarburst population.
A comparison of the region containing the \ion{Ca}{2} K line and the
Balmer lines H8, (H\,+)\,H$\epsilon$, and H$\delta$ with model spectra of
simple stellar populations computed by \citet{bc03} suggests a poststarburst
age of about 300--400 Myr.
Yet, longward of $\lambda$4500 the optical spectrum is increasingly
dominated by \hii-region-like emission lines, suggesting the presence of
an {\em ongoing}, strongly extincted starburst.
Of course, the main evidence for such an obscured central starburst comes
from the infrared (see \S~\ref{sec1}).
The combined evidence suggests, then, that the region of strong starburst
activity has shrunk with time and is now restricted to a highly obscured
central region of $\la$\,1 kpc radius (see \vk\ profile of Fig.~\ref{fig07}).
Presumably, it is from this central region that the gaseous outflow
originates. 

\begin{figure}
  \centering
  \includegraphics[scale=0.55]{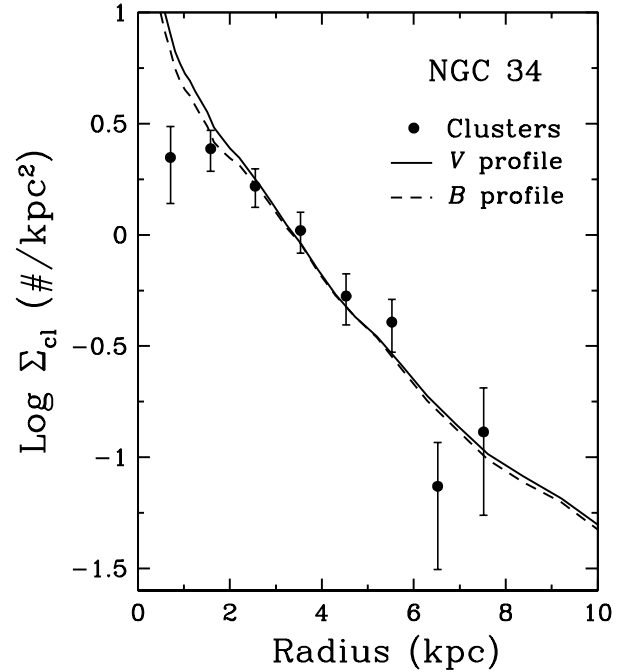}
  \caption{
Radial distribution of candidate clusters in \n34. The surface number
density of clusters, \signcl\ (data points with error bars), is compared
with the mean surface-brightness profiles in $V$ (solid line) and $B$
(dashed) of the underlying galaxy.  The $V$ and $B$ profiles have been
shifted vertically by arbitrary amounts.  Note the good agreement between
the centrally concentrated cluster distribution and the underlying galaxy
light, except within  $r \la 1.5$ kpc where incomplete cluster detection due
to the bright galaxy background becomes significant.
   \label{fig11}}
\end{figure}

\subsection{The System of Young Massive Clusters}
\label{sec35}

The central system of 117 candidate young massive clusters in \n34
(\S~\ref{sec23}) is conspicuous (Figs.~\ref{fig03} and \ref{fig05}) and
contains 17 objects with $-13.0 \geq M_V \geq -15.4$ (Table~\ref{tab04}).
The present subsection describes the clusters' spatial distribution,
luminosity function, known line-of-sight velocities, and spectroscopically
estimated ages. 

\subsubsection{Spatial Distribution}
\label{sec351}

Visual inspection of Fig.~\ref{fig05} suggests that the 117 candidate young
massive clusters lie strongly concentrated toward the center of \n34 and
are distributed approximately uniformly in position angle, though with a
slight excess in numbers toward the NE and a slight deficiency toward the WSW.

Figure~\ref{fig11} shows the radial distribution of the clusters' projected
surface number density, \signcl, plotted vs linear radius.
This surface density, expressed as a number per kpc$^2$, was determined
in successive circular annuli of 1~kpc width centered on the nucleus and
has not been corrected for background objects or completeness.
Given that even the faintest of the 117 clusters are relatively bright,
$V\approx 25.4$, the contribution from background objects over the total
field of view ($34\arcsec\times 34\arcsec$) is negligible, while any
completeness correction will mainly affect \signcl\ in the innermost
1--2 kpc, where the galaxy background is brightest.
For comparison the figure also shows, with arbitrary vertical shifts, the
surface-brightness distribution of the underlying galaxy in $B$ and $V$.

The main result is that over most of the range in radius covered,
$r\approx 1.5$--8 kpc, the surface number density of clusters tracks
the underlying galaxy light remarkably well.  This is similar to the
situation observed in other recent merger remnants (e.g., \n3921:
\citet{schw96b}; \n7252: \citealt{mill97}) and seems to
indicate that violent relaxation redistributed the general luminous matter
of the two merging galaxies in roughly the same manner as it redistributed
the giant molecular clouds
and young clusters forming from them \citep{schw02}.  If this conclusion
is correct, the apparent deficit of clusters within the central $\sim$1.5 kpc
radius relative to the general light distribution is likely due to the
growing incompleteness in cluster detection as the brightness of the galaxy
background increases.

Assuming that, were it not for this central incompleteness, the surface
density \signcl\ would track the mean $V$-profile of \n34 exactly, the
number of candidate clusters within $r \leq 2$ kpc would increase from
30 to about 53, and the total number within the field of view imaged by
the PC of WFPC2 ($r \la 8$ kpc) would increase from 117 to about 140 candidate
clusters with $V \la 25.4$ ($M_V \la -9.3$).

Finally, the effective radius of the candidate-cluster system is
$r_{\rm eff,\,cl} = 3.1$ kpc without incompleteness correction and
$r_{\rm eff,\,cl} \approx 2.6$ kpc with the above estimated correction.

\begin{figure}
  \centering
  \includegraphics[scale=0.555555]{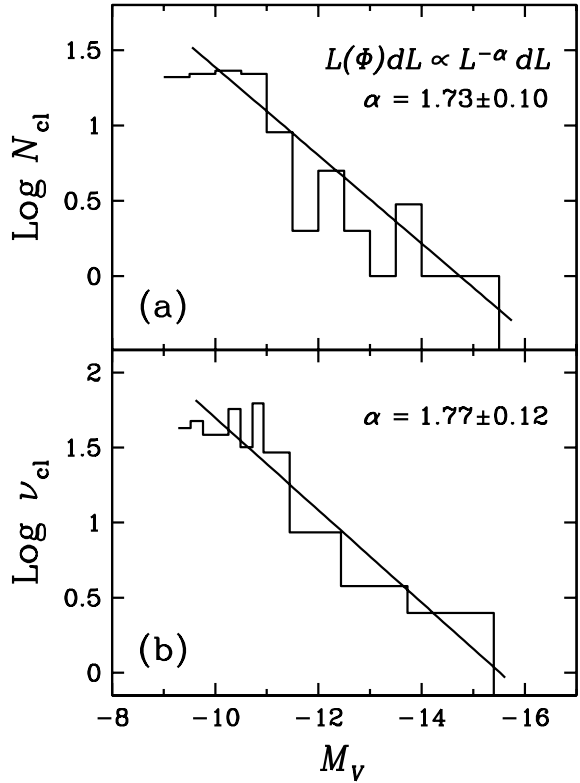}
  \caption{
Luminosity function of candidate young massive clusters in \n34.
(a) Logarithm of number of clusters per 0.5-mag bin, \nclust,  plotted
vs.\ $M_V$.
(b) Logarithm of number of clusters per magnitude interval, \nuclust,
derived from constant-number bins of varying widths and plotted vs.\
$M_V$.
Straight lines show power laws fit by least squares, with exponent $\alpha$
marked (for details, see text).
   \label{fig12}}
\end{figure}

\subsubsection{Luminosity Function}
\label{sec352}

Of the 117 candidate young massive clusters, 87 are more luminous than
$M_V = -10.0$.
The most luminous cluster has $M_V = -15.36$, while the faintest candidate
cluster detected on the one \hst/WFPC2 500s exposure has $M_V = -9.32$.

The cluster luminosity function (LF) was derived as follows.
To minimize effects of incompleteness, which are most severe near the
galaxy's center (\S~\ref{sec351}), we excluded the central 17 clusters
with $r < 1.25$ kpc from the analysis, leaving 100 clusters from which
to determine the LF.
Then we binned the clusters in bins of 0.5 mag width centered at
$M_V = -9.25$, $-$9.75, \dots, $-15.25$.
Figure~\ref{fig12}a shows the resulting LF plotted logarithmically versus
$M_V$ (histogram) and---superposed---a best-fit power law,
$L(\Phi) dL \propto L^{-1.73\pm 0.10} dL$ (straight line).
The fit was computed by least squares in $(M_V, \log\nclust)$ space
with weights proportional to \nclust, the number of clusters per bin.
The lowest-luminosity bin was given zero weight because of the incomplete
number of clusters in it.

As a check and because of the small numbers of clusters involved, we
also binned clusters into constant-number bins of varying widths (2 bins
with 5 clusters each for the brightest clusters, plus 9 bins with 10
clusters each for the remainder). 
For each bin, the number of clusters per unit magnitude, \nuclust, was
computed by dividing the number of clusters by the bin width. 
Figure~\ref{fig12}b shows the resulting LF plotted logarithmically vs
$M_V$ and the best-fit power law derived from it in a similar fashion as
above, $L(\Phi) dL \propto L^{-1.77\pm 0.12} dL$.
Obviously, the two derived power laws agree to well within the errors
of their exponents, and we adopt a mean value of $-1.75\pm 0.1$ for
the exponent of the best fit.

The resulting LF for the young massive clusters of \n34,
$L(\Phi) dL \propto L^{-1.75\pm 0.1} dL$, is in good accord with the
power laws found for cluster systems in many other merging and starburst
galaxies.  These power laws have an average exponent of $-1.93$ with a
scatter of $\pm 0.18$ \citep{whit03}.

\subsubsection{Cluster Velocities}
\label{sec353}

Table~\ref{tab06} gives the heliocentric radial velocities \czhel\
measured for five of the star clusters.
These velocities represent averages of velocities measured by two
different methods.
First, we determined a mean velocity from 5--12 absorption lines
(see $N_{\rm abs}$ in Col.\ 3) measured individually in each cluster
spectrum.
And second, we also determined a mean velocity for each cluster via
cross-correlation.
Since none of the velocity standards observed during the run matched
the A-type cluster spectra well, we used as a template the spectrum of
the cluster \n7252:\,W3 obtained years earlier with the Blanco 4-m
telescope at similar resolution \citep{ss98}.
This cluster has a similar age and spectrum, and its radial velocity
is known with high accuracy from {\em VLT}/UVES observations,
\czhel~= $4822.5 \pm 1.0$ \kms\ (\citealt{mara04}; cf.\ with
$4821 \pm 7$ \kms\ by \citealt{ss98}).
Results from the two methods agreed to within the combined errors, and
the velocities given in Table~\ref{tab06} represent a weighted average
of the two mean velocities measured for each cluster.

Also given in Table~\ref{tab06} are the cluster velocities $\Delta v$
relative to the nucleus of \n34.
These relative line-of-sight (LOS) velocities were computed from
$$\Delta v = (cz_{\rm hel} - cz_{\rm hel,sys})/(1+z_{\rm hel,sys}),$$
where the denominator is a relativistic correction and the systemic
velocity of \n34 is $cz_{\rm hel,sys} = 5870 \pm 15$ \kms\ (\S~\ref{sec33}).
The values of $\Delta v$ for the five clusters lie in the range
$-210 \la \Delta v \la +80$ \kms, indicating that all five clusters are
physically associated with \n34.

The estimated systemic velocity of the five clusters is $5841 \pm 48$ \kms\
and agrees to within $<\,1\sigma$ with the above systemic velocity of \n34. 
The cluster velocity dispersion estimated via the \citet{pryo93}
maximum-likelihood method is at least $100\pm 36$ \kms, yet remains very
uncertain due to the small number of measured clusters and their one-sided
projected spatial distribution.

\begin{figure*}
  \centering
  \includegraphics[scale=0.9]{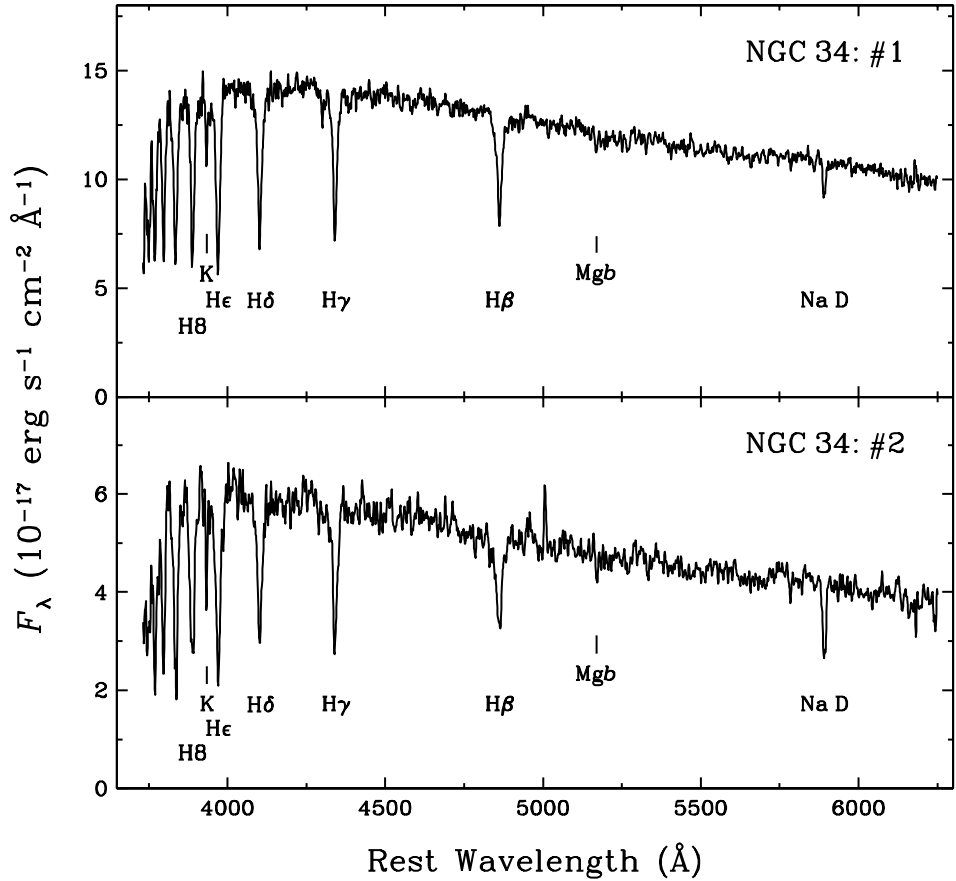}
  \caption{
Ultraviolet-to-red spectra of clusters 1 and 2 in \n34, obtained with the
Baade 6.5-m telescope and LDSS-2 spectrograph through a
$1\farcs03\times 330\arcsec$ slit oriented at various position angles (see
Table~\ref{tab01}). 
The spectra are flux calibrated and plotted vs.\ rest wavelength.
The spectrum of Cluster~2 has been slightly Gaussian smoothed 
($\sigma = 1.0$ \AA) for better display.  Note in both spectra the
strong Balmer absorption lines indicative of A-type main-sequence stars.
Cluster~2 also shows some residual [O~III]$\lambda\lambda$4959,\,5007 line
emission and strong Na\,D absorption, both indicative of gas in its
environment.
   \label{fig13}}
\end{figure*}

The LOS velocities and apparent spatial alignment of clusters 3, 7, 13, and
2 provide strong evidence against these clusters rotating in a disk and some
evidence in favor of halo-type motions.  These four clusters lie along an
apparent line extending from 2.3 kpc NNW of the nucleus to 2.7 kpc SE and
reaching its closest distance of 0.8 kpc from the nucleus between clusters
7 and 13.  Sequentially, the LOS velocities of clusters 3, 7, 13, and 2 are
$+$80, $-$210, $+$79, and $-$20 \kms, respectively, suggesting the presence
of random motions larger than any possible disk-rotation components.

\subsubsection{Cluster Ages and Metallicities}
\label{sec354}

The two brightest of the five spectroscopically observed clusters yielded
spectra of sufficient quality to permit a simple analysis of their likely
ages and metallicities.
Figure~\ref{fig13} displays the flux-calibrated spectra of these two
clusters, numbers 1 and 2 in Table~\ref{tab04}, plotted vs the rest
wavelength.
As pointed out above (\S~\ref{sec24}), the strong Balmer absorption
lines are indicative of A-type main-sequence stars, suggesting that both
clusters have ages roughly in the range 10$^8$--10$^9$ yr.

Given the spectra's relatively low signal-to-noise ratios, we have found
it difficult to measure metal lines with the accuracy required for a
reliable determination of the metallicities.
Yet the Balmer absorption lines, which contain most of the age information,
are strong and easily measured.
Hence, we adopt a two-step procedure similar to the one used by
\citet{ss98}.
First, we assume that the clusters have approximately solar metallicities
and determine their ages from three Balmer-line indices via comparison
with model spectra computed for clusters of $Z = \zsun$.
Then, we estimate the metallicities as best we can and check whether the
assumption of solar metallicity is consistent with the observations.

To determine cluster ages and estimate metalliticies, we measured 12 Lick
linge-strength indices---including \hbet, \hgama, \hdela, \mgb, and
several Fe indices \citep{fabe85,wort97,trag98}---from appropriately
smoothed versions of the observed cluster spectra (i.e., smoothed to
the $\sim$8--11 \AA\ resolution of original Lick spectra).
For comparison, we also measured indices for the nuclear spectrum.
Table~\ref{tab07} presents the values of seven indices used in the final
analysis, plus the index combination \mgfe\ (see below) and logarithmic
ages derived for the two clusters.

\begin{figure}
  \centering
  \includegraphics[scale=0.50]{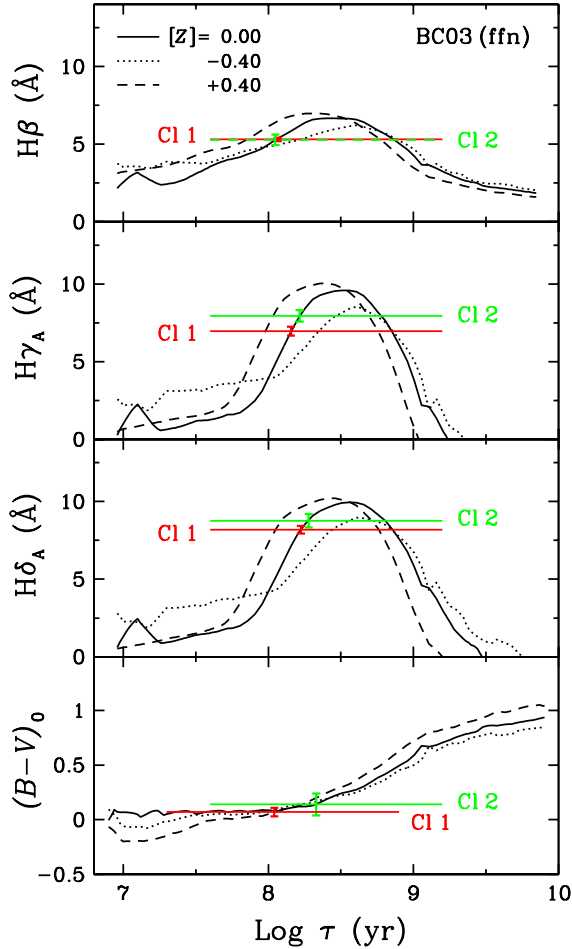}
  \caption{
Evolution of the Lick indices \hbet, \hgama, \hdela\ and color index
\bvzero\ of model star clusters as a function of age $\tau$ for three
metallicities ({\em solid, dotted, and dashed curves}), compared with
values measured for clusters \n34:\,1 and 2 ({\em horizontal lines}).
Note that the color index \bvzero\ favors the lower of two possible
Balmer-line ages for each cluster and line index.
   \label{fig14}}
\end{figure}

\begin{figure*}
  \centering
  \includegraphics[angle=-90,scale=0.65]{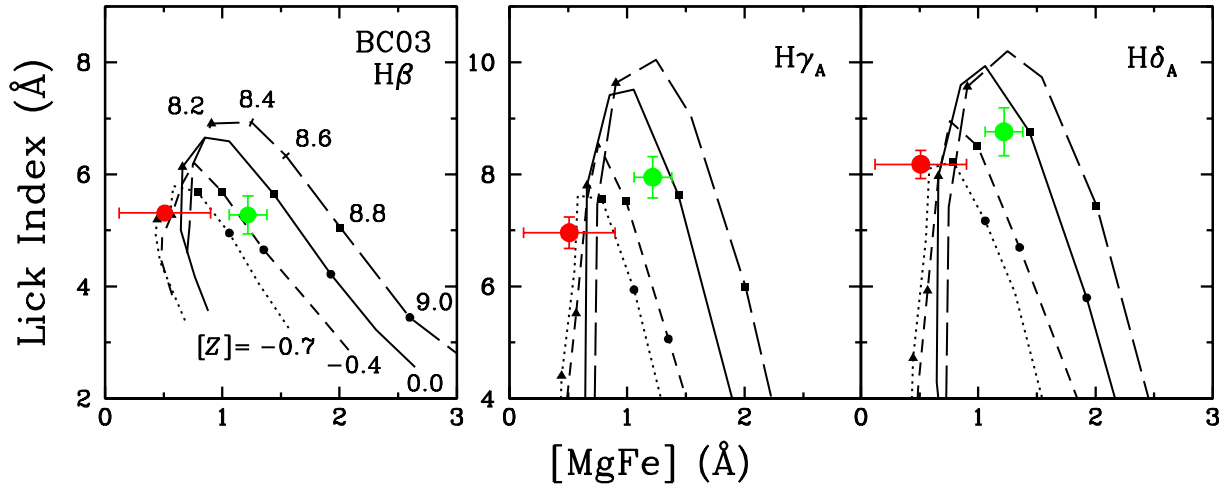}
  \caption{
\hbet, \hgama, and \hdela\ vs \mgfe\ diagrams for clusters 1 ({\em red
data points}) and 2 ({\em green}) in \n34.  Evolutionary tracks computed
by BC03 for model star clusters of metallicities $[Z] = -0.7$, $-$0.4,
0.0, and $+$0.4 ({\em long dashes}) are shown for comparison, with
selected logarithmic ages marked along each track.  From these diagrams
Cluster 1 appears to be definitely young ($<$300 Myr), though of ill
determined metallicity, while Cluster 2 appears to be about 600 Myr
old and of slightly subsolar metallicity.
   \label{fig15}}
\end{figure*}

\begin{figure}
  \centering
  \includegraphics[scale=0.75]{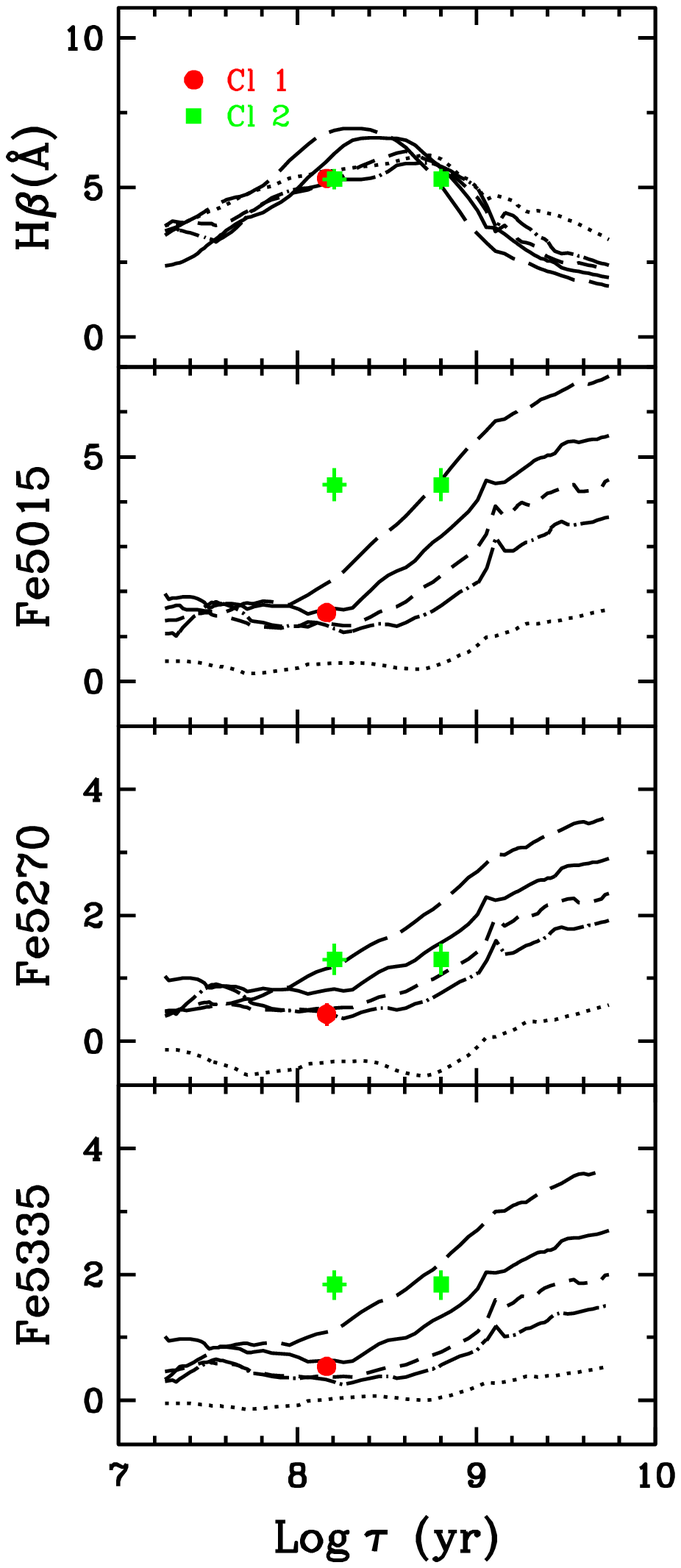}
  \caption{
Lick spectral indices \hbet, Fe5015, Fe5270, and Fe5335 measured for
clusters 1 and 2 in \n34 and compared
with spectral evolution models by BC03 for clusters of five different
metallicities: $[Z] = -1.7$ ({\em dotted curves}), $-$0.7, $-$0.4, 0.0,
and $+$0.4 ({\em long-dashed curves}).  The data points for Cluster~1 are
plotted at the logarithmic age given in Table~\ref{tab07}, while those
for Cluster~2 are plotted at both the low and high logarithmic ages
given in that table.  If one adopts the high age for Cluster~2, both
clusters have roughly solar metallicities to within a factor of about
2--3;  for details, see \S~\ref{sec354}.
   \label{fig16}}
\end{figure}

Figure~\ref{fig14} illustrates our derivation of cluster ages from the
measured three Balmer-line indices \hbet--\hdela\ and the broad-band
color \bvzero.
The plotted curves show the predicted evolution of these four quantities
as a function of logarithmic age, as computed by
\citet[hereafter BC03]{bc03} for model star clusters of logarithmic
metallicities $[Z] \equiv \log(Z/\zsun) = -0.4$, 0.0, and $+$0.4
relative to the sun.\footnote{
Using cluster models computed by \citet{korn05} instead of BC03
yields similar results.}
To compare the observations with the models, the figure also shows the
line indices and colors measured for clusters 1 and 2 as horizontal lines.
The ages of these two clusters were then determined from the figure as
follows.
Logarithmic ages were read off at the intersections between the horizontal
lines marking the measurements and the curves representing the model
indices and color for $[Z] = 0$.
Since for each cluster the Balmer-line indices yield two possible age 
values, the measured color index \bvzero---corrected for Milky-Way
foreground reddening---was used to try to select the more likely of the
two values.
In each case, the color index seems to favor the younger of the two
possible ages (especially when one considers that any correction for
internal reddening in \n34 would yield a bluer intrinsic cluster color).

For Cluster~1, the value of \bvzero\ and its small error favor the younger
age so clearly that we simply computed a weighted mean logarithmic age
from the individual ages obtained for \hbet, \hgama, and \hdela.
This mean logarithmic age, based on an assumed solar metallicity, is given
in the last column of Table~\ref{tab07} and corresponds to a (linear) age
of $150\pm 20$ Myr.

For Cluster~2, the situation is more complicated.  The \bvzero\ color
also favors the younger of two possible ages, yet various metal-line
indices seem to point strongly toward the older age.  We illustrate this
with two figures containing diagnostic diagrams.

Figure~\ref{fig15} shows each of the three age-sensitive Lick indices
\hbet--\hdela\ plotted vs the metallicity index \mgfe\ \citep{gonz93}.
This index is defined through
[MgFe]~$\equiv [{\rm Mg} b\times \case{1}{2} ({\rm Fe5270} +
{\rm Fe5335})]^{1/2}$,
where \mgb, Fe5270, and Fe5335 are Lick indices expressed in angstroms;
it is nearly insensitive to variations in $\alpha$/Fe, the ratio of
$\alpha$-elements to iron \citep{tmb03}.  Besides the data points for
the two clusters, each panel also displays evolutionary tracks for model
star clusters computed by BC03 for metallicities of $[Z] = -0.7$, $-$0.4,
0.0, and +0.4.\footnote{
Note that for clusters of age $\la$500 Myr, the traditional grids of
isochrones and isometallicity lines become severely jumbled, whereas
evolutionary tracks yield a relatively clear picture.}
Whereas Cluster~1 appears to be young ($<$300 Myr) no matter what its
exact metallicity is, Cluster~2 in these diagrams appears to be
nearly 630 Myr old ($\log\tau \approx 8.8$) and of slightly subsolar
metallicity. 

A similarly high age for Cluster~2 seems to be indicated by the observed
strengths of three individual Fe-line indices, two of which have already
been used in combination with the \mgb\ index above.
Figure~\ref{fig16} plots the predicted and observed Lick indices \hbet,
Fe5015, Fe5270, and Fe5335 vs logarithmic cluster age.
Whereas at its younger age Cluster~1 falls near or slightly below the
solar-metallicity track on all three Fe-index diagrams, Cluster~2 falls
well outside any evolutionary tracks in two of the three Fe-index diagrams.
However, when plotted at its older age (as inferred from the Balmer-line
indices) Cluster~2 falls within about a factor of two of the
solar-metallicity tracks in all three Fe-index diagrams.
Hence, for this cluster the higher mean logarithmic age given in
Table~\ref{tab07}, corresponding to an age of $640\pm 40$ Myr, seems to
be at least as likely as the lower age of $160\pm 30$ Myr favored
by the \bvzero\ index.

In conclusion, Cluster~1 seems definitely young ($\sim$150 Myr), while
Cluster~2 could be similarly young but appears, from the strengths of
its Mg and Fe lines, more likely to be about four times as old
($\sim$640 Myr).
Both clusters seem to have metallicities compatible with our assumption
of roughly solar metallicity.\footnote{
Note that we have avoided drawing any conclusions from the \ion{Ca}{2} K
and \ion{Na}{1} D lines which may, especially in Cluster~2, feature
significant nonstellar components due to \n34's gaseous outflow
(\S~\ref{sec34}).}

\subsection{Parameters of \n34}
\label{sec36}

To summarize the main results of \S~\ref{sec3}, Table~\ref{tab08} presents
the relevant parameters of \n34 collected into one place.

\section{DISCUSSION}
\label{sec4}

The present discussion addresses various issues concerning the structure
of \n34, the nature of its young massive clusters, and its gaseous outflow.
It concludes with a description of the sequence of events that may have
unfolded during the merger up to its present stage.

\subsection{\n34: Remnant of a ``Wet'' Unequal-Mass Merger}
\label{sec41}

\n34 exhibits several signatures that we have come to associate with mergers
of gas-rich disks:
two long main tidal tails, despite the galaxy's relatively isolated position
(\S~\ref{sec31});
a single main body with an apparently single nucleus;
plenty of \hi\ and molecular gas, $M_{\rm H\,I\,+\,H_2} = 1.2\times
10^{10}\,\msun$ (\S~\ref{sec1} and Table~\ref{tab08});
a starburst with a SFR of 50--90\,\msunyr\ \citep{vald05,prou04};
and, as an extra, a likely AGN. 
In addition, \n34 also features a remarkably blue, exponential stellar disk
(\S~\ref{sec32}) with a scale length of about 1.6 kpc (Table~\ref{tab05})
and extending out to a radius of at least 10 kpc (Fig.~\ref{fig08}).

Three questions of interest concerning the past merger are:
(1) Is this blue exponential disk a surviving disk, or is it a new disk
that formed from gas pooled during or after the merger?
(2) What was the approximate mass ratio of the two disk galaxies that
merged?
And (3), has this merger essentially run to completion?

\begin{figure}
  \centering
  \includegraphics[scale=0.6]{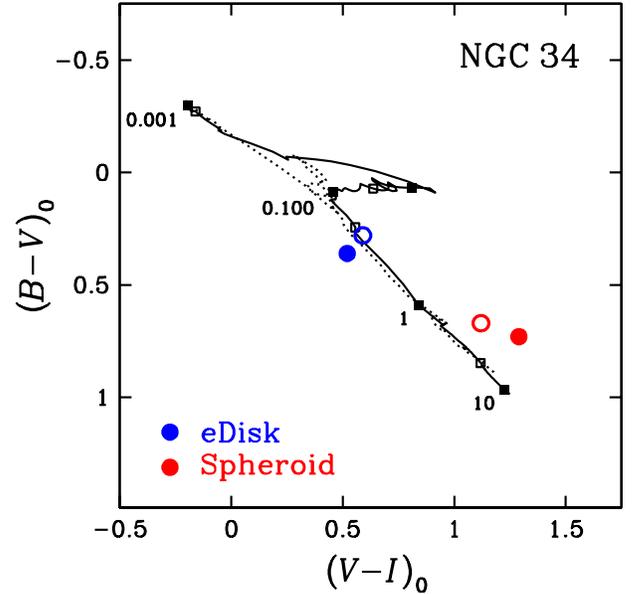}
  \caption{
Two-color diagram comparing the \bvzero,\vizero\ colors of \n34's
exponential disk and spheroid with evolutionary tracks for single-burst
stellar populations of metallicity $[Z]=0$ ({\em solid line}) and
$-$0.4 ({\em dotted line}) computed by BC03.
The filled squares along the solar-metallicity track mark population
ages of 0.001, 0.010, 0.100, 1, and 10 Gyr, while the open squares
mark ages of 0.003 to 3 Gyr.
The two filled color dots indicate the locations of the best-fit model
spheroid and disk without central hole, while the two open circles 
indicate those for the best-fit model spheroid and disk {\em with} a
central hole.
(The dot sizes correspond to about $\pm$2-$\sigma$ errors in the color
indices.) 
Note that regardless of the absence or presence of a central hole, the
disk colors suggest a single-burst disk age of approximately 400 Myr.
   \label{fig17}}
\end{figure}

In trying to address the first question, we note that the prominence of
the exponential disk diminishes strongly from $B$ to $I$, and the disk is
undetectable in the $K$ surface-brightness profile measured by RJ04
(\S~\ref{sec32}).
This suggests that the disk consists mainly of stars younger than
$\sim$1~Gyr, since any significant older stellar component of the disk
should show in the $K$ profile.
The position of the disk in the \bvzero\ vs \vizero\ two-color diagram
shown in Fig.~\ref{fig17} supports this hypothesis and allows us to estimate
the age of the stellar component that dominates the optical light of the
disk. 

Shown in this diagram are evolutionary tracks for single-burst stellar
populations (SSP) of metallicities $[Z]=0$ and $-$0.4 computed by BC03
and the colors of the best-fit exponential disk ({\em blue filled and open
circle}) and spheroid ({\em red circles}).
The filled circles mark the disk and spheroid colors for the best-fit
model with a full exponential disk, while the open circles mark the
corresponding colors for the best-fit model with a disk that has a
central hole (see \S~\ref{sec32}).
The fact that in both cases the location of the disk in the two-color
diagram falls close to the SSP tracks suggests that the disk light may,
indeed, be dominated by a postburst population with perhaps a relatively
narrow age spread.
Projecting the two disk locations perpendicularly onto the model evolutionary
tracks yields single-burst disk ages of $\tau_{\rm eD} \approx 380\pm 50$ Myr
for $[Z]=0$ and $440\pm 50$ Myr for $[Z]=-0.4$, respectively, where the
estimated errors reflect uncertainties in metallicity and extinction, but
not in the systematics of the model.

How massive is this disk, if indeed it formed mainly over a relatively short
period about 400 Myr ago?
With a mass-to-light ratio of $M/L_V=0.22\pm 0.02\,(M/L_V)_{\odot}$ for
an SSP of $[Z]=0$ and the above age $\tau_{\rm eD}$ (BC03), the absolute
magnitudes of the best-fit model disk translate into disk masses of
$M_{\rm eD} = (3.3\pm 0.3)\times 10^9\,\msun$ for the full disk and
$(2.1\pm 0.2)\times 10^9\,\msun$ for the disk with a central hole,
respectively.
These estimated disk masses represent only about 27\% and 17\%, respectively,
of the total mass of cold gas still present in \n34 ($M_{\rm H\,I\,+\,H_2}
= 1.2\times 10^{10}\,\msun$, see \S~\ref{sec1}).
Hence, this blue disk may well have formed from gas pooled in the late
stages of the merger.

This brief discussion is not meant to exclude the possibility that the
stellar disk may, at a diminished rate, continue to grow even at the present
time.
Upon visual inspection of the \hst-archival $V$ image one can see a close
correspondence between some of the young star clusters and their surrounding
dust lanes and knotty luminous filaments.
This correspondence suggests that, especially near the two major dust lanes
that {\em may} connect the inner disk to the outer tails (\S~\ref{sec31}),
stars and clusters may still be forming and, therefore, adding to the
disk's stellar mass.

In trying to address the second question, concerning what the approximate
mass ratio of the two input disk galaxies may have been, we note that
there are at least three clues.

First, the blue exponential disk---being most likely newly formed rather
than a surviving structure---suggests that the merger leading to \n34
must have been close to being a major merger (here defined as $m/M\ga 1/3$),
or else the dominant of the two old-star disks would have partially survived
\citep[e.g.,][]{barn98,naab03}.
Therefore, a clearly minor merger with $m/M < 1/4$ seems unlikely, and
the lack of a moderately massive old disk places a lower limit of
$\sim$0.3 on the mass ratio of the two input galaxies.

Second, the fact that the N tail appears significantly brighter than the
S tail argues against a major merger of two nearly equal-mass disks,
making a mass ratio close to 1:1 seem unlikely.
By exclusion, then, the likely mass ratio of the two input disks must
have been somewhere in the range\ \ $1/3\la m/M\la 2/3$.

Third, another morphological feature of \n34 also seems to argue for a
mass ratio approximately in the above range.
On high-contrast displays (e.g., Figs.~\ref{fig01}a and \ref{fig01}e)
\n34 appears unlike either of the two classical remnants of nearly
equal-mass mergers, \n7252 \citep{schw82,himi95} and \n3921 \citep{schw96a}.
Rather, \n34 appears to have grafted onto it a low-surface-brightness
cloud of debris (the NW Cloud, see \S~\ref{sec31}), from which emerge
at least one, and possible two, faint tail-like features.
Model simulations of unequal-mass disk mergers with gas
\citep[e.g.,][]{barn02,naab06} suggest that up to four tidal tails may
form and coexist in direct $m/M=1/3$ mergers, with ``each [tail]
originating from a different disk at a different passage''
\citep[][esp.\ Fig.\ 5]{barn02}.
This situation may, in fact, correspond to the present configuration in
\n34 and would seem to support a recent merger of mass ratio $\sim$1/3.

Independent support for this conclusion comes from the global kinematics
of the remnant.
In a diagram plotting a measure of rotational support,
$(V_{\rm rot}/\sigma)^{\ast}$, versus the isophotal-shape parameter
$\overline{a_4}/a$, \n34 clearly falls in the quadrant of rotationally
supported, disky merger remnants \citep[][esp.\ Figs.\ 4 and 5]{rj06b}.
Simulations of disk--disk mergers with gas by \citet{naab06} suggest that
remnants of mergers with mass ratios of $m/M\approx 1/3$ populate the region
of the diagram containing \n34, whereas remnants of 1:1 mergers do not.

Finally, the third question of interest concerns the current stage of
the disk--disk merger in \n34.  Specifically, has the merger in this
galaxy essentially run to completion?
Given the observed strong nuclear outflow (\S~\ref{sec34}) and general
evidence that starburst-induced outflows may peak around the time when the
two nuclei coalesce \citep[e.g.,][]{cox06}, the presence or absence of a
second nucleus should help us discern just how advanced the current stage
of this merger is.

Despite a past claim for a second nucleus in \n34 \citep{mile96},
we have found no evidence for it and can put some strong limits on its
maximum brightness in the $K$ band.
Early observations at 10~\micron\ showed a very compact, slightly
North--South elongated mid-IR source at the nucleus with a full width
at half maximum of $\la 1\farcs7\times 2\farcs3$ \citep{keto91} or
$\la 700\times 950$ pc at our adopted distance.
A radio continuum map obtained at 8.44 GHz with the VLA and a
$0\farcs25\times 0\farcs40$ beam shows an essentially unresolved central
source, with possibly a second, much weaker source about $1\farcs2$ south
(Fig.~1 in \citealt{cond91}).
This second source is not explicitly mentioned by Condon et al.\ and,
being $\sim$300$\times$ weaker than the primary source, can hardly be
counted as evidence for a second nucleus.

The only positive evidence for a second nucleus stems from mid-IR
observations at 8.8~\micron\ and 12.5~\micron\ carried out with
SpectroCam-10, a diffraction-limited mid-IR camera, at the Palomar 5-m
telescope \citep{mile96}.
These observations suggested the presence of a second mid-IR source about
$1\farcs2$ south of the primary nucleus, with a claimed brightness ratio
of $\sim$0.6 at 8.8~\micron\ and $\sim$0.4 at 12.5~\micron\ relative to
the primary nucleus.
Deconvolutions seemed to enhance the secondary nucleus and its similarity
with the possible secondary 8.44 GHz radio source (see Figs.~4 and 5 in
\citeauthor{mile96}), yet the data clearly had a low signal-to-noise ratio
and the reality of the second source seems doubtful.

We have conducted a search of NICMOS $K$-band images of \n34 available
via the \hst\ Archive and have found no trace of any secondary nucleus
$\sim${}$1\farcs2$ south of the primary one.
Specifically, measurements on the $K$-band exposure N49J21080 (camera
NIC2, filter F205W, exposure time 512 s, program GO-7268, PI: van der
Marel), which nicely shows many of the young massive star clusters,
demonstrate that any secondary nucleus---if present---must be $\ga\,$7.0
mag fainter in $K$ than the primary nucleus is.
Thus, if really a second nucleus existed and was of an instrinsic
luminosity comparable to that of the primary nucleus, it would have to
suffer an additional extinction of $A_K\ga 7.0$ mag, corresponding to
an additional visual extinction of $A_V\ga 65$ mag.

Although we cannot exclude this possibility, we conclude that---at least
at present---there is no good evidence for a second nucleus of any
significance.
The merger in \n34 likely has run past the stage of the two nuclei
coalescing and is, therefore, essentially complete.\footnote{
The NW Cloud, regarded by some as a second galaxy, contains $\ll$1\% of
the total luminosity of \n34 in the $I$ band and has no visible nucleus.
Likely, it is either a tidal fragment of one of the two previous disk
galaxies or else an interacting dwarf companion galaxy unrelated to
the main recent merger event in \n34.}

In short, as best as we can tell \n34 appears to be the remnant of a
gas-rich (``wet'') merger between two disk galaxies with a mass ratio
likely in the range $1/3\la m/M\la 2/3$.
Out of the pooled gas of the two input galaxies a new exponential disk
of stars has formed, and perhaps continues forming.
As judged by the absence of any detectable second nucleus, the merger
must essentially have run to completion.

\subsection{Nature of \n34's Young Massive Clusters}
\label{sec42}

Another signature of a recent gas-rich merger in \n34 are the many luminous
young star clusters (\S~\ref{sec35}).  As a whole, these clusters resemble
the young globular-cluster systems observed in other recent merger remnants,
including \n3597 \citep{holt96,carl99}, \n3921 \citep{schw96b,schw04}, and
\n7252 \citep{mill97,ss98,mara01}. 
The system includes about 140 luminous clusters ($M_V\la -9.3$) whose surface
number density tracks the underlying galaxy light in $V$ remarkably well.
The system's effective radius, $\reffcl\approx 2.6$--3.1 kpc
(\S~\ref{sec352}), is typical for globular-cluster systems both young and old,
while the clusters' power-law LF, $L(\Phi) dL \propto L^{-1.73\pm 0.10} dL$,
is typical for young cluster systems (e.g., \citealt{whit03}).
Given the presence of a newly-formed, young stellar disk in \n34
(\S~\ref{sec41}), an interesting question is whether this cluster system
displays disk or halo kinematics.
From five clusters with measured velocities, there is tentative evidence
that the kinematics is that of a halo population (\S~\ref{sec353}), but
many more cluster velocities are needed to answer the question in a
definitive manner.

The age distribution of the luminous clusters of \n34 is of great
interest in trying to understand the star formation history of this
recent merger remnant.
In the absence of measured color indices for most clusters, our current
information on ages stems mainly from the spectra available for five
clusters.
The strong Balmer absorption lines in these spectra suggest that all
five clusters have ages in the range of about 0.1--1.0 Gyr and
are, therefore, young.
Young ages for most clusters are also indicated by the fact that nearly
3/4 of the detected candidate clusters (87 out of 117) have absolute
magnitudes in the range $-10.0\geq M_V\geq -15.4$, while at most a few
percent of {\em old} globular clusters are typically more luminous than
$M_V = -10.0$.

At present, we cannot determine whether---within the 0.1--1.0 Gyr
age interval---the young clusters of \n34 have a relatively wide or narrow
age distribution.
Of all candidate clusters, only the two most luminous have individually
determined ages.
As described in \S~\ref{sec354}, there is a possibility that both of
these clusters have ages of about 130--190 Myr, although it looks more
likely that Cluster 1 is about $150\pm 20$ Myr old and Cluster 2 about
$640\pm 40$ Myr old.
The first possibility would suggest that cluster formation may have peaked
sharply in the recent past, while the more likely second possibility
suggests that cluster formation may have been drawn out over $\sim$0.5 Gyr
or may have occurred in two or more bursts.
Additional photometry and spectroscopy are needed to shed light on this
important issue.

What are the likely masses of the young star clusters?
For the two most luminous clusters, we can estimate photometric masses
from their absolute visual magnitudes and spectroscopic ages.
For Cluster 1, the photometric mass is $(15\pm 1)\times 10^6\,\msun$ when
derived from BC03 models with a \citet{chab03} IMF and $(19\pm 2)\times
10^6\,\msun$ when derived from \citet{mara05} models with a \citet{krou01}
IMF.
Hence, this cluster is about 3--4 times more massive than $\omega$\,Cen,
the most massive GC in the Milky Way [$M_{\omega\,{\rm Cen}}\approx (4\pm
1)\times 10^6\,\msun$, \citealt{meyl02}].
For Cluster 2, the corresponding photometric masses are about
(20--24)$\times 10^6\,\msun$ (or 5--6 $M_{\omega\,{\rm Cen}}$) if
the cluster is $\sim$640 Myr old, and about (9--11)$\times 10^6\,\msun$
(or 2--3 $M_{\omega\,{\rm Cen}}$) if it is only $\sim$160 Myr old.
Hence, both clusters clearly are young {\em massive} clusters (YMCs),
often understood to be clusters of mass $\ga$10$^5\,\msun$.
By association, presumably most of the 87 candidate clusters more luminous
than $M_V = -10$ are YMCs as well.

Are the YMCs of \n34 gravitationally bound and, therefore, genuine young
globular clusters?

Although we have not attempted to measure effective radii for these
clusters from the one available archival \hst/WFPC2 $V$ exposure, we have
measured their apparent full widths at half maximum with the IRAF task
{\em imexamine} and have compared the values with widths similarly measured
for the young globular clusters S1 and S2 in \n3921 \citep{schw96b}, a
merger remnant at nearly the same redshift distance as \n34.
In both cases, the measurements were made from $V$ exposures taken with
the Planetary Camera (PC) of WFPC2 within eight months of each other.
The apparent widths of the five \n34 clusters observed spectroscopically
are a few percent smaller than the apparent widths of clusters S1 and S2 in
\n3921, which both have measured effective radii of $\reff\la 5$ pc
\citep{schw04}.
Hence, the YMCs of \n34 very likely have effective radii of $\la$\,5 pc
as well, radii that are typical for globular clusters.

Given these normal effective radii and the estimated ages of $\ga$10$^8$ yr,
corresponding to at least 25--50 internal crossing times
($t_{\rm cr}\approx 2$--4 Myr), all five \n34 clusters observed
spectroscopically must be gravitationally bound and are, therefore,
{\em massive} ($\ga{}2\times 10^6\,\msun$) {\em young globular clusters}.

These clusters are likely to be long-lived.
Most of the early mass loss due to supernovae and stellar evolution, which
can disrupt very young clusters, occurs during the first $\sim$100 Myr
(e.g., \citealt{fall01,boil03}).
Since the five \n34 clusters are all older than 100 Myr, yet of normal
compactness, they have already survived their most disruption-prone
period and are likely to survive for at least several Gyr and perhaps
a Hubble time or longer (e.g., \citealt{baum06}, esp.\ Fig.~3).
Their individual lifetimes will depend, in part, on their orbits
and---especially---on whether they reside in the halo or the disk.

In summary, at least the five spectroscopically observed clusters of
\n34 are genuine, massive young globular clusters with long expected
lifetimes.
What fraction of the other 82 YMCs more luminous than $M_V = -10$ can
be described in a similar fashion depends on their individual ages and
thus remains unknown at present.

\subsection{\n34's Outflow}
\label{sec43}

As revealed mainly by the strong blueshifted D lines of \ion{Na}{1}, the
center of \n34 drives a strong outflow of cool, neutral gas with a mean
(i.e., center-of-line) velocity of $\vnad = -620\pm 60$ \kms\ and a maximum
detected velocity of about $-1050\pm 30$ \kms\ (\S~\ref{sec34}).
Much lower net outflow velocities averaging about $-75$ \kms\ are also
observed in the ionized gas.
As far as we can tell from the spectra of clusters 3 and 7, the \ion{Na}{1}
outflow seems to extend northward of the nucleus in a fan-like structure
out to at least $\sim$2.3 kpc NNE in the case of Cluster 3.

Our discovery of this outflow is hardly surprising given the estimated
SFR of $70\pm 20\,\msunyr$ in \n34 (\S~\ref{sec1}).
While most {\em ultraluminous} infrared galaxies (ULIRGs) feature strong
gaseous outflows, many LIRGs also feature similar, though weaker, outflows
\citep[e.g.,][]{heck00,mart05a,veil05}.
It appears that most galaxies with centrally concentrated star formation
in excess of $\sim$10\,\msunyr\ may drive some form of \ion{Na}{1} outflow,
regardless of whether they contain an AGN or not \citep{rupk05}.

\begin{figure}
  \centering
  \includegraphics[scale=0.45]{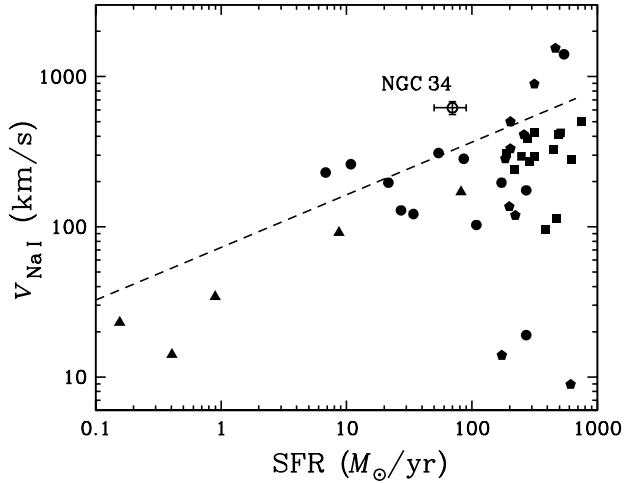}
  \caption{
Mean (center-of-line) velocity of the \ion{Na}{1} outflow in \n34 compared
to similar velocities for a sample of 41 starburst galaxies assembled by
\citet{mart05b}. 
The mean outflow velocities \vnad\ are plotted vs each galaxy's star
formation rate.
The various data points represent ULIRGs ({\em squares} and {\em pentagons}),
LIRGs ({\em filled circles}), and dwarf starburst galaxies ({\em triangles}),
while the dashed line represents an upper-envelope fit taking projection
effects into consideration (for details and references to data, see
\citealt{mart05b}, esp.\ \S~\ref{sec42} and Fig.~\ref{fig06}).
Note the high mean \ion{Na}{1} velocity of \n34's outflow, exceeding the
mean outflow velocities of all but three of the 24 more intensely
star-forming ULIRGs.
   \label{fig18}}
\end{figure}

What is perhaps surprising about the \ion{Na}{1} outflow of \n34 is its
relatively high mean and maximum velocities.
Figure~\ref{fig18} shows the mean outflow velocity (data point with error
bars) plotted in a diagram of \vnad\ vs SFR for 41 starburst galaxies
assembled by \citet[esp.\ Fig.\ 6]{mart05b}.
These galaxies represent the highest surface-brightness objects in their
luminosity class and are thought to outline the highest mean outflow
velocities.
The dashed line represents an upper-envelope fit made via a simple model
of projection effects (for details, see \citealt{mart05b}).
The figure suggests, then, that the neutral-gas outflow of \n34 is
exceptionally strong for the estimated SFR of $70\pm 20\,\msunyr$.
Even when compared to \ion{Na}{1} outflows in ULIRGs, this outflow has
an unusually high mean velocity:
Whereas the SFRs of the 15 ULIRGs with outflows detected by Martin
(squares in Fig.~\ref{fig18}), $190\la {\rm SFR}\la 750\,\msunyr$, are
2--10$\times$ higher than the SFR of \n34, the mean outflow velocities
of these ULIRGs lie in the range 60--500 \kms, while the corresponding
velocity measured for \n34 is\,\ $|\vnad| = 620\pm 60$ \kms.

Despite this ongoing high-velocity outflow there is still plenty of cool
gas left in \n34 ($M_{\rm H\,I\,+\,H_2} = 1.2\times
10^{10}\,\msun$, see \S~\ref{sec1}), and an additional significant amount
of cool gas has already been converted into a young stellar disk
(\S~\ref{sec41}).
Even with its infrared luminosity of $\lir\approx 10^{11.5}\,\lsun$ \n34
is clearly not in the league of the most luminous infrared galaxies
($\lir > 10^{12}\,\lsun$) where AGNs may---according to an increasingly
widely held view---scour remnants free of gas so efficiently as to rapidly
suppress all star formation.
Specifically, the transition to this extreme regime may occur for very
massive remnants with velocity dispersions of $\sigvel\approx 240$\,\kms\
\citep{scha06}, while \n34 has only $\sigvel = 201\pm 8$\,\kms\
\citep{rj06a}.
Hence, the galaxy's cool-gas mass, SFR, and depth of the potential well
are all consistent with each other.

Perhaps the most interesting lesson contributed by \n34 is that the sequence
of various events may matter.
As we are about to discuss, the outflow of cool gas in this galaxy seems
to be occurring toward the end of the main starburst and after the gas has
settled into a major new disk.

\subsection{Sequence of Events}
\label{sec44}

With a strong outflow of cool gas under way in \n34, it is of interest
to establish what events preceded it during the recent merger.

Summarizing our findings, the spatially extended phase of the
merger-induced starburst seems to have peaked more than $\sim$100 Myr
ago, giving birth to a system of genuine young globular clusters with
ages approximately in the range 0.1--1.0 Gyr (\S~\ref{sec42}).
From gas settling into a disk toward the end of the merger
\citep[e.g.,][]{barn02}, a new stellar disk formed perhaps as much as
$\sim$400 Myr ago (\S~\ref{sec41}).
A similar age is inferred from the optical nuclear spectrum, which
suggests a 300--400 Myr old poststarburst population surrounding the
highly obscured nucleus proper.
The lack of any detectable second nucleus suggests that the nuclei
of the two participant galaxies of mass ratio $1/3\la m/M\la 2/3$ have
coalesced.
Yet, there still is a strong gaseous outflow.
Therefore, {\em if}\, this outflow is of relatively short duration (say,
$\la$50 Myr), then it is occurring {\em after} most of the merger
processes have run to completion.

We now elaborate on some details of this sketchy scenario.

First, although the present cluster photometry does not allow us to exclude
that some massive clusters may still be forming, the fact that five of the
most luminous clusters all appear spectroscopically older than $\sim$100 Myr
strongly suggests that the main cluster-formation epoch, and hence the
peak of the galaxy-wide starburst, is over.
If there were younger clusters of similar mass, they should---after
all---be more luminous, yet they do not seem to be numerous among the
brightest clusters.

Second, similar things can be said about the young stellar disk.
Although it may still be forming stars at a diminished rate, it looks
surprisingly smooth and appears to be at a more advanced evolutionary
stage than, e.g., the intensely star-forming central disk (``mini-spiral'')
of \n7252 \citep{schw82,whit93,mill97}.
The sheer existence of this smooth young disk in \n34 and its classical
exponential structure make it seem unlikely that a strong, gas-depleting
outflow coexisted during the disk's main formation period a few 100 Myr ago.
Hence, like cluster formation disk formation seems to have largely
preceded the present-day strong gaseous outflow.

Finally, although the nuclei appear to have coalesced, there is no firm
evidence as to exactly when this final phase of the merging process
occurred.
Yet, the exceptionally high central $H$-band luminosity density measured
in \n34 \citep{vdma00}, supported by a strong trend toward redder central
colors (Figs.~\ref{fig06} and \ref{fig07}), suggests that a highly obscured
central starburst, likely paired with a weak AGN (\S~\ref{sec1}), is
still occurring.
If so, the nuclei may either still be in the last throes of their
coalescence ($<$50 pc apart) or may have coalesced just recently, creating
the final, concentrated starburst (and AGN) that drives the present
outflow.
This, then, seems to us the most likely explanation of the combined
observations.

One interesting speculation is that---dynamical friction being weaker
and acting more slowly in unequal-mass mergers than in equal-mass
mergers---the seemingly late phase of strong gaseous outflow observed in
\n34 may be a relatively normal consequence of the 1/3 -- 2/3 mass ratio
between the two galaxies that merged.

Whatever the exact interpretation and assuming that gaseous outflows peak
for relatively brief periods, the indications seem strong that at least
in \n34 the current outflow follows the main, galaxy-wide starburst, follows
disk rebuilding, and may either immediately follow or be concurrent with
the coalescence of the two nuclei.

\section{SUMMARY}
\label{sec5}

We have described imaging, photometric, and spectroscopic observations
of the merger remnant \n34 (Mrk 938) obtained with the Baade and du Pont
telescopes at Las Campanas, as well as the photometric analysis of an
archival 500 s exposure in $V$ obtained with the \hst/WFPC2 camera.
\n34 has often been classified as a Sey~2, but its nuclear spectrum
is presently thought to be dominated by a highly obscured starburst, with
a likely weak AGN contribution.
The main results of our observations and analysis are as follows:

1. \n34 is relatively luminous ($M_V = -21.6$ for $H_0 = 70$) and situated
at a distance of 85.2 Mpc ($\czhel = 5870\pm 15$ \kms).
The galaxy features a single, very red nucleus, a main spheroid containing a
prominent blue central disk and much outer fine structure, and a pair of
tidal tails of unequal length and surface brightness, suggestive of two
former disk galaxies of unequal mass.
These galaxies appear to have completed merging.
The remnant now features a rich system of young massive clusters,
the above-mentioned blue exponential disk, and a strong gaseous outflow,
all signatures of a recent gas-rich merger accompanied by a strong starburst.

2. \n34's system of candidate young star clusters comprises about 140 objects
more luminous than $M_V\approx -9.3$.
Their surface number density tracks the underlying galaxy light in $V$
closely, and their luminosity function is a power law typical of young
cluster systems, $L(\Phi) dL \propto L^{-1.73\pm 0.10} dL$.
Of the 117 directly detected candidate clusters, 87 have absolute
magnitudes in the range $-10.0 \geq M_V \geq -15.4$.
Among the most luminous clusters, five observed spectroscopically all
feature strong Balmer absorption lines indicative of ages of
$\sim$0.1--1.0 Gyr.
They have estimated photometric masses of $2\times 10^6 \la M \la 2\times
10^7\,\msun$ (for a Chabrier or Kroupa IMF) and are gravitationally
bound young globulars.
Their systemic kinematics seems to be dominated by relatively large random
motions rather than by disk rotation, but velocities of more and spatially
better distributed clusters will be needed to check this tentative result.
Finally, the cluster system's effective radius,
$\reffcl\approx 2.6$--3.1 kpc, is typical for globular-cluster
systems both young and old.

3. \n34's prominent central disk turns out to be exponential and can be
traced, via surface-brightness profile decomposition, out to at least
10 kpc radius.
Its smooth structure and blue colors, $\bvedzero \approx 0.36$ and
$\viedzero \approx 0.52$, suggest that its optical light may be
dominated by a poststarburst stellar population about 400 Myr old.
This disk contributes $\sim$51\%, 43\%, and 27\% of the total $B$, $V$,
and $I$ light, respectively, if one assumes that it has no central
hole, and about 2/3 of these values if one assumes that it does have
a central hole.
In either case, because of its youth this exponential disk contributes
significantly to the total optical luminosity of \n34 even though its
mass may be relatively modest (2--4$\,\times 10^9\,\msun$ if $\sim$400 Myr
old).

4. As discovered from the broad blueshifted D lines of \ion{Na}{1},
the center of \n34 drives a strong outflow of cool, neutral gas.
The center-of-line velocity of this gas is $\vnad = -620\pm 60$ \kms,
while the maximum detected velocity reaches about $-1050\pm 30$ \kms.
These outflow velocities are unusually high for a galaxy with an infrared
luminosity of $\lir\approx 10^{11.5}\,\lsun$ and estimated SFR of
$70\pm 20$ \msunyr.
Much lower net outflow velocities averaging about $-75$ \kms\ are also
observed in the ionized gas.
We note that---since blueshifted Na D lines are also seen in two of the
cluster spectra---future detailed mapping of the cool-gas outflow should
be feasible via multi-object cluster spectroscopy.

5. The available morphological and structural evidence suggests that
\n34 is the likely remnant of two recently merged gas-rich disk galaxies
that had an estimated mass ratio of $1/3 \la m/M \la 2/3$.
While the merger proceeded, a galaxy-wide starburst seems to have formed
the young massive clusters beginning perhaps $\sim$600 Myr ago and the
young exponential disk beginning shortly thereafter.
By now, the two merging galaxies' nuclei appear to have coalesced, the
starburst has shrunk to a highly obscured central region of $\la$\,1 kpc
radius, and there is a strong gaseous outflow.
The outflow's high velocity may indicate that the obscured central
starburst, and perhaps also the AGN activity, may be at or near their
peak strengths.

We close with an educated guess:  Given the presence of a young stellar
disk and of $\sim$\,$1.2\times 10^{10}$\,\msun\ of cool gas that seems
unlikely to be entirely blown away in the future (\S~\ref{sec43}), \n34
may---over the next few Gyrs---turn into a bulge-dominated Sa galaxy
akin to, though less massive than, The Sombrero (\n4594).

\acknowledgments

We thank Christopher Mihos for drawing our attention to \n34;
Herman Olivares, Felipe S\'anchez, and Hern\'an Nu\~nez for
expert assistance at the telescopes;
Frank Valdes and Michael Fitzpatrick from the former IRAF Help Desk at
NOAO for their steady support;
Ivo Busko for his kind help with the STSDAS fitting package;
Pieter van Dokkum for making available his cosmic-ray removal software;
Gustavo Bruzual and St\'ephane Charlot for the early release of
their latest cluster-evolution models;
and Mark Phillips, Barry Madore, and Jane Rigby for helpful discussions.
This research has benefited greatly from use of the NASA/IPAC
Extragalactic Database (NED), which is operated by the Jet Propulsion
Laboratory, California Institute of Technology, under contract with NASA.
One of us (F.S.) acknowledges partial support from the NSF through grants
AST-99\,00742 and AST-02\,05994.
This paper is dedicated to the memory of Horace W.\ Babcock, who envisioned,
founded, and developed the Las Campanas Observatory.

%%%%%%%%%%%%%%%%%%%%%%%%%%%%%%%  REFERENCES  %%%%%%%%%%%%%%%%%%%%%%%%%%%%%%%%

%\newpage

%%%%%%%%%%%%%%%%%%%%%%%%%%%%%%%%   TABLES   %%%%%%%%%%%%%%%%%%%%%%%%%%%%%%%%%%%
%
% Tables should be submitted one per page, so put a \clearpage before each.
%
% Two options are available to the author for producing tables:  the
% deluxetable environment provided by the AASTeX package or the LaTeX
% table environment.  Use of deluxetable is preferred.
%
% For samples of three different tables, see:
%	~/LATEX/AAS5.2/sample.tex
%

%% \clearpage
%%
%% \centerline{\bf TABLES}
%%
%% Table 1.  Log of Observations of NGC 34
%%

%%%%%%%%%%%%%%%%%%%%%%%  Table 1  %%%%%%%%%%%%%%%%%%%%%%%%
\clearpage

%  n0034_lc02_tab010.latex

\begin{deluxetable}{llllcrrcl}
\tabletypesize{\footnotesize}
\tablewidth{16cm}
%%\tablenum{1}
\tablecolumns{9}
\tablecaption{Log of Observations of NGC 34\label{tab01}}
\tablehead{
  \colhead{}             &
  \colhead{}             &
  \colhead{}             &
  \colhead{}             &
  \colhead{}             &
  \colhead{}             &
  \colhead{Total\ }      &
  \colhead{Wavelength}   &
                           \\
  \colhead{}             &
  \colhead{}             &
  \colhead{}             &
  \colhead{CCD}          &
  \colhead{}             &
  \colhead{P.A.}         &
  \colhead{\,Expos.}     &
  \colhead{Coverage}     &
                           \\
  \colhead{Date}         &
  \colhead{Telescope}    &
  \colhead{Instrument\tablenotemark{a}} &
  \colhead{Detector}     &
  \colhead{Filter}       &
  \colhead{(deg)}        &
  \colhead{(s)}          &
  \colhead{(\AA)}        &
  \colhead{Notes\tablenotemark{b}\ }
}
\startdata
1995 May 13 ....&   {\em HST}    &  WFPC2& Loral   & F606W  & 282.6&  500& 4800--7200& {\em HST} Archive \\
2000 Sep 30 .....&  du Pont 2.5 m&     DC& Tek \#5 & $B$    &   1.2& 6000& 3850--4850& Seeing $0\farcs7$ \\
                 &               &       &         & $V$    &   1.2& 1500& 4900--5800& Seeing $0\farcs8$ \\
                 &               &       &    & $I_{\rm KC}$&   1.2& 3600& 7300--9300& Seeing $0\farcs7$ \\
2002 Sep 1 .......& Baade 6.5 m  & LDSS-2& SITe \#1& \nodata& 351.9& 1200& 3650--6900& Cl.\ 1            \\
                 &               &       &         &        &   8.1& 2400& 3650--6900& Cl.\ 1,\,2        \\
2002 Oct 31 .....&  Baade 6.5 m  & LDSS-2& SITe \#1& \nodata&   8.1& 2400& 3650--6850& Cl.\ 1,\,2        \\
2002 Nov 1 ......&  Baade 6.5 m  & LDSS-2& SITe \#1& \nodata& 144.5& 2400& 3650--6850& Cl.\ 2,\,3,\,7,\,13 \\
                 &               &       &         &        & 144.5&  120& 3650--6850& Nucleus           \\
\enddata
\tablenotetext{a}{DC = Direct camera;\ \ LDSS-2 = Low-Dispersion Survey Spectrograph 2, with\ \
$1\farcs03\times 330\arcsec$\ \ slit.}
\tablenotetext{b}{Cluster numbers refer to objects in Table~\ref{tab04}.}
\end{deluxetable}

%%%%%%%%%%%%%%%%%%%%%%%  Table 2  %%%%%%%%%%%%%%%%%%%%%%%%
%\clearpage

%  n0034_lc02_tab020.latex

\begin{deluxetable}{cccccccc}
%%\tablenum{2}
\tablecolumns{8}
\tablewidth{9cm}
\tablecaption{\bvi\,\ Aperture Photometry of NGC 34\label{tab02}}
\tablehead{
  \colhead{Aper.\tablenotemark{a}} &
  \colhead{Radius}               &
  \colhead{$V$}                  &
  \colhead{$\sigma_V$}           &
  \colhead{\bv}                  &
  \colhead{$\sigma_{B\!-\!V}$}   &
  \colhead{\vi}                  &
  \colhead{$\sigma_{V\!-I}$}
                                   \\
  \colhead{(\arcsec)}            &
  \colhead{(kpc)}                &
  \colhead{(mag)}                &
  \colhead{(mag)}                &
  \colhead{(mag)}                &
  \colhead{(mag)}                &
  \colhead{(mag)}                &
  \colhead{(mag)}
}
\startdata
\phn\phn5.....& \phn1.03    &  14.818 &  0.001 &  0.725 &  0.002 &  1.325 &  0.001 \\
\phn10.....   & \phn2.06    &  14.117 &  0.001 &  0.676 &  0.002 &  1.194 &  0.001 \\
\phn15.....   & \phn3.10    &  13.749 &  0.001 &  0.633 &  0.002 &  1.126 &  0.001 \\
\phn20.....   & \phn4.13    &  13.551 &  0.001 &  0.608 &  0.002 &  1.096 &  0.001 \\
\phn25.....   & \phn5.16    &  13.436 &  0.001 &  0.595 &  0.002 &  1.084 &  0.001 \\
\phn50.....   &    10.3\phn &  13.224 &  0.004 &  0.580 &  0.005 &  1.085 &  0.004 \\
\phn75.....   &    15.5\phn &  13.168 &  0.008 &  0.579 &  0.010 &  1.090 &  0.009 \\
100.....      &    20.6\phn &  13.140 &  0.014 &  0.581 &  0.018 &  1.088 &  0.015 \\
125.....      &    25.8\phn &  13.115 &  0.023 &  0.586 &  0.028 &  1.086 &  0.024 \\
150.....      &    31.0\phn &  13.098 &  0.032 &  0.589 &  0.039 &  1.087 &  0.034 \\
\enddata
\tablenotetext{a} {Aperture diameter (in arcsec).}
\end{deluxetable}

%%%%%%%%%%%%%%%%%%%%%%%  Table 3  %%%%%%%%%%%%%%%%%%%%%%%%
\clearpage

%  n0034_lc02_tab030.latex

\begin{deluxetable}{ccccccccc}
\tabletypesize{\small}
%%\tablenum{3}
\tablecolumns{7}
\tablewidth{10cm}
\tablecaption{\bvi\,\ Surface Photometry of NGC 34\label{tab03}}
\tablehead{
  \colhead{}                     &
  \colhead{}                     &
  \colhead{}                     &
  \colhead{}                     &
  \colhead{{\em HST}/PC}         &
  \colhead{}                     &
  \multicolumn{3}{c}{Du Pont 2.5-m Telescope}
                                   \\[4pt]
  \cline{5-5}
  \cline{7-9}
                                   \\[-7pt]
  \colhead{$r$}                  &
  \colhead{\rdV}                 &
  \colhead{}                     &
  \colhead{}                     &
  \colhead{\muvhst}              &
  \colhead{}                     &
  \colhead{\muvlco}              &
  \colhead{\bv}                  &
  \colhead{\vi}
                                   \\
  \colhead{(arcsec)}             &
  \colhead{(arcsec$^{1/4}$)}     &
  \colhead{\phs$\log r$}         &
  \colhead{}                     &
  \colhead{($\mu$)}              &
  \colhead{}                     &
  \colhead{($\mu$)}              &
  \colhead{(mag)}                &
  \colhead{(mag)}
}
\startdata
\phn0.032   & 0.424 &  $-$1.492 &&  14.44  && \nodata & \nodata & \nodata \\
\phn0.072   & 0.518 &  $-$1.143 &&  14.79  && \nodata & \nodata & \nodata \\
\phn0.116   & 0.584 &  $-$0.935 &&  15.08  && \nodata & \nodata & \nodata \\
\phn0.161   & 0.633 &  $-$0.793 &&  15.33  && \nodata & \nodata & \nodata \\
\phn0.206   & 0.674 &  $-$0.686 &&  15.66  &&  16.60  &  0.784  &  1.792  \\
\phn0.297   & 0.738 &  $-$0.527 &&  16.04  && \nodata & \nodata & \nodata \\
\phn0.388   & 0.789 &  $-$0.411 &&  16.43  && \nodata & \nodata & \nodata \\
\phn0.433   & 0.811 &  $-$0.363 &&  16.63  &&  16.90  &  0.799  &  1.580  \\
\phn0.526   & 0.852 &  $-$0.279 &&  16.91  && \nodata & \nodata & \nodata \\
\phn0.635   & 0.893 &  $-$0.197 &&  17.18  && \nodata & \nodata & \nodata \\
\phn0.698   & 0.914 &  $-$0.156 &&  17.27  &&  17.30  &  0.795  &  1.368  \\
\phn0.842   & 0.958 &  $-$0.074 &&  17.49  && \nodata & \nodata & \nodata \\
\phn0.926   & 0.981 &  $-$0.033 &&  17.69  &&  17.62  &  0.771  &  1.219  \\
\phn1.017   & 1.004 & \phn0.007 &&  17.86  && \nodata & \nodata & \nodata \\[7pt]
\phn1.18\phn& 1.042 & \phn0.072 &&  18.08  &&  17.95  &  0.711  &  1.161  \\
\phn1.44\phn& 1.095 & \phn0.158 &&  18.26  &&  18.22  &  0.683  &  1.131  \\
\phn1.70\phn& 1.142 & \phn0.230 &&  18.46  &&  18.44  &  0.657  &  1.118  \\
\phn1.96\phn& 1.183 & \phn0.292 &&  18.65  &&  18.64  &  0.662  &  1.108  \\
\phn2.22\phn& 1.221 & \phn0.346 &&  18.89  &&  18.77  &  0.667  &  1.100  \\
\phn2.48\phn& 1.255 & \phn0.394 &&  18.91  &&  18.89  &  0.653  &  1.085  \\
\phn3.01\phn& 1.318 & \phn0.479 &&  19.10  &&  19.08  &  0.631  &  1.045  \\
\phn3.65\phn& 1.382 & \phn0.562 &&  19.38  &&  19.32  &  0.630  &  1.019  \\
\phn4.41\phn& 1.449 & \phn0.645 &&  19.69  &&  19.61  &  0.613  &  1.001  \\
\phn5.34\phn& 1.520 & \phn0.728 &&  19.90  &&  19.83  &  0.553  &  0.955  \\
\phn6.46\phn& 1.594 & \phn0.810 &&  20.24  &&  20.16  &  0.517  &  0.922  \\
\phn7.82\phn& 1.672 & \phn0.893 && \nodata &&  20.59  &  0.501  &  0.913  \\
\phn9.46\phn& 1.754 & \phn0.976 && \nodata &&  21.09  &  0.493  &  0.943  \\
   11.45\phn& 1.839 & \phn1.059 && \nodata &&  21.62  &  0.471  &  0.976  \\
   13.85\phn& 1.929 & \phn1.141 && \nodata &&  22.16  &  0.516  &  1.026  \\
   16.76\phn& 2.023 & \phn1.224 && \nodata &&  22.83  &  0.519  &  1.101  \\
   20.28\phn& 2.122 & \phn1.307 && \nodata &&  23.41  &  0.537  &  1.144  \\
   24.54\phn& 2.226 & \phn1.390 && \nodata &&  24.01  &  0.526  &  1.149  \\
   29.69\phn& 2.334 & \phn1.473 && \nodata &&  24.85  &  0.557  &  1.192  \\
   32.66\phn& 2.390 & \phn1.514 && \nodata &&  25.23  &  0.564  &  1.177  \\
   39.52\phn& 2.507 & \phn1.597 && \nodata &&  25.78  & \nodata & \nodata \\
   47.81\phn& 2.630 & \phn1.680 && \nodata &&  26.23  & \nodata & \nodata \\
\enddata
\tablecomments{Partial listing of surface photometry performed.
Measurements at $r \leq 1\farcs02$ are on {\em HST}/PC radius grid, with
\muvlco\ (in $\mu \equiv$ mag arcsec$^{-2}$) interpolated to that
grid, while measurements at $r > 1\farcs02$ are on LCO radius
grid, with \muvhst\ interpolated to this latter grid.
Estimated errors due to uncertainties in the measured
sky level are small at most radii, but exceed $\pm$0.10 mag
(1$\,\sigma$) in \muvlco, \bv, and \vi\ at $r > 30\arcsec$ and
reach 0.29 mag in \muvlco, 0.124 mag in \bv, and
0.135 mag in \vi\ for the last listed values.}
\end{deluxetable}\notetoeditor{Please note that in the above Table~3 some
space after the 14th data line (radius = 1.017) is important, for reasons
clear from the second sentence of `tablecomments'.}

%%%%%%%%%%%%%%%%%%%%%%%  Table 4  %%%%%%%%%%%%%%%%%%%%%%%%
\clearpage

%  n0034_lc02_tab040.latex

\begin{deluxetable}{cccccccccccl}
\tabletypesize{\small}
%%\rotate
%%\tablenum{4}
\tablecolumns{12}
\tablewidth{15cm}
\tablecaption{Positions and Magnitudes of the 20 Brightest Candidate Clusters in NGC 34\label{tab04}}
\tablehead{
  \colhead{}                          &
  \colhead{}                          &
  \colhead{}                          &
  \colhead{}                          &
  \colhead{}                          &
  \multicolumn{2}{c}{$r_{\rm proj}$\tablenotemark{a}} &
  \colhead{}                          &
  \colhead{}                          &
  \colhead{}                          &
  \colhead{}                          &
  \colhead{}
                                        \\
  \colhead{}                          &
  \colhead{}                          &
  \colhead{}                          &
  \colhead{}                          &
  \colhead{}                          &
  \multicolumn{2}{c}{---------------------} &
  \colhead{}                          &
  \colhead{$V$\tablenotemark{b}}      &
  \colhead{$M_V$\tablenotemark{c}}    &
  \colhead{\dvindex\tablenotemark{d}} &
  \colhead{}
                                        \\
  \colhead{Object}                    &
  \colhead{}                          &
  \colhead{R.A.(2000)}                &
  \colhead{Decl.(2000)}               &
  \colhead{}                          &
  \colhead{(arcsec)}                  &
  \colhead{(kpc)}                     &
  \colhead{}                          &
  \colhead{(mag)}                     &
  \colhead{(mag)}                     &
  \colhead{(mag)}                     &
  \colhead{Notes\tablenotemark{e}}
}
\startdata
\phn1....... &&  00 11 06.997 & $-$12 06 21.76 &&  8.82 & 3.64 && $19.38\pm0.01$ & $-$15.36 &  2.03 &  Sp, colors\tablenotemark{f} \\
\phn2....... &&  00 11 06.903 & $-$12 06 31.55 &&  6.78 & 2.80 && $20.04\pm0.01$ & $-$14.70 &  1.85 &  Sp, colors\tablenotemark{g} \\
\phn3....... &&  00 11 06.441 & $-$12 06 22.07 &&  5.53 & 2.28 && $20.47\pm0.01$ & $-$14.27 &  2.00 &  Sp        \\
\phn4....... &&  00 11 06.561 & $-$12 06 25.54 &&  1.92 & 0.79 && $20.65\pm0.02$ & $-$14.09 &  1.75 &            \\
\phn5....... &&  00 11 06.542 & $-$12 06 26.60 &&  0.82 & 0.34 && $20.71\pm0.07$ & $-$14.03 &  2.52 &  Extended? \\  
\phn6....... &&  00 11 06.297 & $-$12 06 21.88 &&  6.56 & 2.71 && $20.77\pm0.01$ & $-$13.97 &  1.94 &            \\
\phn7....... &&  00 11 06.588 & $-$12 06 24.90 &&  2.64 & 1.09 && $20.80\pm0.02$ & $-$13.94 &  1.82 &  Sp        \\
\phn8....... &&  00 11 06.374 & $-$12 06 23.37 &&  4.70 & 1.94 && $20.82\pm0.01$ & $-$13.92 &  1.65 &            \\
\phn9....... &&  00 11 06.491 & $-$12 06 26.85 &&  0.87 & 0.36 && $20.84\pm0.06$ & $-$13.90 &  2.02 &            \\
   10....... &&  00 11 06.429 & $-$12 06 26.20 &&  1.99 & 0.82 && $21.07\pm0.04$ & $-$13.67 &  1.75 &            \\
   11....... &&  00 11 06.314 & $-$12 06 31.87 &&  5.51 & 2.28 && $21.14\pm0.02$ & $-$13.60 &  2.01 &            \\
   12....... &&  00 11 06.554 & $-$12 06 24.94 &&  2.49 & 1.03 && $21.18\pm0.02$ & $-$13.56 &  1.93 &            \\
   13....... &&  00 11 06.707 & $-$12 06 27.02 &&  2.54 & 1.05 && $21.24\pm0.02$ & $-$13.50 &  1.85 &  Sp        \\
   14....... &&  00 11 06.701 & $-$12 06 17.03 && 10.67\phn& 4.40&& $21.42\pm0.01$& $-$13.32&  1.95 &            \\
   15....... &&  00 11 06.369 & $-$12 06 28.53 &&  2.70 & 1.11 && $21.44\pm0.02$ & $-$13.30 &  1.69 &            \\
   16....... &&  00 11 06.438 & $-$12 06 28.10 &&  1.59 & 0.66 && $21.64\pm0.06$ & $-$13.10 &  1.64 &            \\
   17....... &&  00 11 06.427 & $-$12 06 27.50 &&  1.60 & 0.66 && $21.74\pm0.06$ & $-$13.00 &  1.88 &            \\
   18....... &&  00 11 06.447 & $-$12 06 30.49 &&  3.34 & 1.38 && $21.77\pm0.03$ & $-$12.97 &  1.80 &            \\
   19....... &&  00 11 06.643 & $-$12 06 22.77 &&  4.91 & 2.03 && $22.00\pm0.03$ & $-$12.74 &  1.85 &            \\
   20....... &&  00 11 06.288 & $-$12 06 24.65 &&  4.57 & 1.89 && $22.31\pm0.04$ & $-$12.43 &  1.82 &            \\
\enddata
\tablecomments{Units of right ascension are hours, minutes, and seconds, and units of
    declination are degrees, arcminutes, and arcseconds.}
\tablenotetext{a}{Projected distance from nucleus at R.A.(2000) = 00 11 06.536, Decl.(2000) = $-$12 06 27.42.}
\tablenotetext{b}{Approximate apparent $V$ magnitude on Johnson system, transformed from F606W system
    as described in text.}
\tablenotetext{c}{Absolute visual magnitude, corrected for Milky Way extinction of
    $A_V=0.089$ and computed for $D = 85.2$ Mpc ($H_0=70$ km s$^{-1}$\,Mpc$^{-1}$).}
\tablenotetext{d}{Concentration index, which is the difference between
magnitudes within apertures of 0.5 pix and 3 pix radius.}
\tablenotetext{e}{Sp: Spectrum was obtained.}
\tablenotetext{f}{Cluster 1: $\bvzero = 0.07\pm0.04$, $\vizero = 0.61\pm0.03$\,.}
\tablenotetext{g}{Cluster 2: $\bvzero = 0.14\pm0.10$, $\vizero = 0.71\pm0.13$\,.}
\end{deluxetable}

%%%%%%%%%%%%%%%%%%%%%%%  Table 5  %%%%%%%%%%%%%%%%%%%%%%%%
%\clearpage

%  n0034_lc02_tab050.latex

\begin{deluxetable}{lcccccccc}
%%\tablenum{5}
%%\rotate
\tablecolumns{9}
\tablewidth{15cm}
\tabletypesize{\footnotesize}
\tablecaption{Model Fits to \bvi\ Surface-Brightness Profiles of \n34\label{tab05}}
\tablehead{
  \colhead{}               &
  \colhead{}               & 
  \multicolumn{2}{c}{Model Spheorid} &
  \colhead{}               & 
  \multicolumn{3}{c}{Model Exponential Disk} &
  \colhead{}
                             \\[4pt]
  \cline{3-4}
  \cline{6-8}
                             \\[-7pt]
  \colhead{}               &
  \colhead{}               & 
  \colhead{\mueff}         &
  \colhead{\reff}          &  
  \colhead{}               & 
  \colhead{\muzero}        &
  \colhead{$\alpha$}       & 
  \colhead{\rhole}         &
  \colhead{rms Residual}
                             \\
  \colhead{\ \ Fitted Model\ \ \ \ }   &
  \colhead{Passband}       &
  \colhead{(\magarcs)}     &
  \colhead{(arcsec)}       &
  \colhead{}               & 
  \colhead{(\magarcs)}     &
  \colhead{(arcsec)}       &
  \colhead{(arcsec)}       &
  \colhead{(\magarcs)}
}
\startdata
Pure \rdV-law\tablenotemark{a}\dotfill &
   $B$ &  $20.938\pm0.057$ &  $5.88\pm0.50$ &&  \nodata &           \nodata &        \nodata &        0.316 \\
&  $V$ &  $20.231\pm0.047$ &  $5.57\pm0.38$ &&  \nodata &           \nodata &        \nodata &        0.260 \\
&  $I$ &  $19.180\pm0.036$ &  $5.57\pm0.29$ &&  \nodata &           \nodata &        \nodata &        0.195 \\
\sphed & &&&&&&& \\
without hole\tablenotemark{b}\dotfill &
   $B$ &  $22.408\pm0.049$ &  $8.36\pm0.20$ &&  $19.551\pm0.021$ &  $4.07\pm0.04$ &  \nodata &        0.072 \\
&  $V$ &  $21.452\pm0.053$ &  $7.51\pm0.21$ &&  $19.017\pm0.027$ &  $3.80\pm0.05$ &  \nodata &        0.070 \\
&  $I$ &  $19.968\pm0.038$ &  $6.94\pm0.13$ &&  $18.557\pm0.040$ &  $3.96\pm0.08$ &  \nodata &        0.093 \\
\sphed & &&&&&&& \\
with hole\tablenotemark{c}\dotfill &
   $B$ &  $21.509\pm0.026$ &  $6.36\pm0.09$ &&  $19.551\pm0.035$ &  $3.79\pm0.06$ &  $2.56\pm0.08$ &  0.064 \\
&  $V$ &  $20.764\pm0.033$ &  $6.20\pm0.12$ &&  $18.946\pm0.057$ &  $3.41\pm0.08$ &  $2.66\pm0.11$ &  0.066 \\
&  $I$ &  $19.632\pm0.027$ &  $6.29\pm0.09$ &&  $18.521\pm0.072$ &  $3.65\pm0.12$ &  $2.53\pm0.17$ &  0.089 \\
\enddata
\tablenotetext{a}{$\mu (r) = \mueff + 8.325 [(r/\reff)^{1/4} - 1)$\,.}
\tablenotetext{b}{$\mu (r) = \mueff + 8.325 [(r/\reff)^{1/4} - 1)\ +\ \muzero + 1.086 (r/\alpha)$\,.}
\tablenotetext{c}{$\mu (r) = \mueff + 8.325 [(r/\reff)^{1/4} - 1)\ +\ \muzero + 1.086 [(r/\alpha)+(\rhole/r)^3]$\,.}
\end{deluxetable}

%%%%%%%%%%%%%%%%%%%%%%%  Table 6  %%%%%%%%%%%%%%%%%%%%%%%%
\clearpage

%  n0034_lc02_tab060.latex

\begin{deluxetable}{lcccccc}
%%\tablenum{6}
\tablecolumns{5}
\tablewidth{7cm}
\tablecaption{Cluster Radial Velocities\label{tab06}}
\tablehead{
  \colhead{}          &
  \colhead{$V$}       &
  \colhead{}          &
  \colhead{\czhel}    &
  \colhead{\dvlos\tablenotemark{b}}
                         \\
  \colhead{Cluster}   &
  \colhead{(mag)}     &
  \colhead{$N_{\rm abs}$\tablenotemark{a}} &
  \colhead{(\kms)}    &
  \colhead{(\kms)}
                          \\
  \colhead{(1)}       &
  \colhead{(2)}       &
  \colhead{(3)}       &
  \colhead{(4)}       &
  \colhead{(5)}
}
\startdata
\phn1........ &  19.38 &    12 & $5783\pm 16$ & \phn$-85\pm 22$ \\
\phn2........ &  20.04 &    11 & $5850\pm 12$ & \phn$-20\pm 19$ \\
\phn3........ &  20.47 & \phn8 & $5952\pm 27$ & \phn$+80\pm 31$ \\
\phn7........ &  20.80 & \phn5 & $5656\pm 40$ &    $-210\pm 43$ \\
   13........ &  21.24 & \phn6 & $5951\pm 23$ & \phn$+79\pm 28$ \\
\enddata
\tablenotetext{a}{Number of absorption lines measured.}
\tablenotetext{b}{{L}ine-of-sight velocity relative to nucleus,\ \
	\dvlos\ = (\czhel$-5870)/1.019580$\ \ (see text).}
\end{deluxetable}

%%%%%%%%%%%%%%%%%%%%%%%  Table 7  %%%%%%%%%%%%%%%%%%%%%%%%
%\clearpage

% n0034_lc02_tab070.latex

\begin{deluxetable}{lccccccccc}
%%\tablenum{7}
\tablecolumns{10}
\tablewidth{15cm}
\tabletypesize{\footnotesize}
\tablecaption{Lick Line Indices and Cluster Ages\label{tab07}}
\tablehead{
  \colhead{}			&
  \colhead{\hdela}		& 
  \colhead{\hgama}		&
  \colhead{\hbet}		&  
  \colhead{Fe5015}		&
  \colhead{\mgb}		& 
  \colhead{Fe5270}		&
  \colhead{Fe5335}		&
  \colhead{\mgfe}		&
  \colhead{}
				  \\
  \colhead{Object}		&
  \colhead{(\AA)}		&
  \colhead{(\AA)}		&
  \colhead{(\AA)}		&
  \colhead{(\AA)}		&
  \colhead{(\AA)}		& 
  \colhead{(\AA)}		&
  \colhead{(\AA)}		&
  \colhead{(\AA)}		&
  \colhead{$\log \tau$\tablenotemark{a}}
}
\startdata
Cluster 1.....&
	$8.2\pm 0.3$ & $7.0\pm 0.3$ & $5.3\pm 0.1$   & $1.5\pm 0.2$ & $0.5\pm 0.2$ &
	$0.4\pm 0.2$ & $0.5\pm 0.2$ & $0.51\pm 0.39$ & $8.16\pm 0.05$ \\
Cluster 2\tablenotemark{b}...&
	$8.8\pm 0.5$ & $8.0\pm 0.4$ & $5.3\pm 0.4$   & $4.4\pm 0.4$ & $1.0\pm 0.4$ &
	$1.3\pm 0.3$ & $1.8\pm 0.3$ & $1.22\pm 0.16$ & $8.80\pm 0.03$\tablenotemark{c} \\
Nucleus\tablenotemark{b}.....&
	$6.6\pm 0.4$ & $4.6\pm 0.3$ & $2.9\pm 0.2$   & $3.0\pm 0.3$ & $0.6\pm 0.1$ &
	$0.8\pm 0.2$ & $0.3\pm 0.2$ & $0.59\pm 0.24$ &    \nodata \\
\enddata
\tablenotetext{a}{Cluster age $\tau$ expressed in years.}
\tablenotetext{b}{Indices measured after clipping emission lines from
                  spectrum.} 
\tablenotetext{c}{One of two possible values, the other being $8.21\pm 0.07$;
                  for details, see \S~\ref{sec354}.}
\end{deluxetable}

%%%%%%%%%%%%%%%%%%%%%%%  Table 8  %%%%%%%%%%%%%%%%%%%%%%%%
\clearpage

% n0034_lc02_tab090.latex

\begin{deluxetable}{lll}
\tabletypesize{\small}
%%\tablenum{8}
\tablecolumns{3}
\tablewidth{11cm}
\tablecaption{Parameters of NGC 34 (Mrk 938)\label{tab08}}
\tablehead{
\colhead{Parameter\phm{AAA}}  &
\colhead{Symbol\phm{AAA}}     &
\colhead{Value\phm{AAAAAA}}
}
\startdata
Right ascension\tnm{a} &	$\alpha$(J2000) &	$00^{\rm h}11^{\rm m}06\fs54$    \\
Declination\tnm{a} &		$\delta$(J2000) &	$-12\degr\,06\arcmin\,27\farcs4$ \\
Heliocentric velocity of nucleus &      \czhel  &	$5870\pm 15$ \kms	\\
Velocity relative to Local Group &
				\czlg &			5961 \kms		\\
Distance\tnm{b} & 		$\Delta$ &		85.2 Mpc		\\
Distance modulus (true)\tnm{b} &     $(m-M)_0$ &	34.65 mag		\\
Projected scale\tnm{b} &	$s$ &			413 pc arcsec$^{-1}$	\\
Isophotal major diameter at $B=25$ \magarcstab & 
				$D_{25}$ &		$\ga$\,107\arcsec\,= 44 kpc \\
isophotal minor diameter at $B=25$ \magarcstab &
				$d_{25}$ &		$47\farcs3 = 19.5$ kpc	\\
Total apparent blue magnitude\tnm{c} &
				\btot &			$13.75\pm 0.01$ mag	\\
Total apparent visual magnitude\tnm{c} &
				\vtot &			$13.17\pm 0.01$ mag	\\
Total apparent $I$ magnitude\tnm{c} &     \itot &	$12.08\pm 0.01$ mag	\\
Total apparent $K$ magnitude\tnm{c,d} &   \ktot &	$9.94\pm 0.02$ mag	\\
Milky Way foreground extinction\tnm{e} &  $A_V$ &	0.089 mag		\\
Absolute blue magnitude\tnm{b} &	  $M_B$ &	$-$21.02 mag		\\
Absolute visual magnitude\tnm{b} &	  $M_V$ &	$-$21.57 mag		\\
Absolute $K$ magnitude\tnm{b} &		  $M_K$ &	$-$24.72 mag		\\
Color index\tnm{f} &		\bvtotzero &		$0.55\pm 0.02$ mag	\\
Color index\tnm{f} &		\vitotzero &		$1.05\pm 0.01$ mag	\\
Color index\tnm{f} &		\vktotzero &		$3.15\pm 0.03$ mag	\\
Infrared luminosity ($\lambda\lambda$\,10--1300 \micron)\tnm{g} &
				 $\log(\lir/\lsun)$ &	11.54			\\
Mass of neutral hydrogen gas\tnm{b,h} &
				\mhi &			$5.3\times 10^9 \,\msun$ \\
Mass of molecular gas (from CO)\tnm{b,i} &
				\mhtwo &		$(7\pm 3)\times 10^9 \,\msun$ \\
Number of galaxies in \n34 group & $N_{\rm gal}$ &	$\geq$3			\\[4pt]
Main body: &				   &					\\
\ \ Apparent central surface brightness & $V_0$ &	14.44 \magarcstab	\\
\ \ Central velocity dispersion of stars\tnm{j} &
				\sigvel &		$201\pm 8$ \kms		\\
\ \ Effective radius in $V$ passband &	  \reff &	$6\farcs15\pm 0\farcs10 =
							   2.54\pm 0.04$ kpc	\\
\ \ Surface brightness at \reff &
				\veff\ (\beff ) &	20.08 (20.60) \magarcstab \\
\ \ Bulge-to-disk ratio (blue)\tnm{k} &   (B/D)$_B$ &	0.94			\\
\ \ Bulge-to-disk ratio (visual)\tnm{k} & (B/D)$_V$ &	1.32			\\
\ \ Bulge-to-disk ratio ($I$-band)\tnm{k} & (B/D)$_I$ &	2.7			\\[4pt]
Central exponential disk: &	           &					\\
\ \ Apparent semi-major axis &	$a$ &			$\sim$8$\arcsec\approx 3.3$ kpc \\
\ \ Major-axis position angle &	P.A. &			351$\degr$		\\
\ \ Apparent axis ratio &	$b/a$ &			0.72			\\
\ \ Inclination (approx.) &	$i$ &			44$\degr$		\\
\ \ Approximate scale length\tnm{k} &
				$\alpha$ &		$4\farcs0\approx 1.6$ kpc \\
\ \ Maximum extent\tnm{k} &	$r_{\rm max,eD}$ &	$\ga\,25\arcsec\approx 10$ kpc \\
\ \ Absolute visual magnitude\tnm{k} &
				$M_{V,0}$ &		$-$20.61 mag		\\
\ \ Color index\tnm{f,k} &	\bvedzero &		$0.36\pm 0.02$ mag	\\
\ \ Color index\tnm{f,k} &	\viedzero &		$0.52\pm 0.02$ mag	\\
\ \ Estimated mass (if disk 400 Myr old) &
				$M_{\rm eD}$ &		(2--4)$\times 10^9\,\msun$
										\\[4pt]
Cluster system: &			    &					\\
\ \ Number of star clusters of $M_V\la -9.3$&
				\nclust &		$\sim$140		\\
\ \ Effective radius of cluster system &
				\reffcl &		$2.8 \pm 0.3$ kpc	\\
\ \ Power-law exponent of luminosity function &
				$\alpha$ &		$-1.75\pm 0.1$		\\
\ \ Luminosities of 20 brightest clusters &
				$M_V$ &			$-$12.4 to $-$15.4 mag	\\
\ \ Likely ages of two brightest clusters &
				$\tau$ &		150, 640 (or 160)  Myr	\\[4pt]
Outflow: &			           &					\\
\ \ Mean outflow velocity of \ion{Na}{1} &
				$\langle\vnad\rangle$ &	$-620\pm 60$ \kms	\\
\ \ Maximum outflow velocity of \ion{Na}{1} &
				max(\vnad) &		$-1050\pm 30$ \kms	\\
\ \ Mean velocity shift of emission lines &
				$\Delta v_{\rm em}$ &	$-75\pm 23$ \kms	\\[4pt]
Tidal tails: &				   &					\\
\ \ Projected length of N tail\tnm{b} &
				\rmax (N tail) &	$92\arcsec = 38$ kpc	\\
\ \ Projected length of S tail\tnm{b} &
				\rmax(S tail) &	$63\arcsec = 26$ kpc	\\
\enddata
\tnt{a} {Position of nucleus measured from {\it HST}/WFPC2\ \,$V$ image.}
\tnt{b} {For $H_0 = 70$ km s$^{-1}$ Mpc$^{-1}$.}
\tnt{c} {Within $B$~= 26.5 \magarcstab\ isophote.}
\tnt{d} {From \citet{rj04}, corrected to adopted $B$-isophote.}
\tnt{e} {\citet{schl98}.}
\tnt{f} {Corrected for Milky Way foreground reddening, but not internal reddening.}
\tnt{g} {\citet{chin92}.}
\tnt{h} {From compilation by \citet{kand03}.}
\tnt{i} {From \citet{kand03}, \citet{krug90}, and \citet{chin92}.}
\tnt{j} {\citet{rj06a}.}
\tnt{k} {For disk without central hole; see \S~\ref{sec32} and Table~\ref{tab05} for details.}
\end{deluxetable}

\clearpage


\begin{thebibliography}{}

\bibitem[Afanasev et al.(1980)]{afan80} Afanasev, V.\ L., Lipovetskii, V.\ A.,
   Markarian, B.\ E., \& Stepanian, J.\ A. 1980, Astrofizika, 16, 193
   (English transl.\ Astrophysics, 16, 119)

\bibitem[Allington-Smith et al.(1994)]{alli94} Allington-Smith, J., et al.
   1994, \pasp, 106, 983

\bibitem[Argyle \& Eldridge(1990)]{argy90} Argyle, R.\ W., \& Eldridge, P.
   1990, \mnras, 243, 504

\bibitem[Baggett et al.(1998)]{bagg98} Baggett, W.\ E., Baggett, S.\ M., \&
   Anderson, K.\ S.\ J. 1998, \aj, 116, 1626

\bibitem[Barnes(1998)]{barn98} Barnes, J.\ E. 1998, in Galaxies: Interactions
   and Induced Star Formation, ed. D.\ Friedli, L.\ Martinet, \& D.\
   Pfenniger (Berlin: Springer), 275

\bibitem[Barnes(2002)]{barn02} Barnes, J.\ E. 2002, \mnras, 333, 481

\bibitem[Barnes \& Hernquist(1992)]{bh92} Barnes, J.\ E., \& Hernquist, L.
   1992, \araa, 30, 705

\bibitem[Barnes \& Hernquist(1996)]{bh96} Barnes, J.\ E., \& Hernquist, L.
   1996, \apj, 471, 115

\bibitem[Baumgardt(2006)]{baum06} Baumgardt, H. 2006, in Globular
   Clusters---Guides to Galaxies, ed. T.\ Richtler \& S.\ S.\ Larsen
   (Berlin: Springer), in press (astro-ph/0605125)

\bibitem[Boily \& Kroupa(2003)]{boil03} Boily, C.\ M., \& Kroupa, P. 2003,
   \mnras, 338, 673

\bibitem[Bottinelli et al.(1990)]{bott90} Bottinelli, L., Gouguenheim, L.,
   Fouque, P., \& Paturel, G. 1990, \aaps, 82, 391

\bibitem[Bruzual \& Charlot(2003)]{bc03} Bruzual A., G., \& Charlot, S. 2003,
   \mnras, 344, 1000 (BC03)

\bibitem[Buchanan et al.(2006)]{buch06} Buchanan, C.\ L., et al. 2006, \aj,
   132, 401

\bibitem[Busko \& Steiner(1990)]{busk90} Busko, I.\ C., \& Steiner, J.\ E.
   1990, \mnras, 245, 470 

\bibitem[Carlson et al.(1999)]{carl99} Carlson, M.\ N., \etal\ 1999, \aj,
   117, 1700

\bibitem[Chabrier(2003)]{chab03} Chabrier, G. 2003, \pasp, 115, 763

\bibitem[Chini et al.(1992)]{chin92} Chini, R., Kr\"ugel, E., \& Steppe, H.
   1992, \aap, 255, 87

\bibitem[Condon et al.(1991)]{cond91} Condon, J.\ J., Huang, Z.-P., Yin,
   Q.\ F., \& Thuan, T.\ X. 1991, \apj, 378, 65

\bibitem[Corbett et al.(2003)]{corb03} Corbett, E.\ A., et al. 2003, \apj,
   583, 670

\bibitem[Cox et al.(2006)]{cox06} Cox, T.\ J., Di Matteo, T., Hernquist, L.,
   Hopkins, P.\ F., Robertson, B., \& Springel, V. 2006, \apj, 643, 692

\bibitem[da Costa et al.(1998)]{daco98} da Costa, L.\ N., et al. 1998, \aj,
   116, 1

\bibitem[Dahari(1985)]{daha85} Dahari, O. 1985, \apjs, 57, 643

\bibitem[de Vaucouleurs(1953)]{deva53} de Vaucouleurs, G. 1953, \mnras, 113,
   134

\bibitem[de Vaucouleurs et al.(1991)]{rc3} de Vaucouleurs, G., de Vaucouleurs,
   A., Corwin, H.\ G., Buta, R.\ J., Paturel, G., \& Fouqu\'e, P. 1991, Third
   Reference Catalogue of Bright Galaxies (Berlin: Springer)

\bibitem[Faber et al.(1985)]{fabe85} Faber, S.\ M., Friel, E.\ D., Burstein,
   D., \& Gaskell, C.\ M. 1985, \apjs, 57, 711

\bibitem[Fall \& Zhang(2001)]{fall01} Fall, S.\ M., \& Zhang, Q. 2001, \apj,
   561, 751

\bibitem[Ferrarese \& Merritt(2000)]{ferr00} Ferrarese, L., \& Merritt, D.
   2000, \apj, 539, L9

\bibitem[Gebhardt et al.(2000)]{gebh00} Gebhardt, K., et al. 2000, \apj, 539,
   L13

\bibitem[Goldader et al.(1997a)]{gold97a} Goldader, J.\ D., Joseph, R.\ D.,
   Doyon, R., \& Sanders, D.\ B. 1997a, \apjs, 108, 449

\bibitem[Goldader et al.(1997b)]{gold97b} Goldader, J.\ D., Joseph, R.\ D.,
   Doyon, R., \& Sanders, D.\ B. 1997b, \apj, 474, 104

\bibitem[Gon\c calves et al.(1999)]{gonc99} Gon\c calves, A.\ C.,
   V\'eron-Cetty, M.-P., \& V\'eron, P. 1999, \aaps, 135, 437

\bibitem[Gonz\'alez(1993)]{gonz93} Gonz\'alez, J.\ J. 1993, Ph.\ D.\ thesis,
   UC Santa Cruz

\bibitem[Guainazzi et al.(2005)]{guai05} Guainazzi, M., Matt, G., \&
   Perola, G.\ C. 2005, \aap, 444, 119

\bibitem[Hamuy et al.(1992)]{hamu92} Hamuy, M., Walker, A.\ R., Suntzeff,
   N.\ B., Gigoux, P., Heathcote, S.\ R., \& Phillips, M.\ M. 1992, \pasp,
   104, 533

\bibitem[Heckman et al.(2000)]{heck00} Heckman, T.\ M., Lehnert, M.\ D.,
   Strickland, D.\ K., \& Armus, L. 2000, \apjs, 129, 493

\bibitem[Hibbard \& Mihos(1995)]{himi95} Hibbard, J.\ E., \& Mihos, J.\ C.
   1995, \aj, 110, 140

\bibitem[Hibbard \& van Gorkom(1996)]{hibb96} Hibbard, J.\ E., \& van
   Gorkom, J.\ H. 1996, \aj, 111, 655

\bibitem[Holtzman et al.(1995)]{holt95} Holtzman, J.\ A., \etal\ 1995, \pasp,
   107, 1065

\bibitem[Holtzman et al.(1996)]{holt96} Holtzman, J.\ A., \etal\ 1996, \aj,
   112, 416

\bibitem[Hopkins et al.(2006)]{hopk06} Hopkins, P.\ F., Hernquist, L., Cox,
   T.\ J., Di Matteo, T., Robertson, B., \& Springel, V. 2006, \apjs, 163, 1

\bibitem[Imanishi \& Alonso-Herrero(2004)]{iman04} Imanishi, M., \&
   Alonso-Herrero, A. 2004, \apj, 614, 122

\bibitem[Kandalyan(2003)]{kand03} Kandalyan, R.A. 2003, \aap, 398, 493

\bibitem[Keto et al.(1991)]{keto91} Keto, E., Jernigan, G., Ball, R.,
   Arens, J., \& Meixner, M. 1991, \apj, 374, L29

\bibitem[Kormendy(1977)]{korm77} Kormendy, J. 1977, \apj, 217, 406

\bibitem[Korn et al.(2005)]{korn05} Korn, A.\ J., Maraston, C., \&
   Thomas, D. 2005, \aap, 438, 685

\bibitem[Kroupa(2001)]{krou01} Kroupa, P. 2001, \mnras, 322, 231

\bibitem[Kr\"ugel et al.(1990)]{krug90} Kr\"ugel, E., Steppe, H., \& Chini, R.
   1990, \aap, 229, 17

\bibitem[Landolt(1973)]{land73} Landolt, A.\ U. 1973, \aj, 78, 959

\bibitem[Landolt(1992)]{land92} Landolt, A.\ U. 1992, \aj, 104, 340

\bibitem[Larson \& Tinsley(1978)]{lt78} Larson, R.\ B., \& Tinsley, B.\ M.
   1978, apj, 219, 46

\bibitem[Malkan et al.(1998)]{malk98} Malkan, M.\ A., Gorjian, V., \&
   Tam, R. 1998, \apjs, 117, 25

\bibitem[Maraston(2005)]{mara05} Maraston, C. 2005, \mnras, 362, 799

\bibitem[Maraston et al.(2004)]{mara04} Maraston, C., Bastian, N., 
   Saglia, R.\ P., Kissler-Patig, M., Schweizer, F., \& Goudfrooij, P.
   2004, \aap, 416, 467

\bibitem[Maraston et al.(2001)]{mara01} Maraston, C., Kissler-Patig, M.,
   Brodie, J.\ P., Barmby, P., \& Huchra, J.\ P. 2001, \aap, 370, 176

\bibitem[Martin(2005a)]{mart05a} Martin, C.\ L. 2005a, in ASP Conf.\ Ser.\
   331, Extra-Planar Gas, ed.\ R.\ Brown (San Francisco: ASP), 305

\bibitem[Martin(2005b)]{mart05b} Martin, C.\ L. 2005b, \apj, 621, 227

\bibitem[Mazzarella \& Boroson(1993)]{mazz93} Mazzarella, J.\ M.,\&
   Boroson, T.\ A. 1993, \apjs, 85, 27

\bibitem[Mazzarella et al.(1991)]{mazz91} Mazzarella, J.\ M., Bothun, G.\ D.,
   \& Boroson, T.\ A. 1991, \aj, 101, 2034

\bibitem[Meylan(2002)]{meyl02} Meylan, G. 2002, in IAU Symp.\ 207,
   Extragalactic Star Clusters, ed.\ D.\ Geisler, E.\ K.\ Grebel, \&
   D.\ Minniti (San Francisco: ASP), 555

\bibitem[Mighell \& Rich(1995)]{migh95} Mighell, K.\ J., \& Rich, R.\ M. 1995,
   \aj, 110, 1649

\bibitem[Mihos(2001)]{miho01} Mihos, C. 2001, in ASP Conf.\ Ser.\ 240,
   Gas and Galaxy Evolution, ed.\ J.\ E.\ Hibbard, M.\ P.\ Rupen, \&
   J.\ H.\ van Gorkom (San Francisco: ASP), 143

\bibitem[Mihos \& Hernquist(1996)]{mh96} Mihos, J.\ C., \& Hernquist, L. 1996,
   \apj, 464, 641

\bibitem[Miles et al.(1996)]{mile96} Miles, J.\ W., Houck, J.\ R.,
   Hayward, T.\ L., \& Ashby, M.\ L.\ N. 1996, \apj, 465, 191

\bibitem[Miller et al.(1997)]{mill97} Miller, B.\ W., Whitmore, B.\ C.,
   Schweizer, F., \& Fall, M. 1997, \aj, 114, 2381

\bibitem[Mulchaey et al.(1996)]{mulc96} Mulchaey, J.\ S., Wilson, A.\ S.,
   \& Tsvetanov, Z. 1996, \apjs, 102, 309

\bibitem[Naab \& Burkert(2003)]{naab03} Naab, T., \& Burkert, A. 2003,
   \apj, 597, 893

\bibitem[Naab et al.(2006)]{naab06} Naab, T., Jesseit, R., \& Burkert, A.
   2006, \mnras, 372, 839

\bibitem[Osterbrock \& Dahari(1983)]{oste83} Osterbrock, D.\ E., \& Dahari, O.
   1983, \apj, 273, 478

\bibitem[Prouton et al.(2004)]{prou04} Prouton, O.\ R., Bressan, A., Clemens,
   M., Franceschini, A., Granato, G.\ L., \& Silva, L. 2004, \aap, 421, 115

\bibitem[Pryor \& Meylan(1993)]{pryo93} Pryor, C., \& Meylan, G. 1993, in
   ASP Conf.\ Ser.\ 50, Structure and Dynamics of Globular Clusters, 
   ed.\ S.\ G.\ Djorgovski \& G.\ Meylan (San Francisco: ASP), 357

\bibitem[Riffel et al.(2006)]{riff06} Riffel, R., Rodr\'iguez-Ardila, A.,
    \& Pastoriza, M.\ G. 2006, \aap, 457, 61

\bibitem[Rothberg \& Joseph(2004)]{rj04} Rothberg, B., \& Joseph, R.\ D. 2004,
   \aj, 128, 2098

\bibitem[Rothberg \& Joseph(2006a)]{rj06a} Rothberg, B., \& Joseph, R.\ D.
   2006a, \aj, 131, 185

\bibitem[Rothberg \& Joseph(2006b)]{rj06b} Rothberg, B., \& Joseph, R.\ D.
   2006b, \aj, 132, 976

\bibitem[Rupke et al.(2005)]{rupk05} Rupke, D.\ S., Veilleux, S., \&
   Sanders, D.\ B. 2005, \apj, 632, 751

\bibitem[Sandage \& Tammann(1975)]{sand75} Sandage, A., \& Tammann, G.A. 1975,
   \apj, 197, 265

\bibitem[Schawinski et al.(2006)]{scha06} Schawinski, K., et al. 2006,
   \nat, 442, 888

\bibitem[Schlegel, Finkbeiner, \& Davis(1998)]{schl98} Schlegel, D.\ J.,
   Finkbeiner, D.\ P., \& Davis, M. 1998, \apj, 500, 525

\bibitem[Schweizer(1982)]{schw82} Schweizer, F. 1982, \apj, 252, 455

\bibitem[Schweizer(1983)]{schw83} Schweizer, F. 1983, in IAU Symp.\ 100,
   Internal Kinematics and Dynamics of Galaxies, ed.\ E.\ Athanassoula
   (Dordrecht: Reidel), 319

\bibitem[Schweizer(1996)]{schw96a} Schweizer, F. 1996, \aj, 111, 109

\bibitem[Schweizer(2002)]{schw02} Schweizer, F. 2002, in IAU Symp.\ 207,
   Extragalactic Star Clusters, ed.\ D.\ Geisler, E.\ K.\ Grebel, \&
   D.\ Minniti (San Francisco: ASP), 630

\bibitem[Schweizer(2003)]{schw03} Schweizer, F. 2003, in ASP Conf.\ Ser.\
   296, New Horizons in Globular Cluster Astronomy, ed.\ G.\ Piotti et al.\
   (San Francisco: ASP), 467

\bibitem[Schweizer et al.(1996)]{schw96b} Schweizer, F., Miller, B.\ W.,
   Whitmore, B.\ C., \& Fall, S.\ M. 1996, \aj, 112, 1839

\bibitem[Schweizer \& Seitzer(1998)]{ss98} Schweizer, F., \& Seitzer, P.
   1998, \aj, 116, 2206

\bibitem[Schweizer et al.(2004)]{schw04} Schweizer, F., Seitzer, P., \&
   Brodie, J.\ P. 2004, \aj, 128, 202

\bibitem[Soifer et al.(1987)]{soif87} Soifer, B.\ T., et al. 1987, \apj,
   320, 238

\bibitem[Springel et al.(2005)]{spri05} Springel, V., Di Matteo, T., \&
   Hernquist, L. 2005, \mnras, 361, 776

\bibitem[Stetson(1987)]{stet87} Stetson, P.\ B. 1987, \pasp, 99, 191

\bibitem[Stetson(2000)]{stet00} Stetson, P.\ B. 2000, \pasp, 112, 925

\bibitem[Thomas et al.(2003)]{tmb03} Thomas, D., Maraston, C., \& Bender,
   R. 2003, \mnras, 339, 897

\bibitem[Toomre \& Toomre(1972)]{tt72} Toomre, A., \& Toomre, J. 1972, \apj,
   178, 623

\bibitem[Trager et al.(1998)]{trag98} Trager, S.\ C., Worthey, G., Faber,
   S.\ M., Burstein, D., \& Gonz\'alez, J.\ J. 1998, \apjs, 116, 1

\bibitem[Tremaine et al.(2002)]{trem02} Tremaine, S., et al. 2002, \apj,
   574, 740

\bibitem[Valdes et al.(2005)]{vald05} Vald\`es, J.\ R., Berta, S., Bressan,
   A., Franceschini, A., Rigopoulou, D., \& Rodighiero, G. 2005, \aap, 434,
   149

\bibitem[van der Marel \& Zurek(2000)]{vdma00} van der Marel, R.\ P., \&
   Zurek, D. 2000, in ASP Conf.\ Ser.\ 197, Dynamics of Galaxies: From the
   Early Universe to the Present, ed.\ F.\ Combes, G.\ A.\ Mamon, \&
   V.\ Charmandaris (San Francisco: ASP), 323

\bibitem[van Dokkum(2001)]{vdok01} van Dokkum, P.\ G. 2001, \pasp, 113, 1420

\bibitem[Veilleux et al.(2005)]{veil05} Veilleux, S., Cecil, G., \&
   Bland-Hawthorn, J. 2005, \araa, 43, 769

\bibitem[Veilleux \& Osterbrock(1987)]{veil87} Veilleux, S., \&
   Osterbrock, D.\ E. 1987, \apjs, 63, 295

\bibitem[V\'eron-Cetty \& V\'eron(1986)]{vero86} V\'eron-Cetty, M.-P., \&
   V\'eron, P. 1986, \aaps, 65, 241

\bibitem[Voit(1998)]{voit98} Voit, M. 1998, in {\em HST} Data Handbook
   (Baltimore: STScI), Vol. 1, Version 3.1, Table 28.1

\bibitem[White \& Rees(1978)]{white78} White, S.\ D.\ M., \& Rees, M.\ J.
   1978, \mnras, 183, 341

\bibitem[Whitmore(2003)]{whit03} Whitmore, B.\ C. 2003, in A Decade of
   {\em Hubble Space Telescope} Science, ed.\ M.\ Livio, K.\ Noll, \&
   M.\ Stiavelli (Cambridge: Cambridge Univ.\ Press), 153

\bibitem[Whitmore et al.(2002)]{whit02} Whitmore, B.\ C., Schweizer, F.,
   Kundu, A., \& Miller, B.\ W. 2002, \aj, 124, 147

\bibitem[Whitmore et al.(1993)]{whit93} Whitmore, B.\ C., Schweizer, F.,
   Leitherer, C., Borne, K., \& Robert, C. 1993, \aj, 106, 1354

\bibitem[Worthey \& Ottaviani(1997)]{wort97} Worthey, G., \& Ottaviani, D.\
   L. 1997, \apjs, 111, 377

\end{thebibliography}
\end{document}